\documentclass[preprint,review,12pt,authoryear]{elsarticle}

\usepackage{amssymb}
\usepackage{booktabs}
\usepackage[export]{adjustbox}

\usepackage{lineno}

\usepackage{graphicx}
\usepackage{multirow}
\usepackage[dvipsnames]{xcolor}
\usepackage{balance}
\usepackage{bm}
\usepackage[hyphens]{url}
\usepackage[font=footnotesize,labelfont=bf]{caption}
\usepackage{array}
\usepackage{amsmath}
\usepackage{placeins}
\usepackage{caption}
\usepackage{subcaption}

\journal{Expert Systems with Applications}

\usepackage{etoolbox}
\patchcmd{\pprintMaketitle}
 {\ifvoid\absbox\else\unvbox\absbox\par\vskip10pt\fi}
 {\ifvoid\absbox\else\clearpage\unvbox\absbox\par\vskip30pt\fi}
 {}{}
\patchcmd{\pprintMaketitle}
 {\hrule\vskip12pt}
 {}
 {}{}
\patchcmd{\pprintMaketitle}
 {\hrule\vskip12pt}
 {}
 {}{}
\appto{\pprintMaketitle}{\clearpage}

\begin{document}

\begin{frontmatter}



\title{Using consumer feedback from location-based services in PoI recommender systems for people with autism}


\author{Noemi Mauro, Liliana Ardissono, Stefano Cocomazzi, Federica Cena}
\address{noemi.mauro@unito.it, liliana.ardissono@unito.it, stefano.cocomazzi@edu.unito.it, federica.cena@unito.it}
\address{Computer Science Department, University of Turin \\ Corso Svizzera, 185, I-10149, Turin,  Italy}

Published in Expert Systems with Applications, Elsevier. \\
DOI: \url{https://doi.org/10.1016/j.eswa.2022.116972}\\
\newline
Link to the page of the paper on Elsevier web site:
\\
\url{https://www.sciencedirect.com/science/article/abs/pii/S0957417422003980}
\\
\newline
This work is licensed under the Creative Commons Attribution-NonCommercial-NoDerivatives 4.0 International License. To view a copy of this license, visit
\\
http://creativecommons.org/licenses/by-nc-nd/4.0/ or send a letter to Creative Commons, PO Box 1866, Mountain View, CA 94042, USA.

\begin{abstract}
When suggesting Points of Interest (PoIs) to people with autism spectrum disorders, we must take into account that they have idiosyncratic sensory aversions to noise, brightness and other features that influence the way they perceive places. Therefore, recommender systems must deal with these aspects.
However, the retrieval of sensory data about PoIs is a real challenge because most geographical information servers fail to provide this data. Moreover, {\em ad-hoc} crowdsourcing campaigns do not guarantee to cover large geographical areas and lack sustainability. Thus, we investigate the extraction of sensory data about places from the consumer feedback collected by location-based services, on which people spontaneously post reviews from all over the world.
Specifically, we propose a model for the extraction of sensory data from the reviews about PoIs, and its integration in recommender systems to predict item ratings by considering both user preferences and compatibility information. 
We tested our approach with autistic and neurotypical people by integrating it into diverse recommendation algorithms. For the test, we used a dataset built in a crowdsourcing campaign and another one extracted from TripAdvisor reviews. The results show that the algorithms obtain the highest accuracy and ranking capability when using TripAdvisor data. Moreover, by jointly using these two datasets, the algorithms further improve their performance. 
These results encourage the use of consumer feedback as a reliable source of information about places in the development of inclusive recommender systems. 
\end{abstract}


\end{frontmatter}


\section{Introduction}
\label{sec:introduction}
Most personalized recommender systems consider the individual user's preferences and contextual conditions to select the Points of Interest (PoIs) that are suitable to the individual user \citep{Adomavicius-Tuzhilin:15}. However, when suggesting PoIs to people with Autism Spectrum Disorders (ASD), these systems should take into account that users have idiosyncratic sensory aversions to noise, brightness, and other features, which influence the way they perceive items, especially places \citep{robertson2013relationship}. Aversions should therefore be considered to suggest PoIs that are at the same time interesting and compatible with the target user. This is crucial because what bothers people with autism has great importance in their daily choices and can determine a high level of stress and anxiety \citep{simm2016anxiety}. 

\cite{Mauro-etal:20,Mauro-etal:22} propose to distinguish the role of user preferences and compatibility in PoI suggestion. The idea is to estimate the suitability of a place $p$ for a user $u$ by evaluating how much $u$ is expected to like $p$, and how compatible $p$ is with $u$, depending on $u$'s sensory aversions.
However, retrieving sensory data about PoIs is a real challenge because most geographical information servers, like OpenStreetMap\footnote{\url{https://www.openstreetmap.org/}} and Google Maps,\footnote{\url{https://www.google.it/maps/}} only provide data about properties of places such as their category, address and accessibility. Indeed, the crowdsourcing paradigm \citep{sui2012crowdsourcing}, where people actively provide information about places, can be used to gather the missing data. 
However, that approach covers limited geographical areas and requires a community willing to participate in the data collection, that is not simple to achieve.
Therefore, to identify a sustainable information source, we investigate the usefulness of the reviews available in services such as Yelp\footnote{\url{https://www.yelp.it/}} and TripAdvisor\footnote{\url{https://www.tripadvisor.com/}} to extract sensory data about places. 
Reviews report people's experience with items \citep{Ghose-Ipeirotis:11} and come as a by-product of the increasing usage of location-based services. However, to the best of our knowledge, they have always been employed to mine consumers' opinions about the quality of services and products, overlooking their potential to provide sensory data about items. Moreover, existing feature extraction approaches focus on the identification of the most frequent opinions while we have to adopt a pessimistic feature identification approach to guarantee that people with autism are not disturbed by sensory characteristics which might be rarely reported. 

In this work we propose a model to extract sensory data about places for inclusive recommendation and we pose two research questions: 

RQ1: {\em Does the feedback available in online item reviews collected by a location-based service provide useful sensory information about PoIs?} 

RQ2: {\em How does the sensory information extracted from reviews impact recommendation performance in the personalized suggestion of places?} 

To answer these questions, we developed a model for the extraction of sensory features from consumer feedback and we used it to build a dataset of sensory information about places from TripAdvisor reviews. The present paper describes this model and its integration within a recommender system by predicting the compatibility of sensory features with the user. This work also compares the performance achieved by different recommender systems when they employ crowdsourced data, our TripAdvisor dataset, or both to suggest items to two user groups: ASD people, and people who did not previously receive an autism diagnosis (we denote the latter as neurotypical).
The evaluation results show that, with both groups, consumer feedback supports higher recommendation performance than crowdsourced information. The accuracy (Precision, Recall, and F1) and ranking capability (MAP, MRR) of the algorithms is almost always higher when using TripAdvisor data. Moreover, accuracy, ranking capability, and rating prediction error (MAE, RMSE) decrease when jointly using the two datasets. Furthermore, the recommender systems that deal with both preferences and compatibility outperform those that only take preferences into account.
These results encourage the use of consumer feedback as a reliable source of information in PoI recommendation. They also show that it helps improving suggestions to both autistic and neurotypical people. This is relevant to the development of inclusive recommender systems and paves the way toward sustainable information acquisition models for PoI recommendation.

This work is framed in the PIUMA (Personalized Interactive Urban Maps for Autism)\footnote{PIUMA involves a collaboration among the Computer Science and Psychology Departments of the University of Torino and the Adult Autism Center of the City of Torino.} project, which has the aim to develop novel digital solutions to help people with autism spectrum disorders in their everyday movements \citep{DBLP:conf/chi/RappCBACBTKCB17}. 
Sections \ref{sec:perceptualNeeds} and \ref{sec:related} present the perceptual needs of autistic people and the related work. Section \ref{sec:data} describe the data collection and sensory feature extraction model. Section \ref{sec:model} outlines the recommendation algorithms we tested and Section \ref{sec:validation} describes the validation method we applied. Sections \ref{sec:results} and \ref{sec:discussion} present and discuss the experimental results. Section \ref{sec:limitations} describes limitations and future work, and Section \ref{sec:conclusions} concludes the paper.

\section{Sensory issues of people with autism}
\label{sec:perceptualNeeds}
People with autism spectrum disorders differ in terms of cognitive ability. However, almost all of them show substantial hypo and hypersensitivity to environmental stimuli (Sensory Processor Disorder \citep{matsushima2013,robertson2013relationship}). 
These stimuli can be auditory, olfactory, and tactile. The brain seems unable to appropriately balance the senses \citep{robertson2013relationship}. This means that people with autism appear to react differently to sensory stimulations. A majority of them may be overwhelmed by environmental features that are easily managed by neurotypical subjects. For example, many ASD people are hyper-sensitive to bright lights, or to certain light wavelengths, such as fluorescent lights. Several of them find some sounds, smells, and tastes overwhelming. Certain types of touch (light or deep) can cause uncomfortable feelings, as well. 
Thus, a person with autism might want to avoid places that negatively impact her/his senses \citep{robertson2013relationship}. 
These sensory aversions can cause negative feelings like anxiety, fatigue, sense of oppression  \citep{rapp2019spatial}. 
Due to these features, and to other peculiar characteristics, such as atypical social functioning, autistic people tend to have a reduced range of activities and are less likely to explore new environments \citep{smith2015spatial}.   
Therefore, they need a careful selection of places when moving in their city, or in a different area \citep{rapp2018designing}. It is crucial to find places that satisfy their sensory needs, focusing on aversions derived from their high sensitivity to sensory stimulation. 
The technology could be used to support them because they have a positive attitude towards it, due to the predictability of the interaction. However, most ICT-based solutions assist people in organizing their daily activities \citep{10.1145/1414471.1414475},  helping them in social interactions \citep{kientz2013interactive,grynszpan2014innovative},  and in emotion management \citep{simm2016anxiety,boyd2016saywat} but those solutions overlook space and sensory issues. 

Most services that aim at supporting people with autism in moving around are simple informative websites. Autistic Globetrotting\footnote{\url{http://autisticglobetrotting.com.}} and the Toerisme voor Autisme\footnote{\url{https://www.toerismevoorautisme.be/}} provide information about places that is useful to ASD people. Moreover, recent research highlights the beneﬁts of Virtual Reality interventions, such as computer-based simulations of reality where users can train specific skills needed to move around and travel, e.g., taking a bus \citep{bernardes2015serious}, or a plane \citep{DBLP:conf/eurovr/SocciniCC20}.
At the same time, each person with autism has unique sensitivities; thus, there is a high need to personalize solutions.  

\begin{table*}[t!]
\centering
\caption{Models and types of information used to personalize item suggestion. 
K-NN denotes K-Nearest Neighbors algorithm \citep{Desrosiers-Karypis:11}. MF is Matrix Factorization \citep{Koren-Bell:11}. CARS means Context-Aware Recommenders.}
\label{tab:models}
\resizebox{0.5\paperheight}{!}{%
{\def\arraystretch{1.7}
\begin{tabular}{lllll}
\toprule
Citations & Algorithm & Recommendation Model & Evaluation dimensions & 
\multicolumn{1}{l}{
 \renewcommand{\arraystretch}{0.5}
 \begin{tabular}[l]{@{}l@{}} \hspace{-2.4mm}
 Information Sources  \\ \hspace{-2.4mm}
 (other than  \\
  \hspace{-2.4mm}
 item ratings) \\
 \end{tabular}}\\ 
\midrule

\cite{Lops-etal:11} & CBF & vector distance & category, properties &  item descriptions\\


\multicolumn{1}{l}{
 \renewcommand{\arraystretch}{0.5}
 \begin{tabular}[l]{@{}l@{}} \hspace{-2.4mm}
 \cite{Desrosiers-Karypis:11} \\ \hspace{-2.4mm} \cite{Koren-Bell:11}\\
\end{tabular}}
& CF & K-NN, MF & items  & - \\

\multicolumn{1}{l}{
 \renewcommand{\arraystretch}{0.5}
 \begin{tabular}[l]{@{}l@{}} \hspace{-2.4mm}
 \cite{Adomavicius-Kwon:07} \\ \hspace{-2.4mm} \cite{Zheng:17}\\
 \hspace{-2.4mm} \cite{Jannach-etal:14}
\end{tabular}}
& Multi-Criteria & 
\multicolumn{1}{l}{
 \renewcommand{\arraystretch}{0.5}
 \begin{tabular}[l]{@{}l@{}} \hspace{-2.4mm}
K-NN or MF on \\ \hspace{-2.4mm}
multiple dimensions \\
 \end{tabular}}
& properties &  item metadata\\

\multicolumn{1}{l}{
 \renewcommand{\arraystretch}{0.5}
 \begin{tabular}[l]{@{}l@{}} \hspace{-2.4mm}
 \cite{Burke:02} \\ \hspace{-2.4mm} \cite{Gemmel-etal:12}\\ 
  \hspace{-2.4mm} \cite{Cantador-etal:11}\\
\end{tabular}}
& hybrid & 
\multicolumn{1}{l}{
 \renewcommand{\arraystretch}{0.5}
 \begin{tabular}[l]{@{}l@{}} \hspace{-2.4mm}
weighted hybrid  \\ \hspace{-2.4mm} integration  \\
\end{tabular}}

& category, properties &  
\multicolumn{1}{l}{
  \renewcommand{\arraystretch}{0.5}
  \begin{tabular}[l]{@{}l@{}} \hspace{-2.4mm}
  item descriptions, \\ \hspace{-2.4mm}
  metadata, tags
  \end{tabular}}
  \\
 
\cite{Musto-etal:11} & CBF & vector-distance & category, properties &  item descriptions\\

\cite{Ardissono-etal:03} & 
\multicolumn{1}{l}{
 \renewcommand{\arraystretch}{0.5}
 \begin{tabular}[l]{@{}l@{}} \hspace{-2.4mm}
 compatibility \\ \hspace{-2.4mm}
 evaluation \\
 \end{tabular}}
 &
 \multicolumn{1}{l}{
 \renewcommand{\arraystretch}{0.5}
 \begin{tabular}[l]{@{}l@{}} \hspace{-2.4mm}
 T-Norm tuned by \\ \hspace{-2.4mm}
 preference importance \\
 \end{tabular}}
 & category, properties &  item metadata \\

\cite{Dragone-etal:18} & constraint-based & constraint satisfaction & category, properties  & metadata \\

\multicolumn{1}{l}{
 \renewcommand{\arraystretch}{0.5}
 \begin{tabular}[l]{@{}l@{}} \hspace{-2.4mm}
 \cite{Rubio-etal:19} \\ \hspace{-2.4mm} \cite{Chen-etal:15}\\
\end{tabular}}
& review-based & CF, CBF, etc.
& categories, properties &  item reviews  \\

\cite{Dong-etal:16} & review-based & CBF & categories, properties &  
item reviews
\\

\multicolumn{1}{l}{
 \renewcommand{\arraystretch}{0.5}
 \begin{tabular}[l]{@{}l@{}} \hspace{-2.4mm}
 \cite{O'Mahony-Smyth:18} \\ \hspace{-2.4mm} \cite{Bao-etal:14}\\
 \hspace{-2.4mm} \cite{Musat-Faltings:15}\\
 \hspace{-2.4mm}
\cite{Al-Ghossein-etal:18}\\
 \hspace{-2.4mm}
\cite{Zhao-etal:15}\\
\end{tabular}}
& review-based & CF & categories, properties &  item reviews \\

\cite{Musto-etal:17b} & review-based & 
\multicolumn{1}{l}{
 \renewcommand{\arraystretch}{0.5}
 \begin{tabular}[l]{@{}l@{}} \hspace{-2.4mm}
 MF on multiple\\ \hspace{-2.4mm}
 dimensions \\
 \end{tabular}}
 & categories, properties, &  item reviews \\
 
\cite{Chen-etal:19} & review-based & neural networks & properties &  item reviews \\

\multicolumn{1}{l}{
 \renewcommand{\arraystretch}{0.5}
 \begin{tabular}[l]{@{}l@{}} \hspace{-2.4mm}
 \cite{Shalom-etal:19} \\ \hspace{-2.4mm} \cite{Lu-etal:18}\\
\end{tabular}}
& review-based & 
neural networks + CF & properties &  item reviews\\

\cite{Adomavicius-Tuzhilin:15} & CARS & KNN, MF & 
\multicolumn{1}{l}{
 \renewcommand{\arraystretch}{0.5}
 \begin{tabular}[l]{@{}l@{}} \hspace{-2.4mm}
 category, properties, \\ \hspace{-2.4mm} 
 user context\\
\end{tabular}}
& 
\multicolumn{1}{l}{
 \renewcommand{\arraystretch}{0.5}
 \begin{tabular}[l]{@{}l@{}} \hspace{-2.4mm}
 physical, temporal \\ \hspace{-2.4mm} 
 social dimensions\\
\end{tabular}}
\\

\cite{Baltrunas2011} & CARS & MF & 
\multicolumn{1}{l}{
 \renewcommand{\arraystretch}{0.5}
 \begin{tabular}[l]{@{}l@{}} \hspace{-2.4mm}
 category, properties, \\ \hspace{-2.4mm} 
 user context\\
\end{tabular}}
&
\multicolumn{1}{l}{
 \renewcommand{\arraystretch}{0.5}
 \begin{tabular}[l]{@{}l@{}} \hspace{-2.4mm}
 data provided \\ \hspace{-2.4mm} 
 by users\\
\end{tabular}}
 \\

\cite{Biancalana2013} & CARS, review-based & neural networks & 
\multicolumn{1}{l}{
 \renewcommand{\arraystretch}{0.5}
 \begin{tabular}[l]{@{}l@{}} \hspace{-2.4mm}
 category, properties, \\ \hspace{-2.4mm} 
 user context\\
\end{tabular}}
&  
\multicolumn{1}{l}{
 \renewcommand{\arraystretch}{0.5}
 \begin{tabular}[l]{@{}l@{}} \hspace{-2.4mm}
 social networks \\ \hspace{-2.4mm} 
 location-based\\ \hspace{-2.4mm} 
 services, \\ \hspace{-2.4mm} 
 item reviews
\end{tabular}}
\\

\bottomrule
\end{tabular}%
}
}
\label{tab:related-recommenders}
\end{table*}

\section{Background and related work}
\label{sec:related}
This section positions our work in the related one from three points of view: (i) general-purpose recommendation algorithms, (ii) recommender systems targeted to people with autism, and (iii) methods applied to extract information about items from reviews.

\subsection{Recommender systems - algorithms}
\label{sec:related-recommenders}
Recommender Systems are ``software tools and techniques providing suggestions for items to be of use to a user'' \citep{Ricci-etal:11}. They assist users in finding relevant information, products, and services by offering individualized suggestions.
Table \ref{tab:related-recommenders} classifies these systems on the basis of the data about items they manage.
Content-Based Filtering (CBF) \citep{Lops-etal:11}, Collaborative Filtering (CF) \citep{Desrosiers-Karypis:11,Koren-Bell:11},
collaborative multi-criteria \citep{Adomavicius-Kwon:07,Zheng:17,Jannach-etal:14}, and hybrid recommender systems \citep{Burke:02,Gemmel-etal:12,Cantador-etal:11}  estimate item ratings on the sole basis of users' preferences. \cite{Ardissono-etal:03} model the compatibility of items with the user but they unify it with preferences.
Analogously, constraint-based recommender systems  \citep{Dragone-etal:18} model both compatibility and preferences as a Constraint Satisfaction Problem \citep{Brailsford-etal:99}. 
Differently, we separate preferences from compatibility with sensory features of items by modeling the latter as possible sources of discomfort rather than liking or disliking factors. This separation also distinguishes our model from the recommenders that deal with negative preferences \citep{Musto-etal:11}. In fact, it supports the specification of heterogeneous criteria to deal with user preferences and item compatibility.


Review-based recommender systems \citep{Rubio-etal:19,Chen-etal:15} leverage consumer feedback for their suggestions. They apply different methods to match items to users, such as content-based \citep{Dong-etal:16}, collaborative \citep{O'Mahony-Smyth:18,Bao-etal:14,Musat-Faltings:15,Al-Ghossein-etal:18,Zhao-etal:15}, multi-criteria \citep{Musto-etal:17b}, and neural ones \citep{Chen-etal:19}, as well as hybrid solutions \citep{Shalom-etal:19,Lu-etal:18}.
However, they uniformly treat all the item features extracted from the reviews as targets of user preferences.

Context-aware recommenders consider different variables about the user and her/his context, specifically dealing with the time, location, and nearby people to provide just-in-time recommendations \citep{Adomavicius-Tuzhilin:15}. 
\cite{Baltrunas2011} extend Matrix Factorization to recommend music in a car by considering the user's preferences for the driving style, road type, and so forth.
\cite{Biancalana2013} propose a neural recommender system that personalizes the suggestion of PoIs based on the user's preferences, and on her/his location, transportation means, etc.. 
Similar to these works, we use contextual information about PoIs to steer the system's suggestions, and we employ consumer feedback to build rich models of places.
However, we model both user preferences and idiosyncratic aversions. While we use static data about PoIs to generate the recommendations, our model is based on a modular architecture that makes it seamlessly extensible to retrieve data in real-time from external data sources and sensors.

\begin{table*}[t!]
\centering
\caption{Recommender systems for users with autism spectrum disorders.}
\label{tab:rec_aut}
\resizebox{0.5\paperheight}{!}{%
{\def\arraystretch{2}
\begin{tabular}{llllll}
\toprule
Citations & Recommendation Algorithm  & User Features & Item Suggested & Target & Evaluation\\ 
\midrule
\cite{hong2012designing} & no algorithm  & social issues & social behaviors  & teenager & no\\
\cite{costa2017task} & case based   & age, gender & daily activities & children & no\\
\cite{premasundari2019food} & association rules    
&  
 \multicolumn{1}{l}{
 \renewcommand{\arraystretch}{0.5}
 \begin{tabular}[l]{@{}l@{}} \hspace{-2.4mm}
 symptoms (e.g., \\ \hspace{-2.4mm}
 learning difficulties,\\ \hspace{-2.4mm}
 fine motor skill \\ \hspace{-2.4mm}
 dysfunction,\\ \hspace{-2.4mm}
 language disorder,..)  \\
 \end{tabular}} 
& food and therapies & children & usability \\
\cite{Ng-Pera:18} 
&   \multicolumn{1}{l}{
 \renewcommand{\arraystretch}{0.5}
 \begin{tabular}[l]{@{}l@{}} \hspace{-2.4mm}
 hybrid\\ \hspace{-2.4mm}
 (collaborative,\\ \hspace{-2.4mm}
 graph-based) \\
 \end{tabular}} & 
 \multicolumn{1}{l}{
 \renewcommand{\arraystretch}{0.5}
 \begin{tabular}[l]{@{}l@{}} \hspace{-2.4mm}
 interest\\ \hspace{-2.4mm}
 social skills\\ \hspace{-2.4mm}
 emotional state \\
 \end{tabular}} & social games &  adults &
 
 \multicolumn{1}{l}{
 \renewcommand{\arraystretch}{0.5}
 \begin{tabular}[l]{@{}l@{}} \hspace{-2.4mm}
 accuracy \\ \hspace{-2.4mm} (no ASD subj.) \\
\end{tabular}}
  \\
\cite{banskota2020recommending} & collaborative filtering &  interests, weakness  & videogames & adults &  \multicolumn{1}{l}{
 \renewcommand{\arraystretch}{0.5}
 \begin{tabular}[l]{@{}l@{}} \hspace{-2.4mm}
 accuracy \\ \hspace{-2.4mm} (no ASD subj.) \\
\end{tabular}}
  \\
\cite{Mauro-etal:20} & content-based  & 
\multicolumn{1}{l}{
 \renewcommand{\arraystretch}{0.5}
 \begin{tabular}[l]{@{}l@{}} \hspace{-2.4mm}
 interests, \\ \hspace{-2.4mm} sensory aversions \\
\end{tabular}}
& POIs & adults &  \multicolumn{1}{l}{
 \renewcommand{\arraystretch}{0.5}
 \begin{tabular}[l]{@{}l@{}} \hspace{-2.4mm}
 accuracy \\ \hspace{-2.4mm} (no ASD subj.) \\
\end{tabular}}
  \\
\bottomrule
\end{tabular}%
}
}
\end{table*}

\subsection{Recommender systems - applications for autism}
\label{sec:related-autism}
Recommender systems specifically conceived for people with autism spectrum disorders are rare. Table \ref{tab:rec_aut} summarizes the state-of-art. 

\cite{hong2012designing} propose to provide users with suggestions within a social network aimed at supporting the independence of young adults. However, they focus on the organization of the social network, by relying on peer suggestions, instead of generating recommendations. 
\cite{costa2017task} develop a task recommender system that uses case-based reasoning to suggest the child's daily activity to be performed (related to eating, keeping clean, etc.) based on age, gender, and time of day but it does not consider the child's preferences. Moreover, the level of difficulty of the activities is manually set by the therapist. 
\cite{premasundari2019food} propose a food and therapy recommender system for autistic children based on their symptoms in different areas (social interaction and communication problems, speech deficits, etc.). The system 
targets parents and caregivers, rather than children, and has been exclusively evaluated from a usability viewpoint.
\cite{Ng-Pera:18} propose a hybrid game recommender for adult people with autism, based on collaborative and graph-based recommendation techniques. The system is only tested on neurotypical people.
\cite{banskota2020recommending} present, and empirically evaluate, a recommender system that suggests therapeutic games to adults with autism spectrum disorders. The system can improve users' social-interactive skills, and takes their weaknesses into account in the recommendations. 
Our work differs from the above ones in the application domain, and also because it employs aversions to sensory features, besides user preferences, to steer recommendation.

\begin{table*}[t!]
\centering
\caption{Extraction of item features. LDA denotes Latent Dirichlet Allocation.}
\resizebox{0.5\paperheight}{!}{%
{\def\arraystretch{1.5}
\begin{tabular}{lllll}
\toprule

Citations & Purpose & Feature extraction algorithm & Extracted features & Information Sources\\ 
\midrule
\cite{Lops-etal:11} & 
\multicolumn{1}{l}{
 \renewcommand{\arraystretch}{0.5}
 \begin{tabular}[l]{@{}l@{}} \hspace{-2.4mm}
content-based item  \\ 
\hspace{-2.4mm} recommendation \\
\end{tabular}}
& TF-IDF & item properties & item descriptions\\

\cite{Musat-Faltings:15} & \multicolumn{1}{l}{
 \renewcommand{\arraystretch}{0.5}
 \begin{tabular}[l]{@{}l@{}} \hspace{-2.4mm}
review-based item \\ 
\hspace{-2.4mm} recommendation \\
\end{tabular}} & 
\multicolumn{1}{l}{
 \renewcommand{\arraystretch}{0.5}
  \begin{tabular}[l]{@{}l@{}} \hspace{-2.4mm}
faceted opinion \\ 
\hspace{-2.4mm} extraction \\
\end{tabular}}
& item properties & item reviews \\

\cite{Dong-etal:13} & 
\multicolumn{1}{l}{
 \renewcommand{\arraystretch}{0.5}
  \begin{tabular}[l]{@{}l@{}} \hspace{-2.4mm}
review-based item \\ 
\hspace{-2.4mm} recommendation \\
\end{tabular}} & 
\multicolumn{1}{l}{
 \renewcommand{\arraystretch}{0.5}
 \begin{tabular}[l]{@{}l@{}} \hspace{-2.4mm}
bi-gram and  \\ 
\hspace{-2.4mm} tri-gram analysis \\
\end{tabular}}
& item properties & item reviews \\

\cite{Bao-etal:14} & \multicolumn{1}{l}{
 \renewcommand{\arraystretch}{0.5}
 \begin{tabular}[l]{@{}l@{}} \hspace{-2.4mm}
review-based item \\ 
\hspace{-2.4mm} recommendation \\
\end{tabular}} & 
\multicolumn{1}{l}{
 \renewcommand{\arraystretch}{0.5}
 \begin{tabular}[l]{@{}l@{}} \hspace{-2.4mm}
Non-negative \\ \hspace{-2.4mm} Matrix Factorization  \\
\end{tabular}}
 & item properties & item reviews \\

\multicolumn{1}{l}{
 \renewcommand{\arraystretch}{0.5}
 \begin{tabular}[l]{@{}l@{}} \hspace{-2.4mm}
 \cite{McAuley-Leskovec:13} \\ \hspace{-2.4mm} \cite{Al-Ghossein-etal:18}\\
\end{tabular}}
& \multicolumn{1}{l}{
 \renewcommand{\arraystretch}{0.5}
 \begin{tabular}[l]{@{}l@{}} \hspace{-2.4mm}
review-based item \\ 
\hspace{-2.4mm} recommendation \\
\end{tabular}} & LDA & item properties & item reviews \\

\cite{Pena-etal:20} & \multicolumn{1}{l}{
 \renewcommand{\arraystretch}{0.5}
 \begin{tabular}[l]{@{}l@{}} \hspace{-2.4mm}
review-based item  \\ 
\hspace{-2.4mm} recommendation \\
\end{tabular}} & ensemble methods & item properties & item reviews \\


\cite{Qi-etal:16} & 
\multicolumn{1}{l}{
 \renewcommand{\arraystretch}{0.5}
 \begin{tabular}[l]{@{}l@{}} \hspace{-2.4mm}
product properties  \\ \hspace{-2.4mm}  identification \\
\end{tabular}}
 & LDA + PageRank & item properties & item reviews \\

\cite{Korfiatis-etal:19} &
\multicolumn{1}{l}{
 \renewcommand{\arraystretch}{0.5}
 \begin{tabular}[l]{@{}l@{}} \hspace{-2.4mm}
evaluation aspects  \\ \hspace{-2.4mm}  identification \\
\end{tabular}}
  & Structural Topic Models &
\multicolumn{1}{l}{
 \renewcommand{\arraystretch}{0.5}
 \begin{tabular}[l]{@{}l@{}} \hspace{-2.4mm}
evaluation aspects\\ \hspace{-2.4mm}  of items  \\
\end{tabular}}
& item reviews \\

\cite{Paul-etal:17} & 
\multicolumn{1}{l}{
 \renewcommand{\arraystretch}{0.5}
 \begin{tabular}[l]{@{}l@{}} \hspace{-2.4mm}
review  \\ 
\hspace{-2.4mm} recommendation \\
\end{tabular}}
& double propagation & item properties & item reviews \\

\cite{Xu-etal:17} & aspect extraction & 
\multicolumn{1}{l}{
 \renewcommand{\arraystretch}{0.5}
 \begin{tabular}[l]{@{}l@{}} \hspace{-2.4mm}
Latent Semantic  \\ 
\hspace{-2.4mm} Analysis \\
\end{tabular}}
& item properties & item reviews \\

\cite{Tang-etal:19} & aspect extraction & JABST &
\multicolumn{1}{l}{
 \renewcommand{\arraystretch}{0.5}
 \begin{tabular}[l]{@{}l@{}} \hspace{-2.4mm}
 multi-grain aspects \\ \hspace{-2.4mm} and opinions \\
\end{tabular}}
 & item reviews \\
\bottomrule
\end{tabular}%
}
}
\label{tab:related-features}
\end{table*}

\subsection{Extraction of information about item features}
\label{sec:related-feature-extraction}
Table \ref{tab:related-features} classifies the feature extraction and review analysis models relevant to our work.
Content-Based Filtering \citep{Lops-etal:11} leverages item descriptions for feature extraction. The features representing item properties are typically taken from textual catalogs by applying statistical metrics such as TF-IDF to identify relevant characteristics for the generation of vector models describing items.

Review-based recommender systems use consumer feedback as a description of the experience with items \citep{Ghose-Ipeirotis:11}. They extract aspects from reviews to identify both item properties and users' opinions on such properties, based on the sentiment emerging from online comments. These systems adopt opinion mining techniques like faceted opinion extraction \citep{Musat-Faltings:15}, bi-gram and tri-gram analysis \citep{Dong-etal:13}, Non-negative Matrix Factorization \citep{Bao-etal:14}, Latent Dirichlet Allocation (LDA, see \cite{Blei-etal:10}) \citep{McAuley-Leskovec:13,Al-Ghossein-etal:18} and {\em ensemble} methods \citep{Pena-etal:20}. 
Further techniques are applied in review helpfulness analysis and in the extraction of sentiment about products and services. 
\cite{Qi-etal:16} combine LDA with PageRank \citep{Page-etal:99} on terms to find relevant product properties and \cite{Korfiatis-etal:19} apply Structural Topic Models to extract evaluation aspects from reviews. \cite{Paul-etal:17} use double propagation \citep{Guang:2011} and \cite{Xu-etal:17} use Latent Semantic Analysis to derive aspects from reviews as latent topics. \cite{Tang-etal:19} propose the JABST model to extract multi-grained aspects and opinions, and \cite{Mauro-etal:21} analyze user and item biases for helpfulness evaluation.

We cannot adopt any statistical approaches to extract sensory data about places.
In our context, the notion of ``relevance'' differs from the one used in information retrieval because we have to take a cautious approach to item suggestions. Rather than finding the most frequently occurring aspects of an item in its reviews, we aim at identifying specific sensory features, possibly reported by few users, which might reveal issues that dramatically impact ASD people. 
In other words, the notion of conformity, often adopted in the assessment of reliable data \citep{Li-etal:13}, does not apply to our context.

\begin{figure*}[t]
  \centering
  \includegraphics[width=0.9\textwidth]{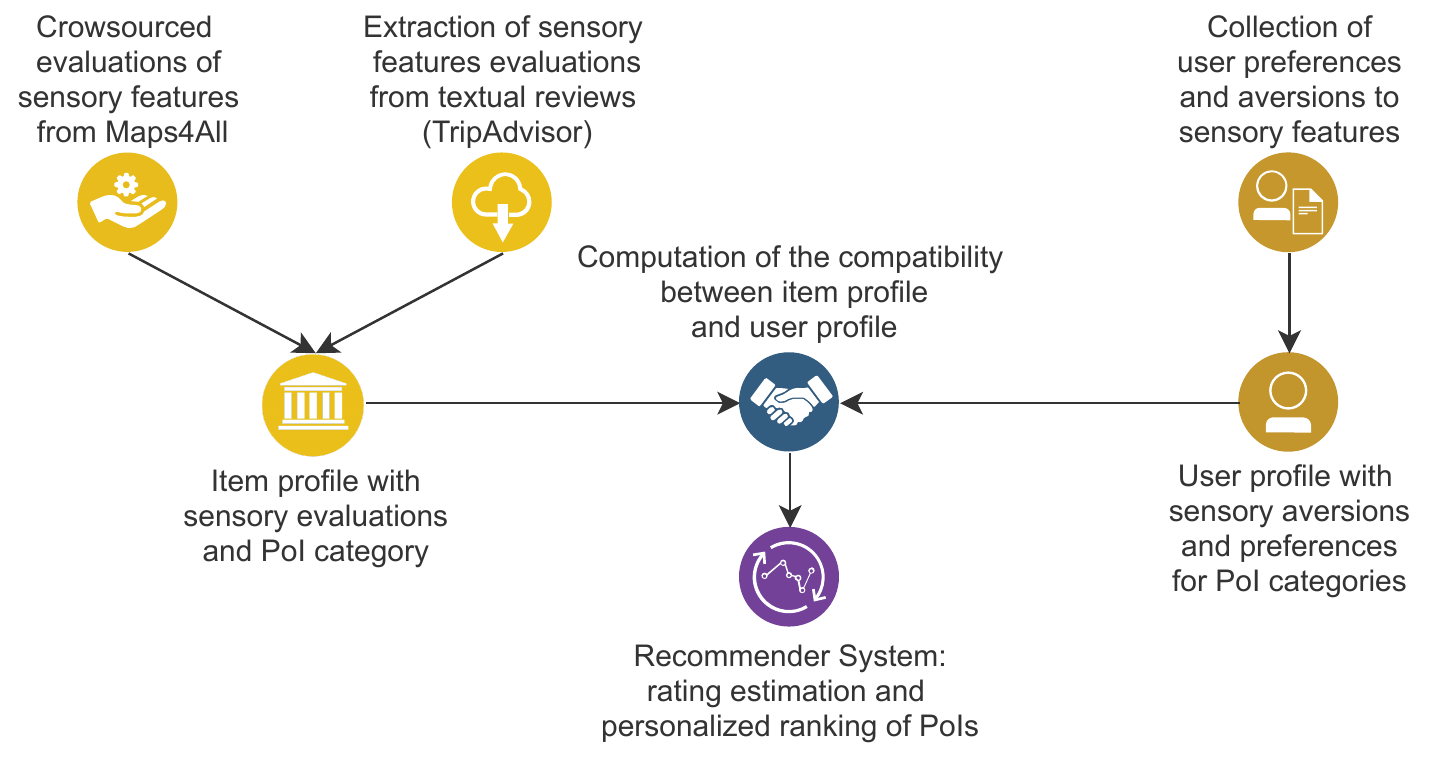}
  \caption{Framework for the compatibility-aware recommendation of places.}
  \label{fig:framework}
\end{figure*} 

\section{Data}
\label{sec:data}
As shown in Figure \ref{fig:framework}, which overviews the framework of our compatibility-aware recommendation model, we base the personalized suggestion of places on the acquisition of item and user profiles that are matched to each other by taking the user's preferences and sensory aversions into account. In the following sections we describe the techniques we developed to acquire the data about users and places, corresponding to the upper layer of the figure.

\subsection{Data about users}
\label{sec:data-users}
Recommender systems suggesting places to autistic people must work under data scarcity. There is a low number of users who can be analyzed to learn their interests: \cite{elsabbagh2012global} indicates that autism affects about 1 in 100 people in Europe. ASD people are hard to contact because they have interaction problems and a tendency to avoid new experiences. Moreover, their attention problems cause difficulties in providing detailed feedback about items \citep{murray2005attention}. These factors hamper both the acquisition of information about individual properties of users, and the execution of massive tests to evaluate the systems targeted to them.
For our work, we employ a dataset that was collected by \cite{Mauro-etal:20}. 
We gathered data by means of a questionnaire in which we asked participants to rate in the [1, 5] Likert scale the following variables:
\begin{itemize}
    \item 
    Preferences for categories of PoIs associated with free time and daily activities, such as places for eating, doing sports, and so forth.\footnote{\label{footnote:categories} The categories are: restaurants, pubs and coffee shops,  ice cream shops, museums and exhibitions, cinemas and theaters, squares, railway stations, malls and markets, comic shops, tech shops, clothing stores, libraries, bookshops.}
    \item 
    Aversions to sensory features of PoIs, and in particular to the
    \linebreak
    \texttt{brightness}, \texttt{crowding}, \texttt{noise}, \texttt{smell} and \texttt{openness} of places. 
\end{itemize}
The questions about aversions derive from the Sensory Perception Quotient test by \cite{tavassoli_sensoriality} that supports the elicitation of basic hyper- and hyposensitivity to external factors from adults with and without autism. The questions have the following format (translated from the Italian language): 
``In a place, how much does it bother you: too much light, very low light, \dots''. Regarding \texttt{brightness} and \texttt{openness}, participants evaluated two extreme conditions, i.e., low or high levels, assuming that the middle ones are not problematic. As far as \texttt{crowding}, \texttt{noise} and \texttt{smell} are concerned, people were asked about their aversion to the highest level because the low levels of these features are usually well tolerated.

Besides the user information derived from the questionnaire, the dataset includes the overall ratings that participants gave to 50 PoIs located in Torino city center, and belonging to the categories of the questionnaire. In the following, we refer to this set of places as $\Pi$. Ratings are in a [1, 5] Likert scale, where 1 represents the lowest value and 5 is the highest one. As in typical user-ratings matrices, we mark unrated features, i.e., features about which the system has no information, with the ``0'' value. 
Two groups of people answered the questionnaire and rated the places:
\begin{itemize}
    \item 20 ASD adults (from 22 to 40 years old, mean age 26.3, median 28; 11 men, 9 women, 0 non-binary and 0 not declared) who are patients of the Autistic Adult center in Torino with medium- and high-functioning. This ratio is roughly consistent with the overall gender ratio of 3:1 (man:woman) diagnosed with autism \citep{loomes2017male}.
    \item 128 neurotypical subjects (from 19 to 71 years old, average age: 28.1, median 23; 63 men, 65 women, 0 non-binary and 0 not declared) who are University students or contacts of this paper's authors.\footnote{We have no mean to know whether the subjects of this group belong to the autism spectrum or not. However, we expect that the neurotypical sample respects the proportion of the entire population. Thus, the group should include no more than 2 ASD people.}
\end{itemize}
The mean number of ratings provided by participants is 31.86 (Standard Deviation - SD=8.07) for autistic subjects and 39.34 (SD=10.52) for neurotypical ones. While the first group was fairly active in rating provision (the minimum number of ratings per user is 25), neurotypical participants varied much more, with a minimum number of ratings equal to 6.
The major contribution of ASD people to data collection can be explained by their higher motivation to actively join in a collective goal that can bring benefits to other people, as well as to themselves, and which also impacts the sense of self-efficacy and empowerment.

\subsection{Crowdsourced data about PoIs}
\label{sec:data-pois}
\cite{Mauro-etal:20} retrieved the data about places from the Maps4All\footnote{\url{https://maps4all.firstlife.org}} crowdsourcing platform, conceived to collect the evaluation of sensory features. 
Maps4All provides ratings in the [1, 5] Likert scale; for each PoI, it returns the mean values of the available ratings.
The platform was used to collect data in two experimental crowdsourcing sessions, during two lessons at the Master degree in ``Social Innovation and ICT" at the University of Torino, in May and December 2019. We involved about 120 students in these sessions, and we asked each of them to anonymously evaluate the sensory features of at least three PoIs in Torino city center. 
Overall, the 50 places of set $\Pi$, which we used in our experiments, received 785 sensory feature evaluations with coverage=49 (the sensory features were evaluated in 49 places of $\Pi$). Henceforth, we denote the dataset we produced as ``Maps4All".

\begin{table}[t]
\centering
\caption{Descriptive statistics of sensory feature evaluations concerning the places of set $\Pi$. The table shows the minimum, maximum and mean (with Standard Deviation) number of evaluations received by features per PoI.}
\label{tab:dataset}
\resizebox{0.7\textwidth}{!}{%
\begin{tabular}{lcccc|cccc}
\toprule
 & \multicolumn{4}{c|}{Maps4All} & \multicolumn{4}{c}{TripAdvisor} \\ \midrule
           & Min & Max & Mean & SD  & Min & Max & Mean & SD \\
        \midrule
\texttt{brightness} &   0 & 9 & 3.14 & 1.26 & 0   &  42 & 2.56  &  6.50 \\
\texttt{crowding}   &   0 & 9 & 3.14 & 1.26 &  0   & 299 & 47.1 & 74.39 \\
\texttt{openness}   &   0 & 9 & 3.14 & 1.26 & 0   & 483 & 80.01 & 118.63\\
\texttt{noise}      &   0 & 9 & 3.14 & 1.26 & 0   & 36 & 3.72 & 6.93\\
\texttt{smell}      &   0 & 9 & 3.14 & 1.26 & 0   & 9 & 0.5 & 1.67\\
\bottomrule
\end{tabular}}
\end{table}

The left portion of Table \ref{tab:dataset} shows the descriptive statistics of Maps4All dataset. The minimum number of ratings received by sensory features is 0 because, in a given place, some features might not have been evaluated.

\subsection{Consumer feedback about PoIs}
\label{sec:feedback}
We also retrieved sensory feature evaluations from consumer feedback extracted from a location-based service, leveraging the spontaneous reviewing activity carried out by its users. 
Specifically, we collected a dataset from TripAdvisor by scraping from its website all the reviews of the places included in set $\Pi$ that were written until June 2020.\footnote{In the analysis of consumer feedback we overlook the identity of the reviews' authors because we are not interested in considering the social relations among TripAdvisor users.}
Only 34 places out of 50 were mapped in the service but we extracted 6696 evaluations of sensory features concerning them. 
The right portion of Table \ref{tab:dataset} shows the statistics about the TripAdvisor dataset. Most sensory features have a definitely higher number of ratings than in Maps4All; for instance, the mean number of ratings of \texttt{crowding} and \texttt{openness} is 80.01 and 47.1, respectively, against 3.14.

\begin{figure}[!t]

     \begin{subfigure}[b]{1\textwidth}
         \centering
    \includegraphics[width=.32\textwidth]{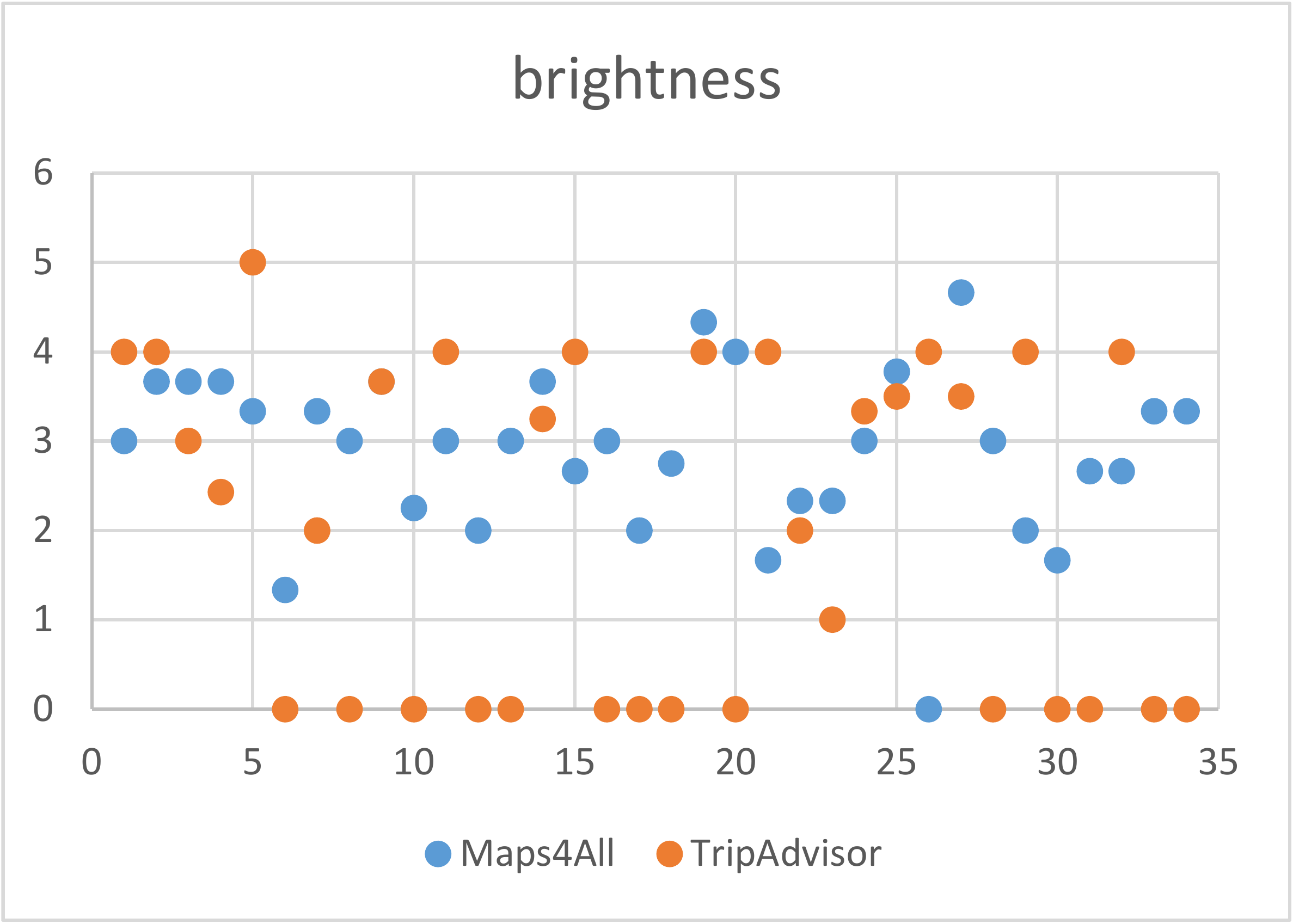}\hfill
    \includegraphics[width=.32\textwidth]{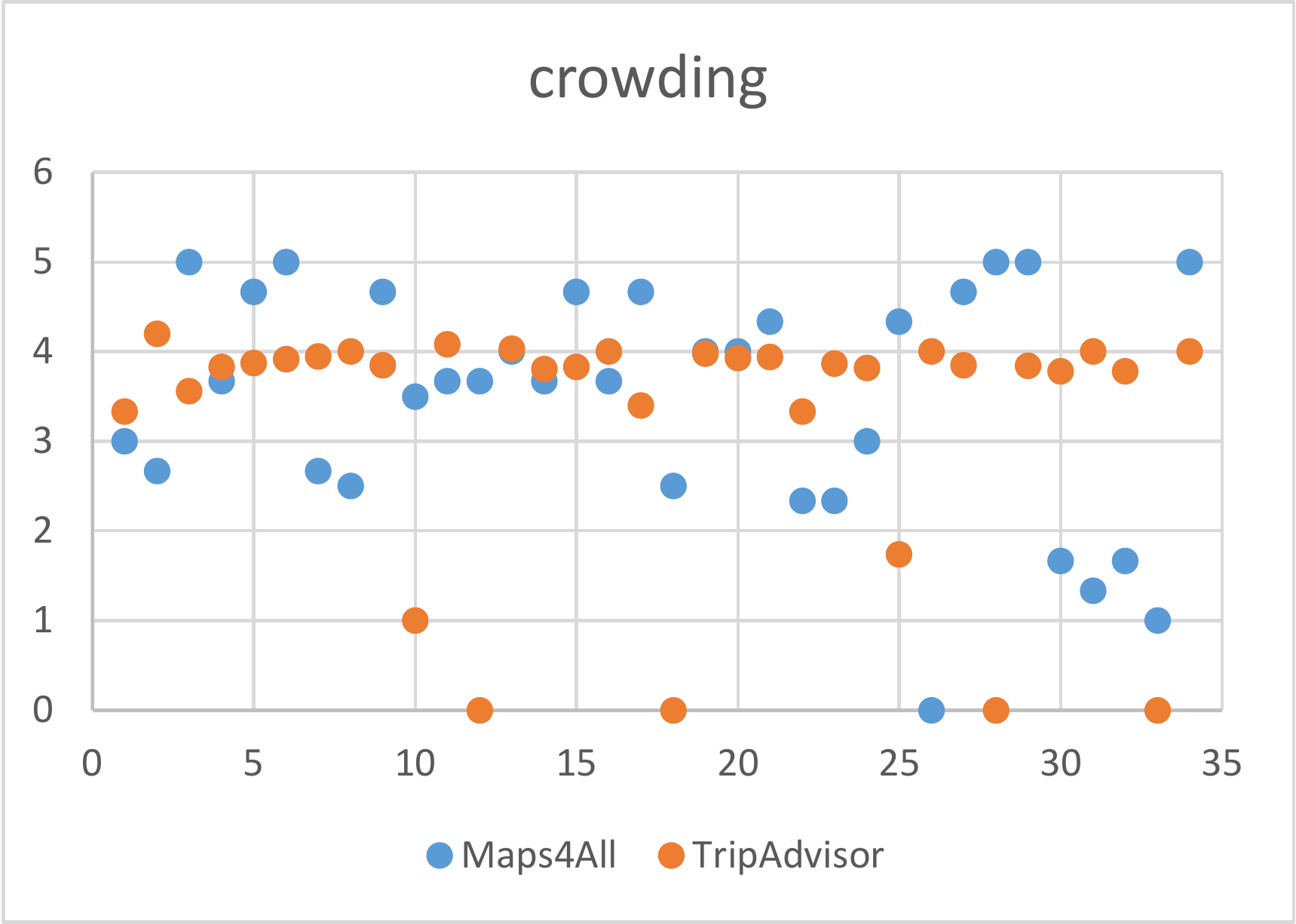}\hfill
    \includegraphics[width=.32\textwidth]{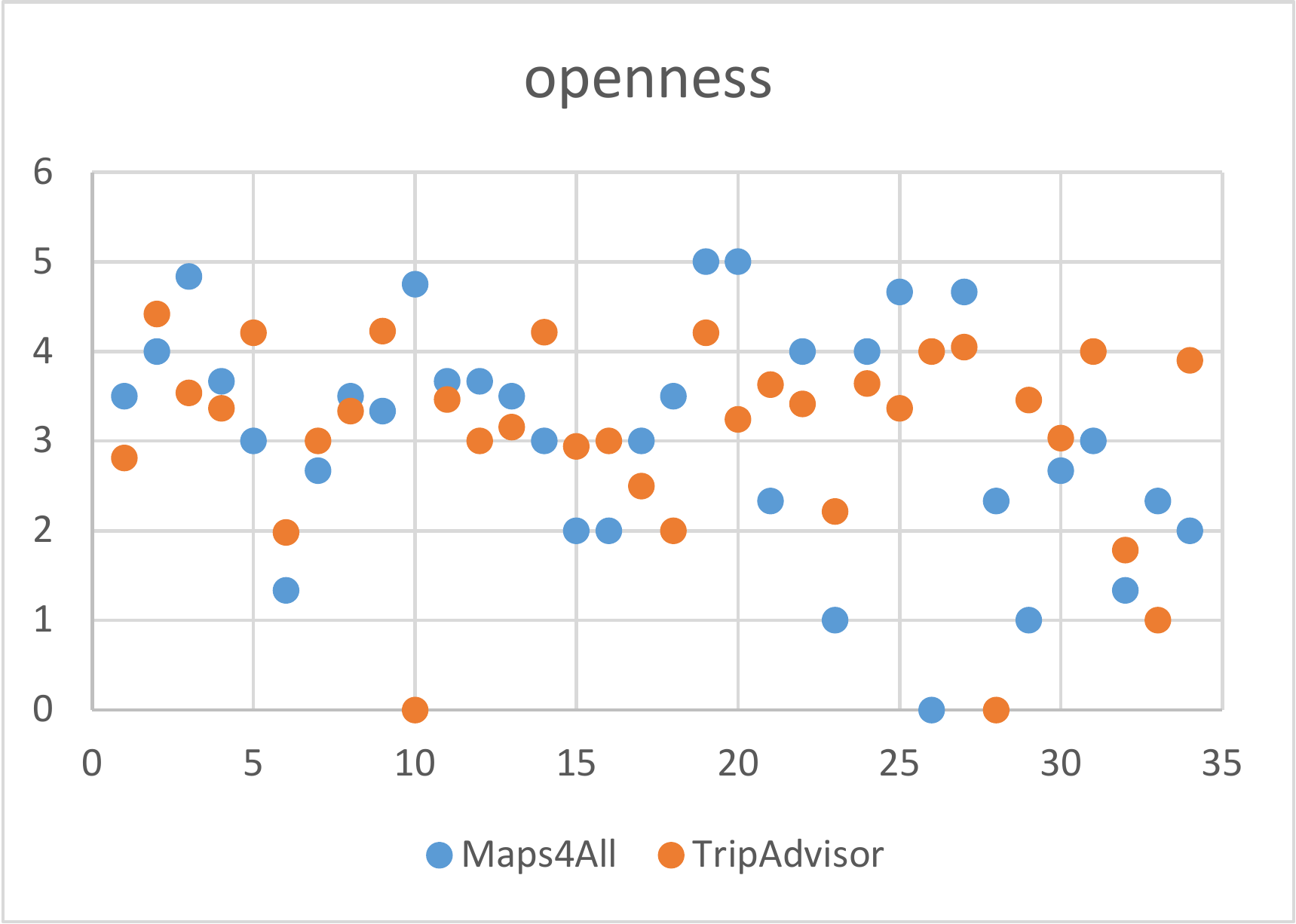}
     \end{subfigure}
   
   \vspace{5mm}
    \begin{subfigure}[b]{1\textwidth}
        \centering  
    \includegraphics[width=.32\textwidth]{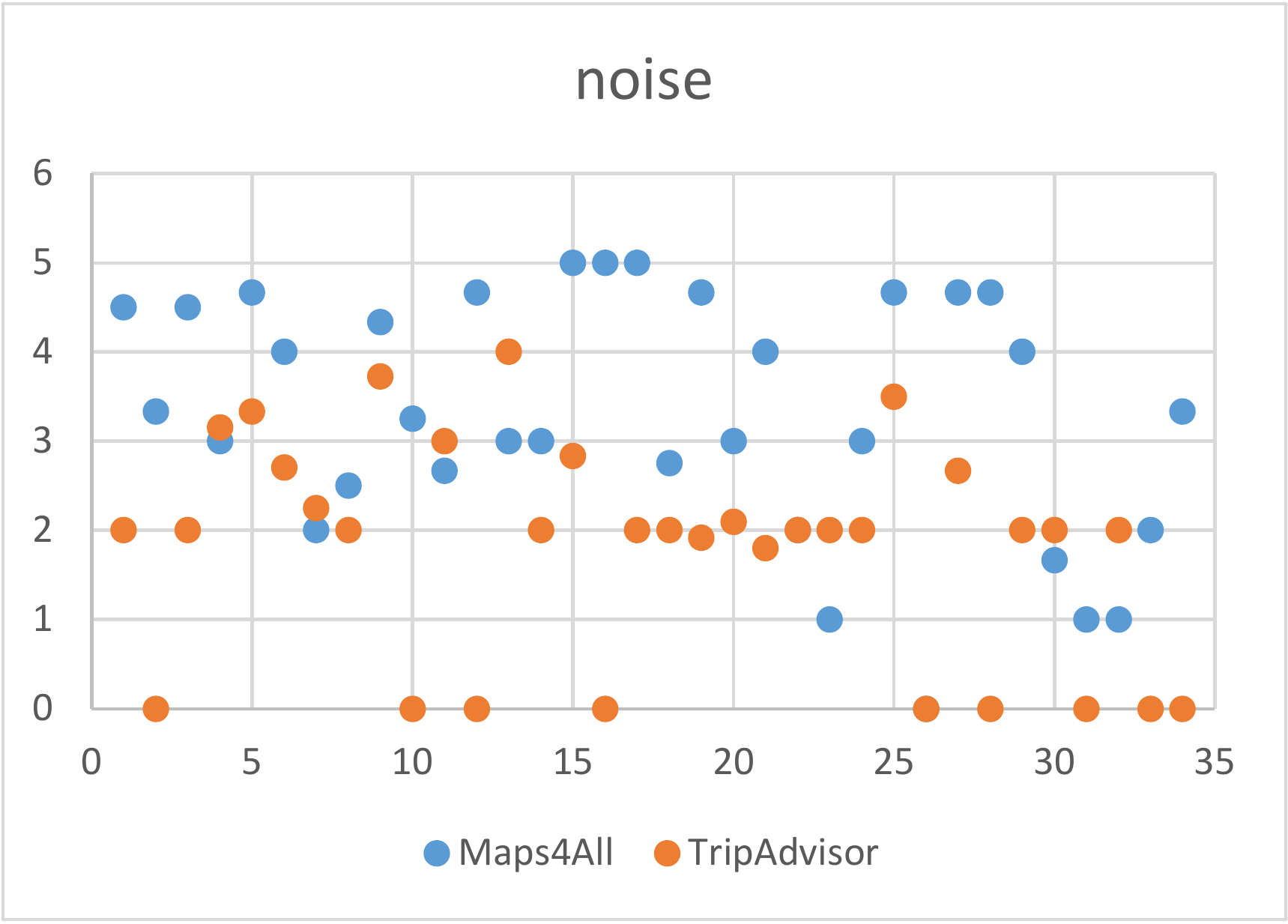}
    \includegraphics[width=.32\textwidth]{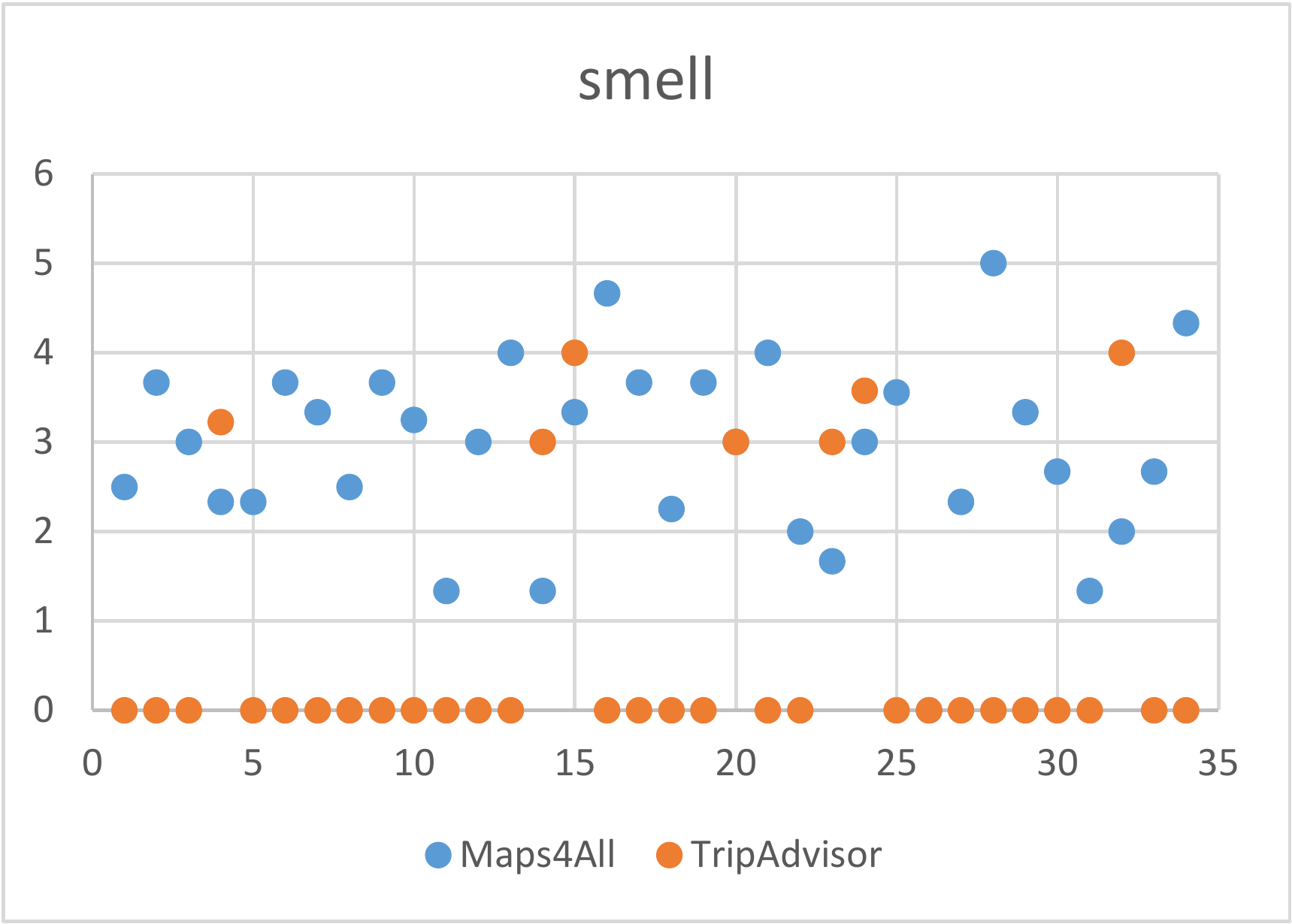}
     \end{subfigure}
\caption{Sensory feature evaluations of the 34 PoIs mapped in both Maps4All and TripAdvisor (MA$\cap$TA). The $X$ axis represents PoIs, the $Y$ axis denotes the mean feature values in [1, 5] obtained from the datasets (0 means unknown value).}
\label{fig:ratings_distribution}
\end{figure}

\begin{table}[t]
\centering
\resizebox{0.7\columnwidth}{!}{%
\begin{tabular}{lllllll}
	\toprule
	                    & Min    & Max    & M\_dist & Standard Deviation     & +ve/-ve & M\_diff \\
	\midrule
	\texttt{brightness} & 0      & 2.3333 & 0.9701  & 0.6448 & +0.1046 & -0.2228 \\
	\texttt{crowding}   & 0.0242 & 2.6667 & 1.0618  & 0.7809 & +0.1250 & -0.0398 \\
	\texttt{openness}   & 0.1667 & 2.4575 & 0.8952  & 0.5486 & +0.2526 & -0.0942 \\
	\texttt{noise}      & 0      & 3      & 1.2698  & 0.8758 & +0.2760 & +0.9442 \\
	\texttt{smell}      & 0      & 2      & 1.0181  & 0.6904 & -0.4647 & -1.0181 \\
	\bottomrule
\end{tabular}}
\caption{Minimum, maximum and mean distance (with Standard Deviation) between the feature evaluations of Maps4All and TripAdvisor for the places of set MA$\cap$TA. Column +ve/-ve reports the correlation values between feature evaluations across datasets. M\_diff shows the difference between the mean values given to features in the datasets.}
\label{tab:value-distances}
\end{table}

TripAdvisor has lower coverage than Maps4All (\texttt{brightness=20}, 
\linebreak
\texttt{crowding}=30, \texttt{noise=25}, \texttt{smell}=7 and \texttt{openness}=32). In other words, in TripAdvisor fewer places received at least one evaluation of their sensory features. The most problematic feature is \texttt{smell}, which is only evaluated in 7 places.
These findings suggest that consumer feedback is a promising source of sensory data but multiple information sources might have to be integrated to extend its coverage of places.

\subsection{Comparison of feature values in the Maps4All and TripAdvisor datasets}
\label{sec:intersection}
We consider the 34 places that are mapped in both datasets. We denote this set of places as MA$\cap$TA. Figure \ref{fig:ratings_distribution} shows the feature values of these PoIs and highlights the data sparsity concerning \texttt{brightness} and \texttt{smell}, and partially \texttt{noise}. Table \ref{tab:value-distances} shows that, on average, the distance between the mean feature values provided by the two datasets is about 1, with a Standard Deviation that ranges from 0.55 to 0.88. Moreover, column M\_diff shows that Maps4All provides higher mean values of \texttt{noise} than TripAdvisor. The opposite holds for \texttt{smell} and \texttt{brightness}, while the values of the other features are balanced.

According to Pearson correlation (column +ve/-ve), most feature values weakly correlate in a positive way in the two datasets.
Differently, \texttt{smell} has a negative correlation (-0.4647) but this is not particularly relevant because TripAdvisor reviews provide little information about this feature.

\subsection{Extraction of sensory features from consumer feedback}
\label{sec:extraction}
\subsubsection{Creation of linguistic resources about sensory features}
\label{sec:sensory-features-resources}
We could not find any linguistic resources for the analysis of sensory features in the Italian language, which is the target of our work. Therefore, three researchers from our University staff collaborated to build a \textit{sensory features dictionary} that associates words to features, and to their values. We consider the following sensory features: \texttt{brightness}, \texttt{crowding}, \texttt{noise}, \texttt{smell}, and \texttt{openness}.
These researchers also defined a \textit{modifiers dictionary} that describes how adverbs and other grade modifiers positively or negatively change the values of features associated with words within the [1, 5] scale adopted in our model. 
When these researchers disagreed with each other, they discussed the outcome with us.

The \textit{sensory features dictionary} is organized as a set of $<w, f, f_w, d_w>$ tuples. Each tuple contains:
\begin{itemize}
    \item 
    A word $w$ referring to a sensory feature $f$ of our model. For instance, adjective ``scuro'' (dark) refers to \texttt{brightness}.
    \item 
    The feature $f$ which $w$ references (\texttt{brightness}).
    \item 
    The basic positioning of $w$ in the [1, 5] scale of the values of $f$, denoted as $f_w$.  
    For example, ``dark'' is associated with a value of \texttt{brightness} equal to 2 ($brightness_{scuro} = 2$) to enable the mapping of expressions such as ``very dark'' to the minimum value of the scale.
    \item 
    The positive or negative direction $d_w$ of change with respect to the basic positioning $f_w$, when $w$ is associated with a grade modifier such as ``little'', ``very'', and so forth.
    For instance, $d_{scuro} = -1$ because a very dark place has lower \texttt{brightness} than a little dark one. Conversely, regarding ``chiaro'' (bright), $d_{chiaro} = 1$ because low values (1, 2) denote dark places, while higher values (3, 4, 5) correspond to brighter places. 
\end{itemize}
The \textit{modifiers dictionary} contains a set of $<m, impact_m>$ pairs. Each pair specifies the impact of a grade modifier $m$ (e.g., ``tanto" - a lot, ``poco" - a little, etc.) on the values of features associated with words. Let us assume that $m$ modifies a word $w$ associated with a feature $f$. Then, $impact_{m}$ indicates how much $m$ changes the value of $w$ with respect to the basic position $f_w$, in the direction specified by $d_w$. The impact of modifiers takes values in the [-2, 2] scale that makes it possible to model positive and negative impact having low (-1, 1) or high (-2, 2) strength. 
For example, $impact_{tanto} = 1$ means that, when this modifier is applied to a word expressing an increasing scale, such as ``bright'', it increments the corresponding feature value by 1. 
Differently, ``a little'' has the opposite behavior, and its impact is -1. 

\subsubsection{Extraction of sensory feature evaluations from reviews}
\label{sec:sensory-features-extraction}
We use standard Natural Language Processing techniques to retrieve the sensory information about places used for recommendation. Starting from the comments about a place $i$, we select the set $REV_i$ of reviews expressed in Italian and we extract the users' perceptions about the sensory features of $i$ in two steps: 1) for each review $rev \in REV_i$, we extract the references to sensory features occurring in $rev$ and their values. 2) for each sensory feature, we assign $i$ the mean value retrieved from all the reviews of $REV_i$.

To extract sensory data (step 1), we analyze each sentence of the reviews by navigating the tree obtained through dependency parsing and we look for the nodes that represent words $w$ of the \textit{sensory features dictionary}. 
For each node $N$ of this type, we compute the value $val_{f_w}$ of feature $f$ as follows:
    \begin{itemize}
        \item 
        If $N$ is a leaf node, or its sub-tree does not include any modifiers, $val_{f_w} = f_w$ as specified by tuple $<w, f, f_w, d_w>$ in the dictionary.
        \item 
        Otherwise (suppose that $w$ is modified by $m$), let $\Delta = impact_m * d_w$ represent the displacement with respect to the basic position of $f_w$. Then, $val_{f_w}$ is obtained by normalizing  $(f_w+\Delta)$ in [1, 5].
    \end{itemize}

\begin{table*}[!t]
\centering
\caption{Notation used to describe the compatibility-aware PoI recommendation model.}
\label{tab:notation}
\resizebox{0.8\textwidth}{!}{%
{\def\arraystretch{1}
\begin{tabular}{ll}
\toprule
Variable & \multicolumn{1}{l}{
  \renewcommand{\arraystretch}{0.4}
  \begin{tabular}[l]{l}Definition  \end{tabular}} 
 \\ \midrule
  $U$       &  \multicolumn{1}{l}{
  \renewcommand{\arraystretch}{0.4}
  \begin{tabular}[l]{l}Set of users $u$  \end{tabular}}
            \\
  $I$       & \multicolumn{1}{l}{
  \renewcommand{\arraystretch}{0.4}
  \begin{tabular}[l]{l}Set of items  $i$ \end{tabular}}
            \\ 
  
  $C$ &  \multicolumn{1}{l}{
  \renewcommand{\arraystretch}{0.4}
  \begin{tabular}[l]{l}Set of item categories $c$  \end{tabular}}
  
   \\
  $L$ & \multicolumn{1}{l}{
  \renewcommand{\arraystretch}{0.4}
  \begin{tabular}[l]{l}Likert scale in $[1, v_{max}]$. In the present work $v_{max}=5$   \end{tabular}}
  
  \\
  $F$ & \multicolumn{1}{l}{
  \renewcommand{\arraystretch}{0.4}
  \begin{tabular}[l]{l}Set of sensory features of items ($f$)   \end{tabular}}
 
   \\
  $F^\uparrow$ & \multicolumn{1}{l}{
  \renewcommand{\arraystretch}{0.4}
  \begin{tabular}[l]{l}
    Set of sensory features such that the higher the value of $f$,\\ the stronger its negative impact on the user (e.g., \texttt{noise})
  \end{tabular}}
   \\
  $F^V$ & 
  \multicolumn{1}{l}{
  \renewcommand{\arraystretch}{0.4}
  \begin{tabular}[l]{l}
    Set of features whose extreme values make people uncomfortable\\ while the middle ones are less problematic (e.g., \texttt{brightness})
  \end{tabular}}  
  \\
   $\vec{\bf {i}}$ &   \multicolumn{1}{l}{
  \renewcommand{\arraystretch}{0.4}
  \begin{tabular}[l]{l}
   Vector storing the value of each feature $f \in F$ of an item $i$
  \end{tabular}}
  
  \\
   $PREF_u$ &   \multicolumn{1}{l}{
  \renewcommand{\arraystretch}{0.4}
  \begin{tabular}[l]{l}
    User preferences for the categories of places
  \end{tabular}}
  
  \\
     $R_u$ &   \multicolumn{1}{l}{
  \renewcommand{\arraystretch}{0.4}
  \begin{tabular}[l]{l}
    Set of ratings that a user $u \in U$ gave to the items of $I$
  \end{tabular}}
   \\
  $a_{ufv}$ &  \multicolumn{1}{l}{
  \renewcommand{\arraystretch}{0.4}
  \begin{tabular}[l]{l}
    A user $u$'s aversion to a value $v$ of a feature $f \in F$
  \end{tabular}} \\
  $comp_{ufi}$ & \multicolumn{1}{l}{
  \renewcommand{\arraystretch}{0.4}
  \begin{tabular}[l]{l}
    Compatibility of item $i$ with $u$ regarding $f$
  \end{tabular}} \\
  
  $\hat{r}_{ui}$ & \multicolumn{1}{l}{
  \renewcommand{\arraystretch}{0.4}
  \begin{tabular}[l]{l}
    Estimation of a user $u$'s rating of item $i$
  \end{tabular}} \\
  
  \bottomrule
\end{tabular}%
}}
\end{table*}

\section{Compatibility-aware PoI recommendation}
\label{sec:model}
This section presents the lower level of the framework for the compatibility-aware recommendation of places shown in Figure \ref{fig:framework}. This portion of the framework is based on the work by \citep{Mauro-etal:20} and we outline it to make the present paper self-contained. Table \ref{tab:notation} shows the notation we use.

\subsection{Item profiles}
Each PoI $i \in I$ (where $I$ = $\Pi$) is described by an item profile that specifies the category of places $c \in C$ to which $i$ belongs, and a vector $\vec{\bf {i}}$ storing its feature values: $\vec{{\bf i}}_f^{\,}$ (in [1, $v_{max}$]) denotes a feature value and we remind that, if that value is unknown, we set $\vec{{\bf i}}_f^{\,}=0$ to denote the lack of knowledge. Feature values are extracted from the Maps4All and/or TripAdvisor datasets.

\subsection{User profiles}
\label{sec:UM}
The information about a user $u \in U$ is stored in a user profile that specifies the following data types, expressed in the $L$ scale: 
\begin{itemize} 
    \item 
    Her/his preferences $PREF_u = \{p_c\,|\,c\in C\}$ for the categories of places.
    \item 
    The sensory aversion to specific values of item features declared by $u$. We denote $u$'s aversion to a value $v$ of a feature $f \in F$ as $a_{ufv}$; e.g., $a_{uf5} = 5$ means that $u$ is very disturbed by an item $i$ such that $\vec{{\bf i}}_f^{\,} = 5$.
    \begin{itemize}
        \item 
        For each $f \in  F^\uparrow$, we assume that $a_{uf1} = 1$. Thus, the user profile only stores a value $a_{ufv_{max}}$ that specifies $u$'s aversion to the maximum value of $f$.
        \item
        For each $f$ in $F^V$, the user profile stores two values that express $u$'s aversion to the minimum and maximum values of $f$. 
    \end{itemize}
\end{itemize}
In our work, the list of sensory aversions of a user $u$ consists of
    $\{a_{u\texttt{brightness}1}, \linebreak a_{u\texttt{brightness}5},$ 
    $a_{u\texttt{crowding}5}, a_{u\texttt{noise}5}, a_{u\texttt{smell}5}, a_{u\texttt{openness}1}, a_{u\texttt{openness}5}\}$.
The user profiles are set to the user data described in Section \ref{sec:data-users}.

\subsection{Evaluation of the compatibility of an individual feature with the user}
\label{sec:featureCompatibility}
The aversion values stored in the user profiles correspond to the extreme values that features can take. Thus, an interpolation method is needed to infer a user $u$'s aversion for the other values of [1, $v_{max}$]. Assuming to represent feature values in the $X$ axis, and aversion in the $Y$ axis of a plane:
\begin{itemize}
    \item
    For each $f \in F^\uparrow$, and given $a_{ufv_{max}}$ in $u$'s profile, we approximate aversion as a line connecting point (1, 1), to point ($v_{max}$, $a_{ufv_{max}}$) to represent the increment of aversion while the value of $f$ increases:
        \begin{equation}
        line^\uparrow(x) = 1 + \frac{(a_{ufv_{max}}-1)(x-1)}{v_{max}-1}
        \label{eq:crescente}
    \end{equation}
    Therefore, $u$'s estimated aversion to $f$ in $i$ is $ea_{ufi} = line^\uparrow(\vec{{\bf i}}_f^{\,})$.
    \item 
    For each $f \in F^V$, and given \{$a_{uf1}, a_{ufv_{max}}$\} in $u$'s profile, $ea_{ufi} = \max (line^\uparrow(\vec{{\bf i}}_f^{\,}), line_\downarrow(\vec{{\bf i}}_f^{\,}))$, where
    \begin{equation}
        line_\downarrow(x) = 1+\frac{(x-v_{max})(1 - a_{uf1})}{v_{max}-1}
        \label{eq:lineDown}
    \end{equation}
    connects (1, $a_{uf1}$) and ($v_{max}$, 1) to represent the decrease in aversion from low to middle values of $f$.
\end{itemize}
Similar to \citep{Mauro-etal:20}, we compute the compatibility of a feature value $\vec{{\bf i}}_f^{\,}$ with a user $u$ as the complement in [1, $v_{max}$] of $u$'s aversion to $f$ because aversion can be described as the opposite of compatibility:
\begin{equation}
    comp_{ufi} = v_{max} + 1 - ea_{ufi}
    \label{eq:individual-feature-compatibility}
\end{equation}
Notice that, if the reviews of $i$ do not mention $f$, we pessimistically set $comp_{ufi} = 1$. 
Even though the lack of references to a feature could be interpreted as a lack of complaints about it, this assumption is reasonable when dealing with neurotypical users who, given the low percentage of autistic people in the population, are plausibly the authors of most reviews. Conversely, we consider the sensory needs of users with autism spectrum disorders, whose sensitivity is much higher. 
To prevent the risk of bothering them, we assume that a feature whose value is unknown is an incompatible one.

\subsection{Aggregation measures}
\label{sec:itemCompatibility}
Before describing the recommendation algorithms we use, we outline the aggregation measures they apply to integrate evaluation components for rating prediction. Depending on the recommendation model, evaluation components can represent the compatibility values of the sensory features or the preference of the user $u \in U$ for the category of the item to be evaluated.
Let us consider a set of evaluation components $\Omega = \{\omega_1, \dots, \omega_k\}$, where $\omega_j$ takes values in $[1, v_{max}]$ and represents an aspect of fit between item and user. We compute the aggregated value $y$ by applying one of the following measures:
\begin{itemize}
    \item
    $Min$: $y$ is the minimum value of set $\Omega$, meaning that the aggregated value corresponds to the worst fit between item and user.
    \item
    $Ave$: $y$ is the mean value of set $\Omega$, denoting average fit.
    \item
    $Cos$: $y$ is a normalization in $[1, v_{max}]$ of Cosine similarity between a vector $\vec{{\bf \omega}}$ representing the values of evaluation components and a vector $\overrightarrow{{\bf{ideal_u}}}$ whose values for the same components best match $u$'s profile.
    The smaller the angle between $\vec{{\bf \omega}}$ and $\overrightarrow{{\bf{ideal_u}}}$, the better $\Omega$ fits $u$. 
    \item
    $RMSD$: the aggregated value is the complement in $[1, v_{max}]$ of the Root Mean Square Deviation between $\vec{{\bf \omega}}$ and $\overrightarrow{{\bf{ideal_u}}}$. This represents the distance between the two vectors ($\overrightarrow{{\bf ideal}_{u\omega}}$ is component $\omega$ of $\overrightarrow{{\bf ideal}_{u}}$):
    \begin{equation}
    \label{eq:rmsd}
        y = 1 + v_{max} - \sqrt{\frac{1} {|\Omega|}* \sum\limits_{\omega\in \Omega} (\omega  - \overrightarrow{{\bf ideal}_{u\omega}} )^2}
    \end{equation}
\end{itemize}

\subsection{Rating prediction}
\label{sec:overallItemEvaluation}
For each $u \in U$ and $i \in I$, we estimate $u$'s evaluation of $i$ ($\hat{r}_{ui}$) by applying the following algorithms described in \citep{Mauro-etal:20,Mauro-etal:22}:\footnote{We did not consider any collaborative recommendation algorithms \citep{Adomavicius-Kwon:07} because our datasets are too small to train them.}
\begin{itemize}
    \item 
    \textsf{Individual (Ind)} estimates item ratings by adapting the relative impact of sensory features compatibility and user preferences to the individual user because it seems that people with autism weight these factors in a personal way \citep{Mauro-etal:20}:
    \begin{equation}
        \hat{r}_{ui} = \alpha * overallComp_{ui} + (1-\alpha) * p_{uci}
        \label{eq:est_rating}
    \end{equation}
    where $p_{uci}$ is $u$'s preference for the category $c$ of item $i$ and $overallComp_{ui}$ is the overall compatibility of $i$ with $u$, given $i$'s sensory features. Moreover, $\alpha$ (in [0, 1]) personalizes the balance between item compatibility and user preferences. Section \ref{sec:validation} describes how $\alpha$ is obtained.
    
    \textsf{Ind} computes $overallComp_{ui}$ by combining the compatibility of the sensory features of $i$ with $u$ using the aggregation measures of Section \ref{sec:itemCompatibility}.
    In $Min$ and $Ave$, 
    $\Omega = \{comp_{u\texttt{brightness}i}, \dots,
   comp_{u\texttt{openness}i}\}$ and its components are defined as in Equation \ref{eq:individual-feature-compatibility}.
    Regarding $Cos$ and $RMSD$, we found that mapping $\Omega$ to feature values improves recommendation performance. Thus, $\vec{{\bf \omega}} = \vec{{\bf i}}$ and $\overrightarrow{{\bf{ideal_{u}}}}$ is an ideal item that minimizes $u$'s aversions. For each $f \in F$, $\overrightarrow{{\bf{ideal_{uf}}}}$ is the most compatible value of $f$, based on $u$'s estimated aversion to $f$. 
    \item 
    \textsf{C-only} is a setting of the \textsf{Ind} algorithm where $\alpha=1$ is used to predict ratings on the basis of its compatibility with the user.
    \item 
    \textsf{Pref-only} is a setting of \textsf{Ind} where $\alpha=0$ is used to evaluate items on the basis of the user's preferences.
    \item 
    \textsf{Multi-Criteria } \textsf{(MC)} computes $\hat{r}_{ui}$ by fusing $u$'s preference for the category of $i$ ($p_{uci}$) with the compatibility of each individual feature ($comp_{ufi}$), managing all such values as independent evaluation factors. It integrates the individual values by applying the aggregation measures of Section \ref{sec:itemCompatibility} by setting $\Omega = \{p_{uci}, comp_{u\texttt{brightness}i}, \dots, comp_{u\texttt{openness}i}\}$. 
    \textsf{MC} differs from \textsf{Ind} because it applies the same aggregation function to all the evaluation parameters, while \textsf{Ind} distinguishes preferences from compatibility and supports the adoption of heterogeneous aggregation criteria to the two types of information. Incidentally, we deal with a single preference for the item category but the preference component might result from the integration of multiple item features.
\end{itemize}

\section{Validation methodology}
\label{sec:validation}
Our experiments pursue two main goals.
Concerning research question RQ1, we are interested in evaluating the usefulness of the sensory data about places gathered from Maps4All and/or from TripAdvisor platforms.
Regarding RQ2, we aim at understanding how the sensory data extracted from consumer feedback impacts recommendation performance and whether, by modeling both user preferences and item compatibility, we obtain higher performance compared to taking only one of these aspects into account.
To satisfy these goals, we compare the performance of the recommendation algorithms by configuring them on each aggregation measure of Section \ref{sec:itemCompatibility}. The algorithms determine whether compatibility and/or user preferences have to be used in rating prediction. The aggregation measures provide alternative data fusion methods.

We are also interested in checking whether the management of compatibility information is relevant to both neurotypical and autistic users. Therefore, we test the algorithms on the datasets of users described in Section \ref{sec:data-users}:
\begin{enumerate}
    \item 
    Users with autism spectrum disorders. We denote this dataset as AUT.
    \item
    Neurotypical users. We denote this dataset as NEU.
\end{enumerate}  
For each recommendation algorithm, we specify the aggregation measure we apply by appending the two names. For example, \textsf{Ind$_{Cos}$} represents the application of the $Cos$ aggregation measure to model \textsf{Ind}.
In addition to the notation of Table \ref{tab:notation}, we define $R$ as the overall set of item ratings provided by the users of $U$ and $\hat{R}$ as the set of estimated ratings. $Relevant_u$ is the set of items that $u \in U$ has positively rated: $Relevant_u = \{i \in I ~|~ r_{ui}>3\}$. $Recomm_u$ is the set of items that the system suggests to $u$: $Recomm_u = \{i \in I ~|~ \hat{r}_{ui}>3\}$, and $k$ denotes the length of the suggestion list. 

We analyze recommendation performance in terms of {\bf Accuracy} (Precision, Recall, and F1 metrics),
{\bf Ranking capability} (MAP and MRR), {\bf Error in rating prediction} (MAE and RMSE) and {\bf User coverage}. The last parameter describes the percentage of users to whom the system recommends items. 
All metrics, except for MAE and RMSE, have to be maximized.

We perform a 5-fold cross-validation in which, for every fold, we use 80\% as training set to find the best $\alpha$ value for each individual user and 20\% as test set. We are interested in optimizing performance with respect to the ranking of items in the recommendation lists. Thus, we run each model to find the best user-specific setting by optimizing its results for MAP using the Exhaustive Grid Search algorithm\footnote{\url{https://scikit-learn.org/stable/modules/grid_search.html#exhaustive-grid-search}.}.
Notice that, to be sure that the other algorithms (\textsf{MC}, \textsf{C-only} and \textsf{Pref-only}, which do not need any training) are consistently evaluated, we run them on the same test sets used for \textsf{Ind}. 

\begin{figure}
    \includegraphics[width=.32\textwidth]{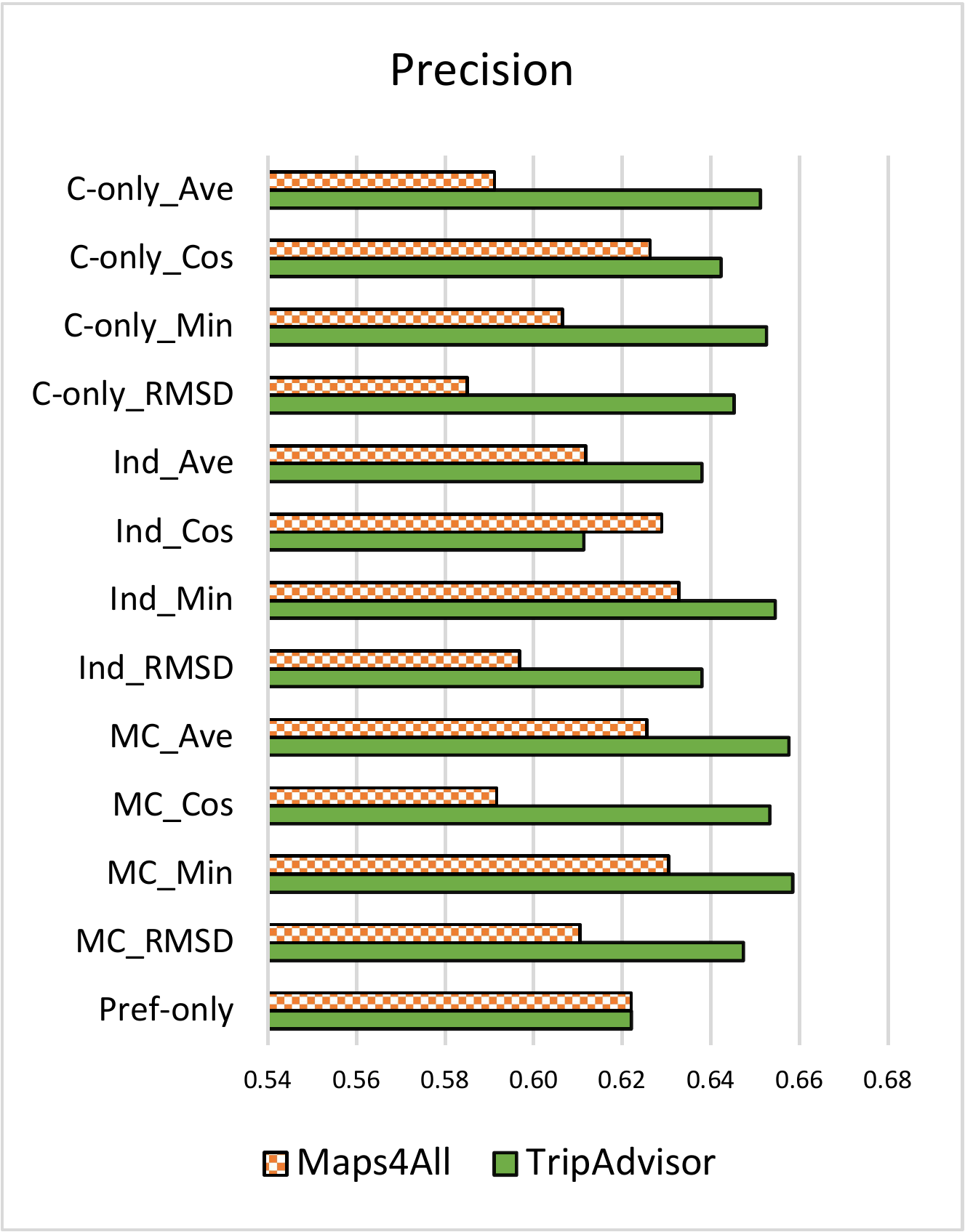}\hfill
    \includegraphics[width=.32\textwidth]{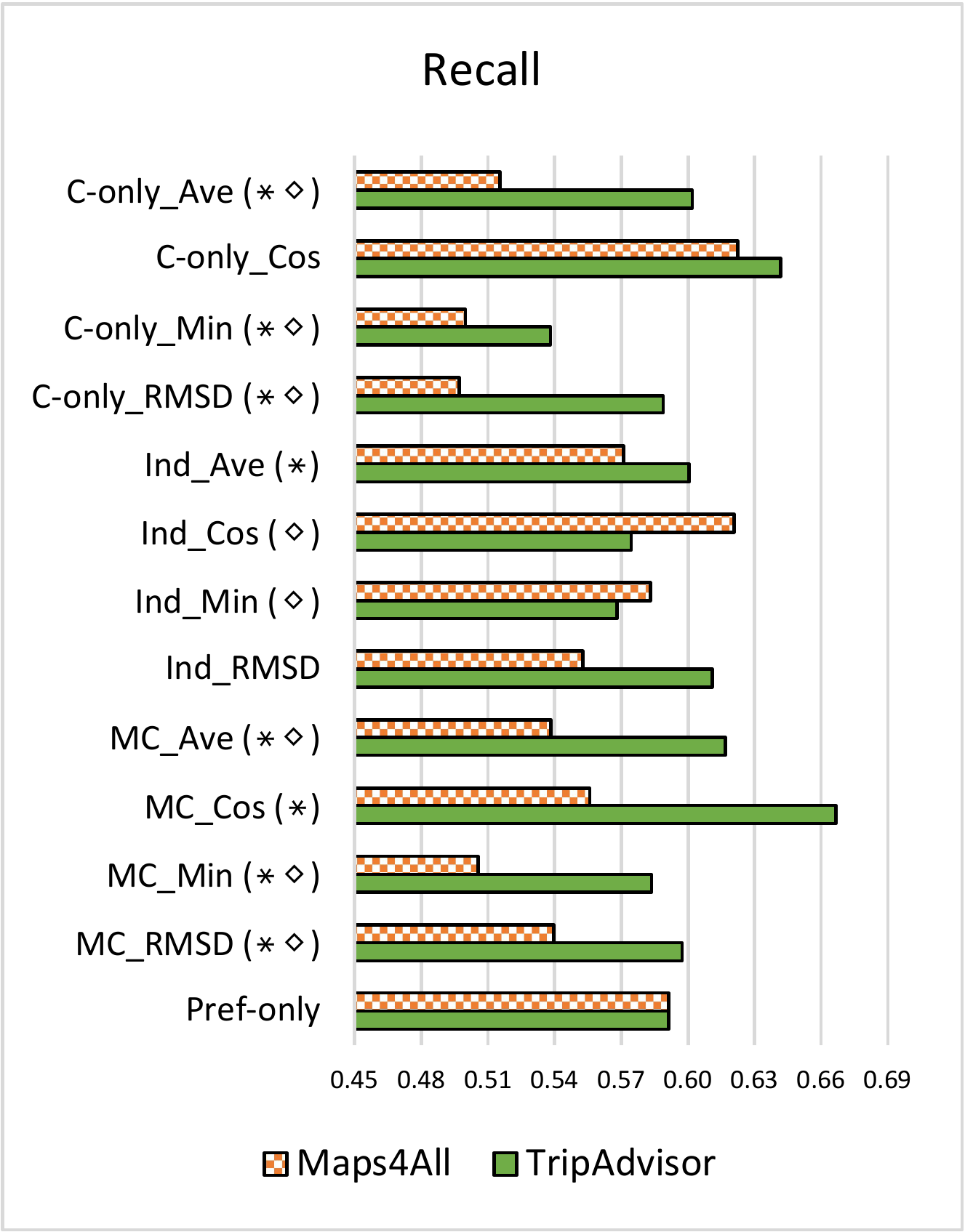}\hfill
    \includegraphics[width=.32\textwidth]{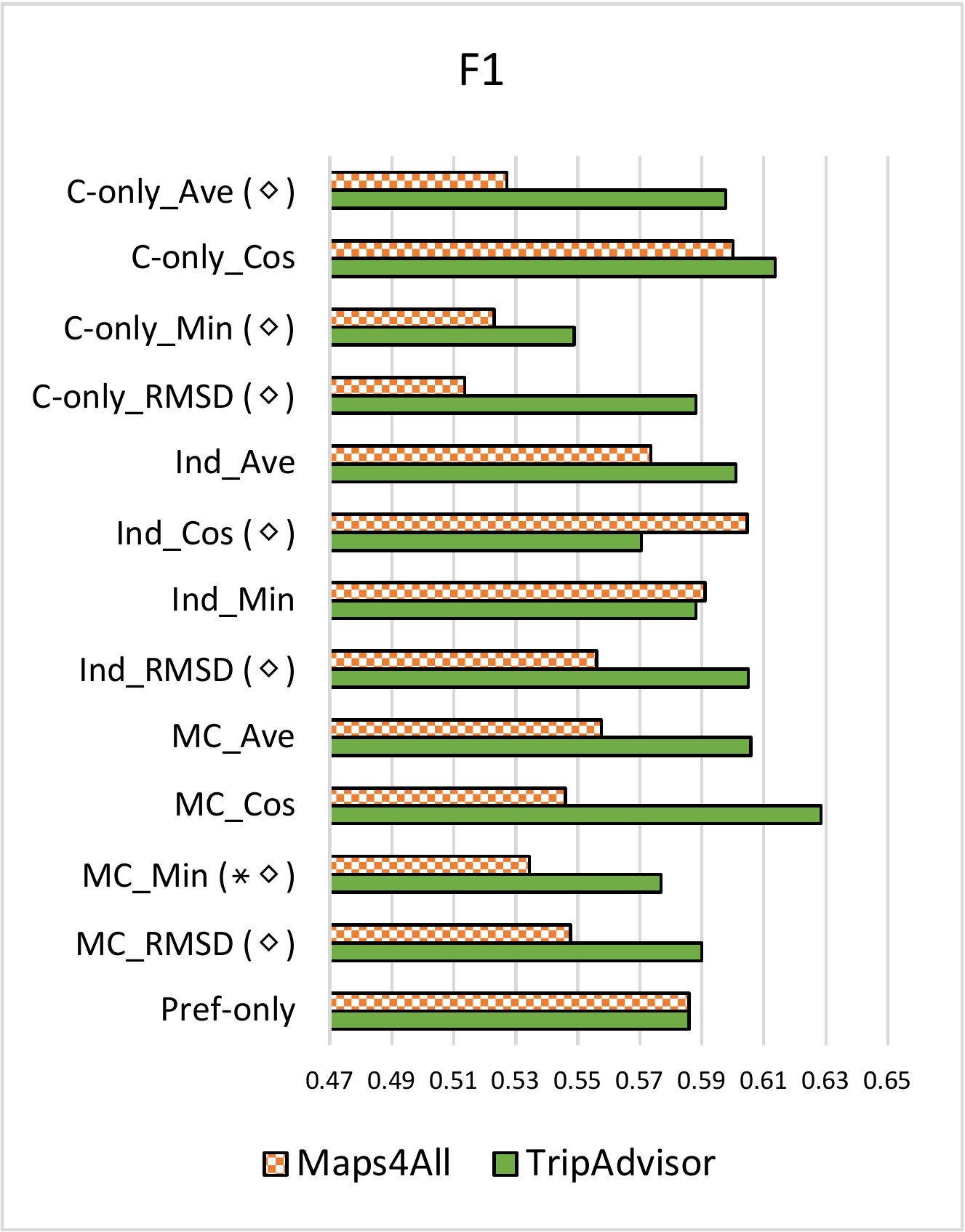}
    \\[\smallskipamount]
    \includegraphics[width=.32\textwidth]{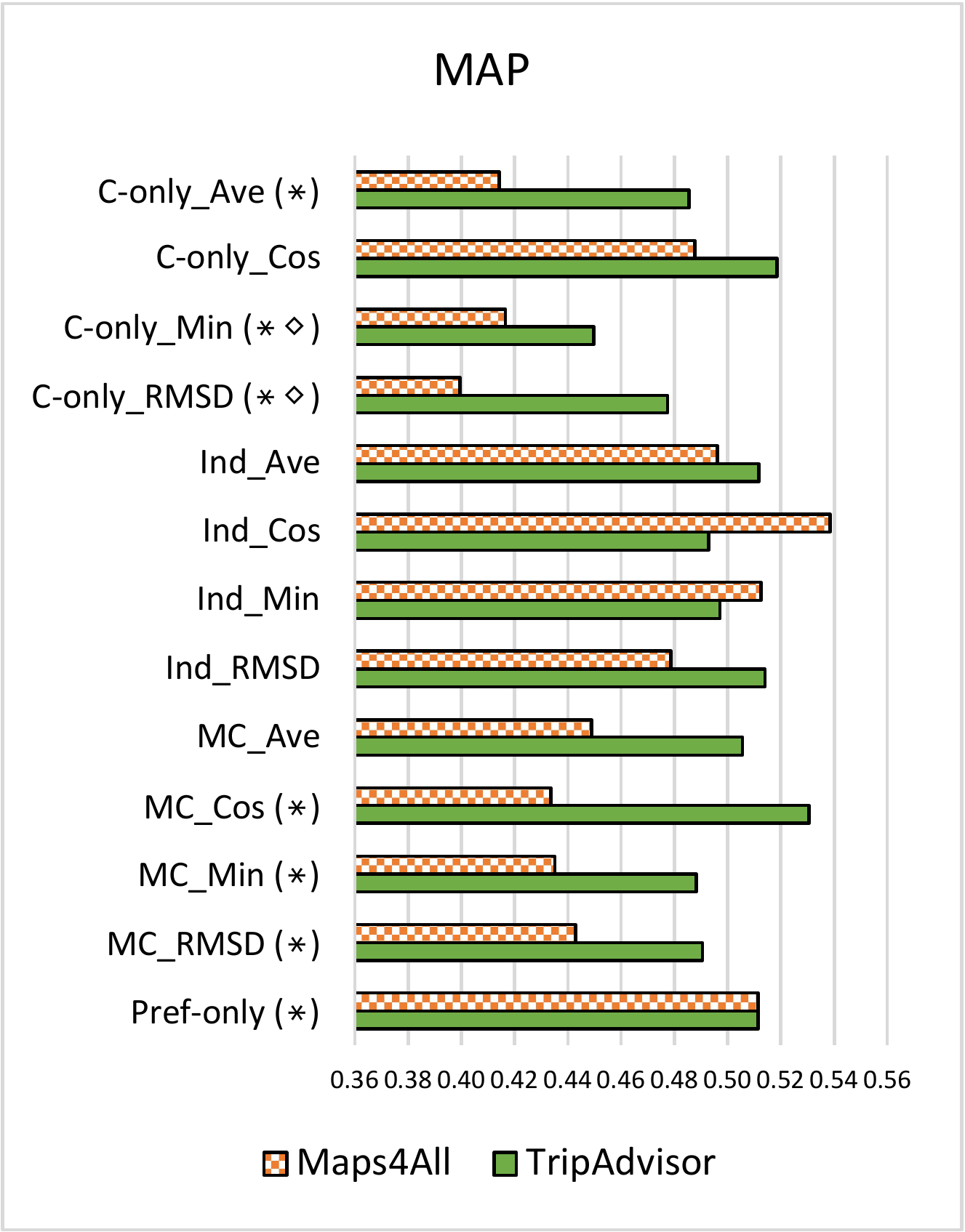}\hfill
    \includegraphics[width=.32\textwidth]{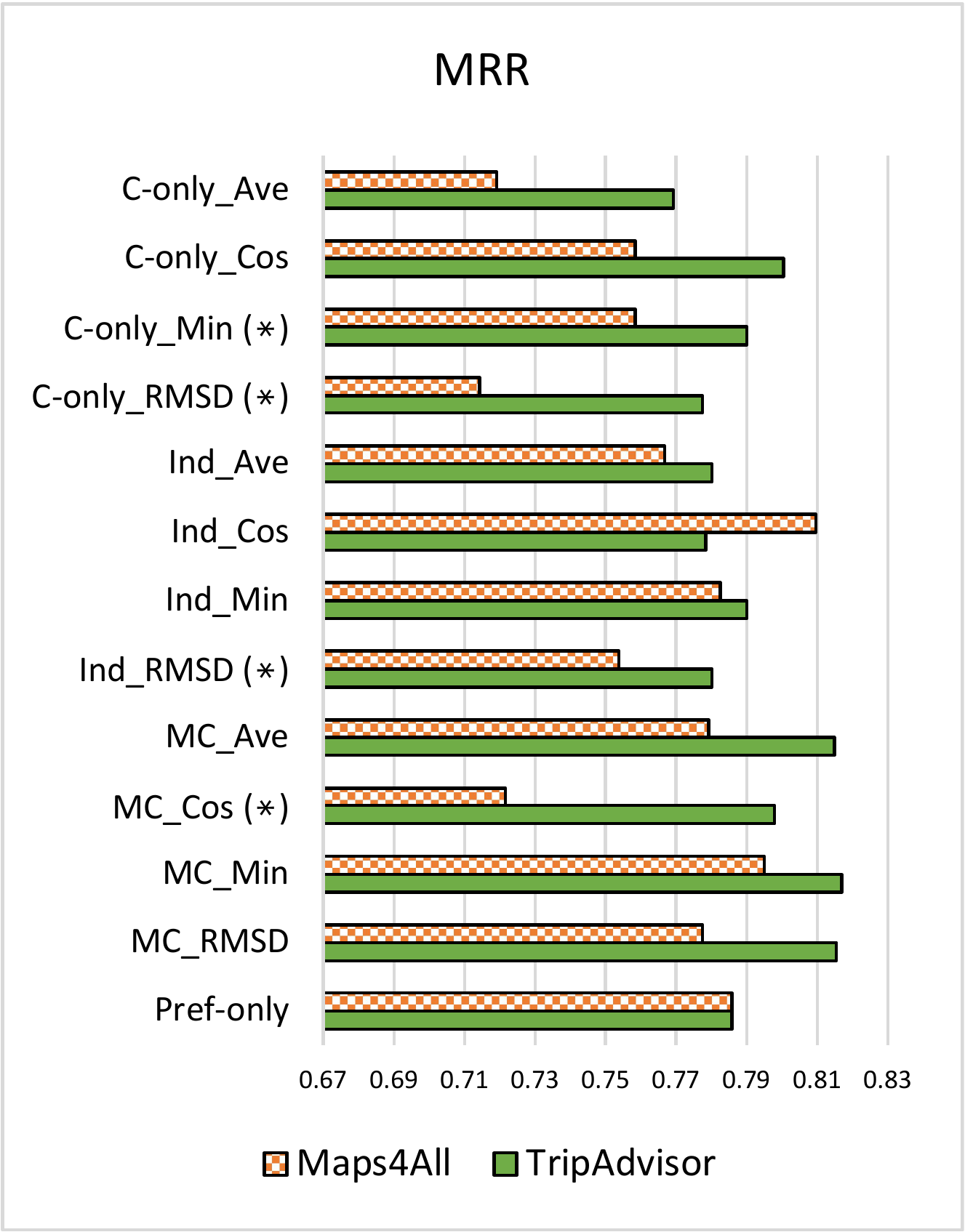}\hfill
    \includegraphics[width=.32\textwidth]{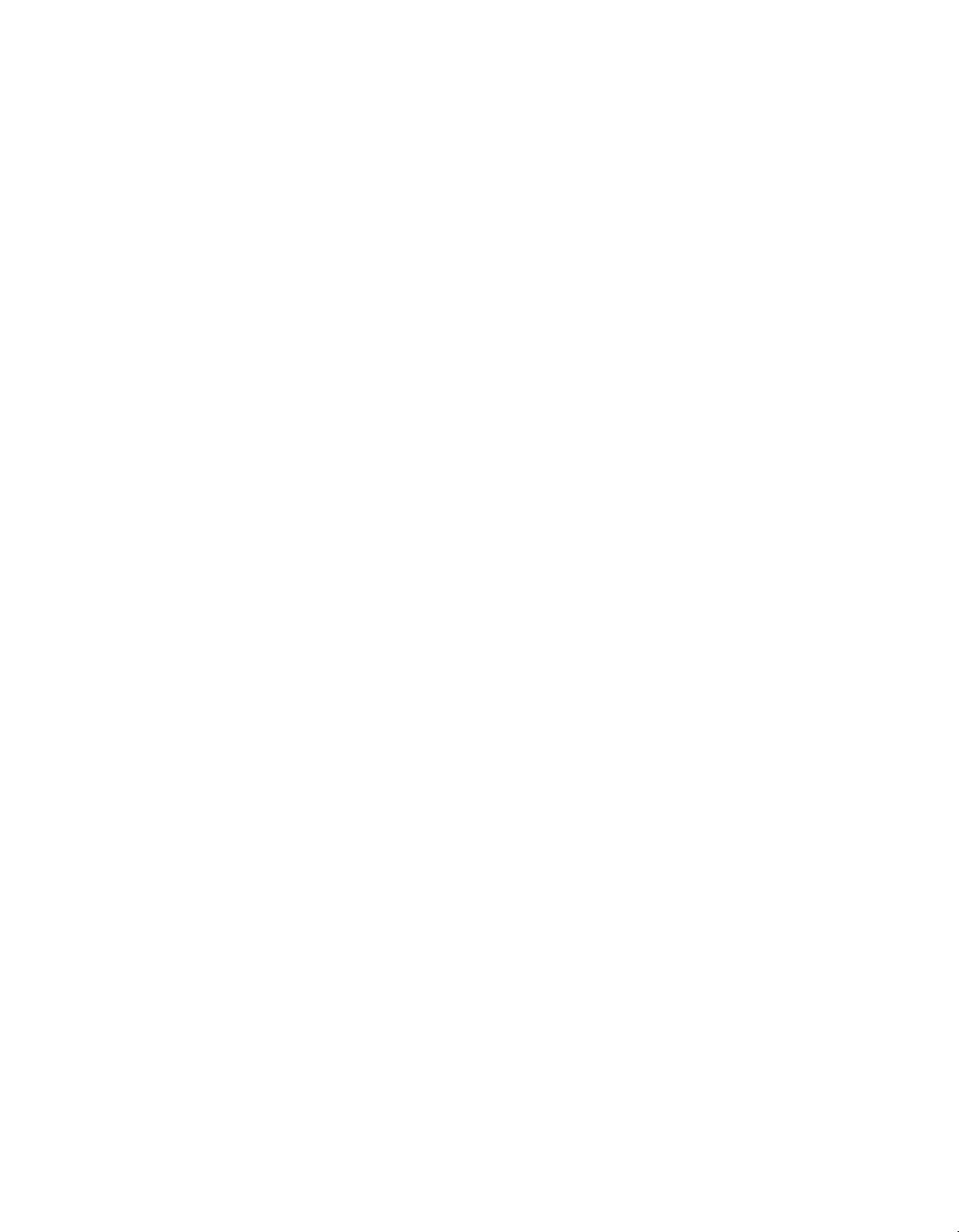}
        \\[\smallskipamount]
    \includegraphics[width=.32\textwidth]{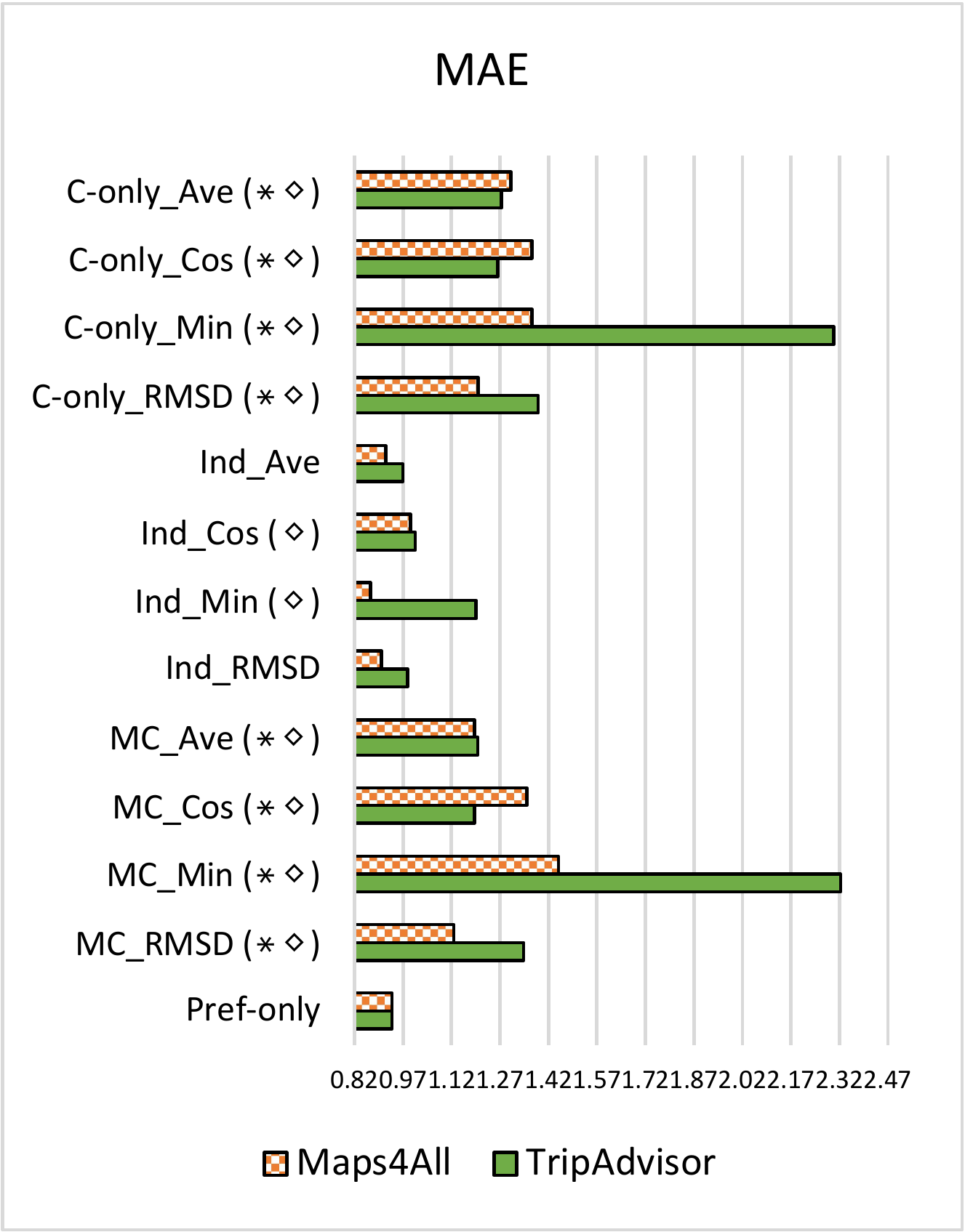}\hfill
    \includegraphics[width=.32\textwidth]{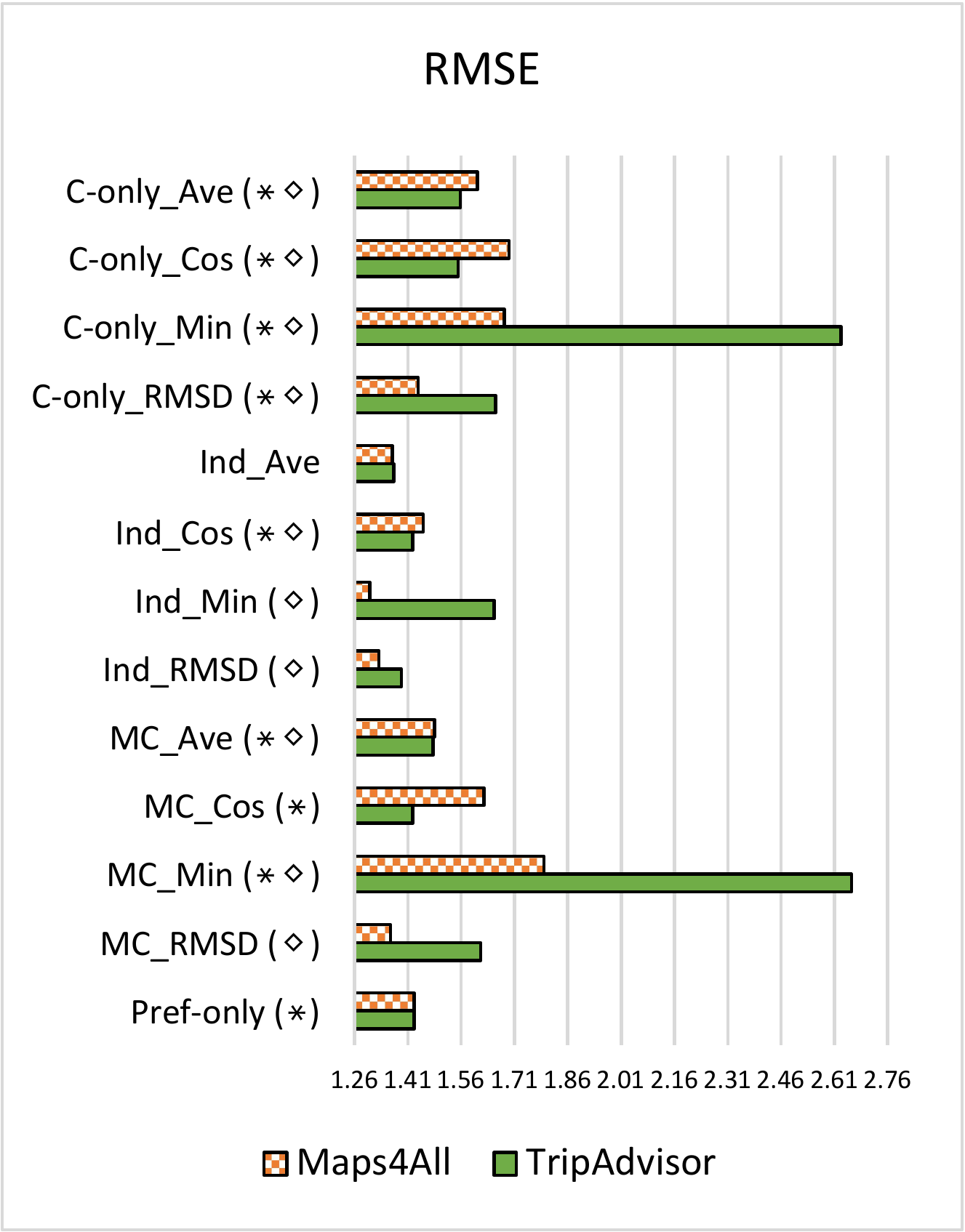}\hfill
    \includegraphics[width=.32\textwidth]{figures/empty.pdf}
\caption{Comparison of performance results using Maps4All and TripAdvisor on the AUT dataset, based on the 50 PoIs of set $\Pi$. 
Symbol ``$\ast$" denotes the statistical significance (t-test, $p<0.05$) of the difference between the best performing algorithm and the other ones on Maps4All. Similarly, ``$\diamond$" denotes significance on TripAdvisor. See Table \ref{tab:autistici}.}
\label{fig:results_aut}
\end{figure}

\begin{figure}

    \includegraphics[width=.32\textwidth]{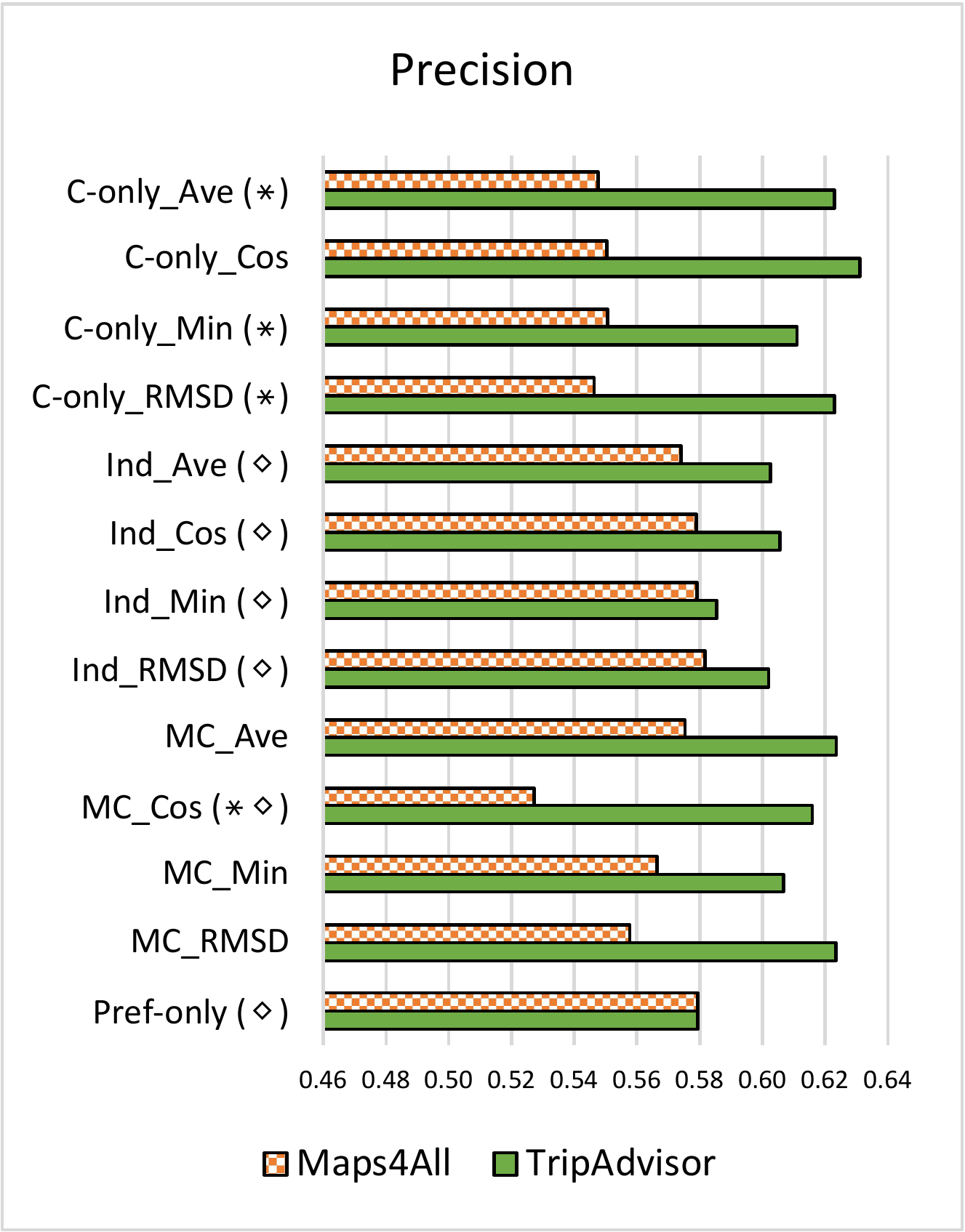}\hfill
    \includegraphics[width=.32\textwidth]{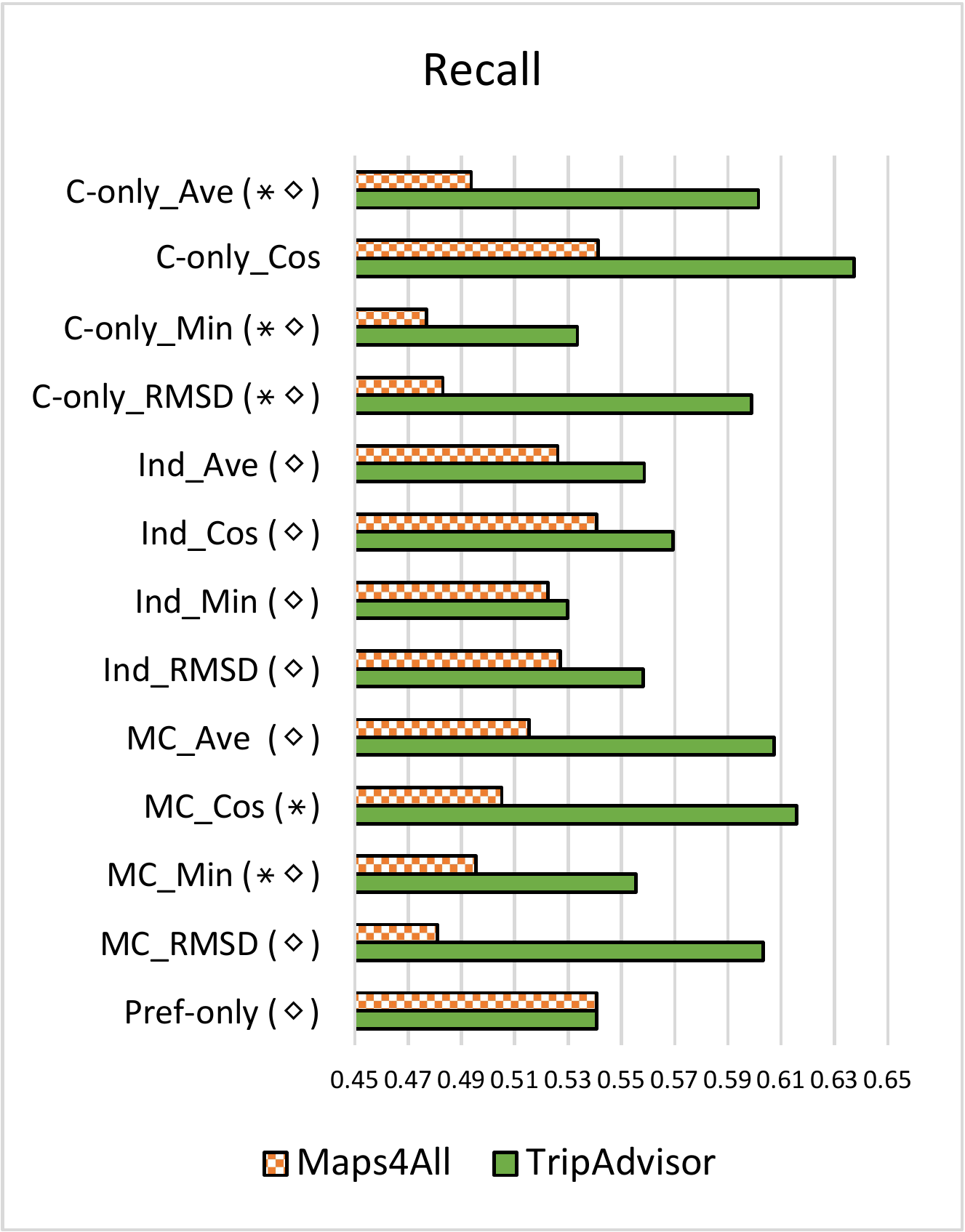}\hfill
    \includegraphics[width=.32\textwidth]{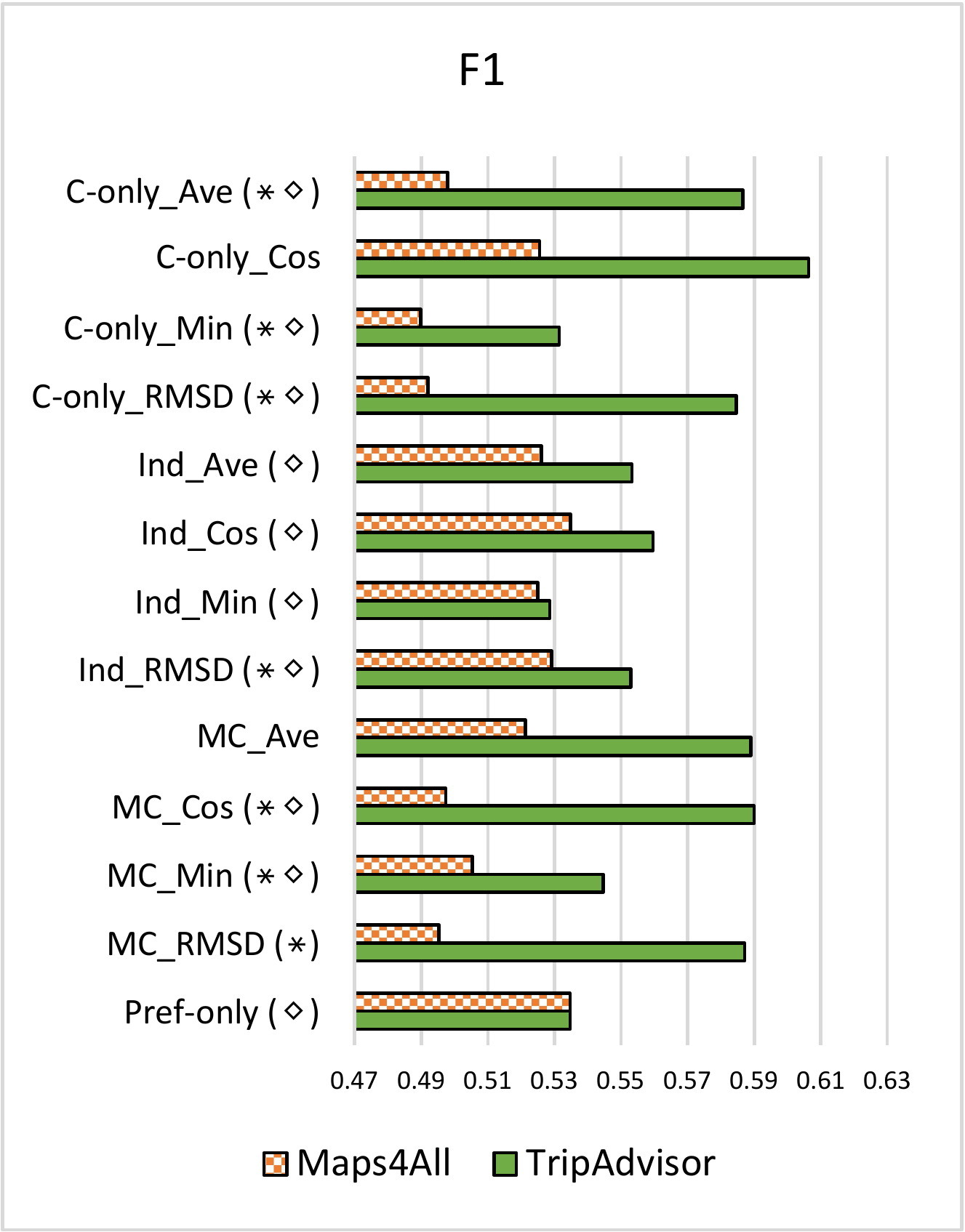}
    \\[\smallskipamount]
    \includegraphics[width=.32\textwidth]{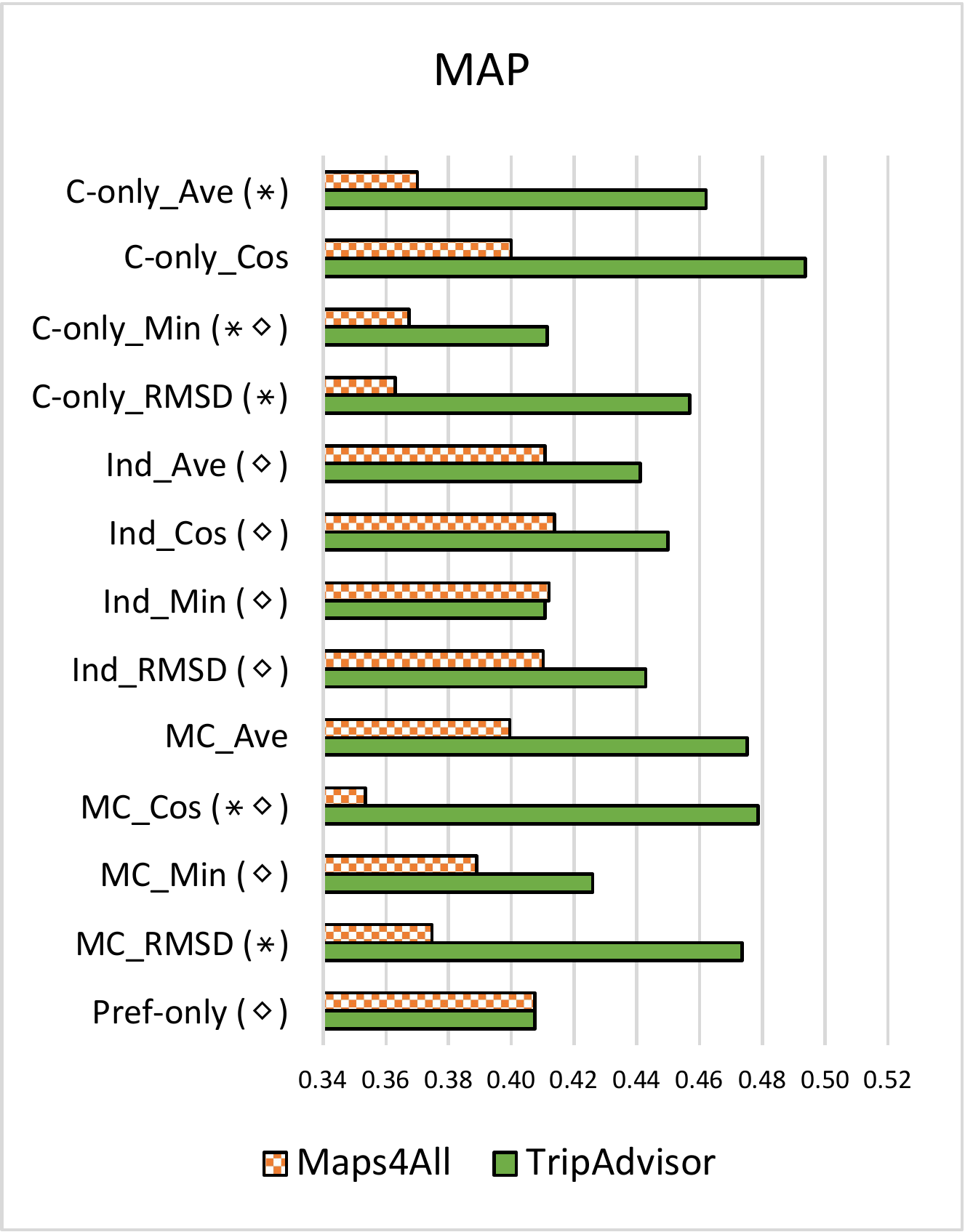}\hfill
    \includegraphics[width=.32\textwidth]{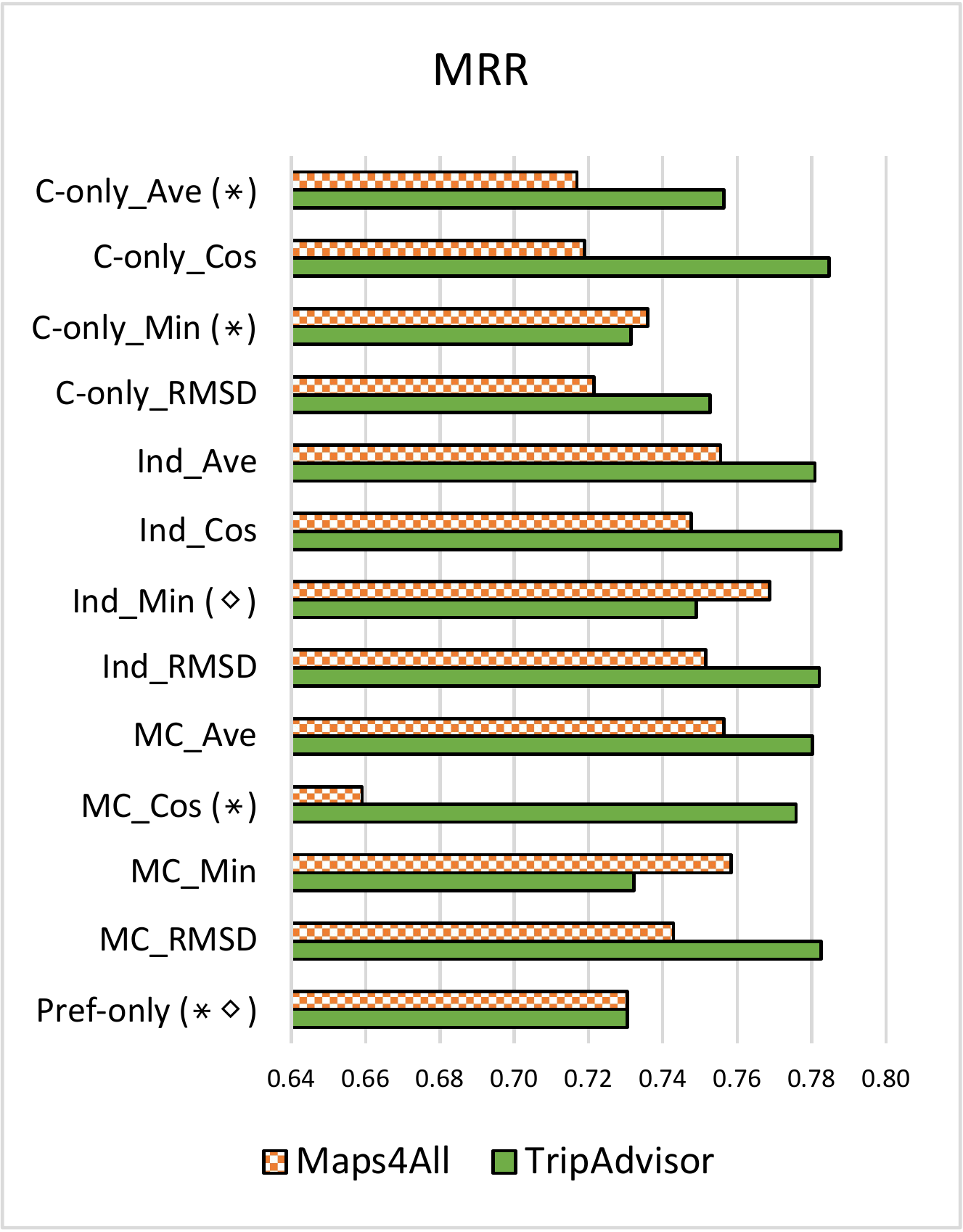}\hfill
    \includegraphics[width=.32\textwidth]{figures/empty.pdf}
        \\[\smallskipamount]
    \includegraphics[width=.32\textwidth]{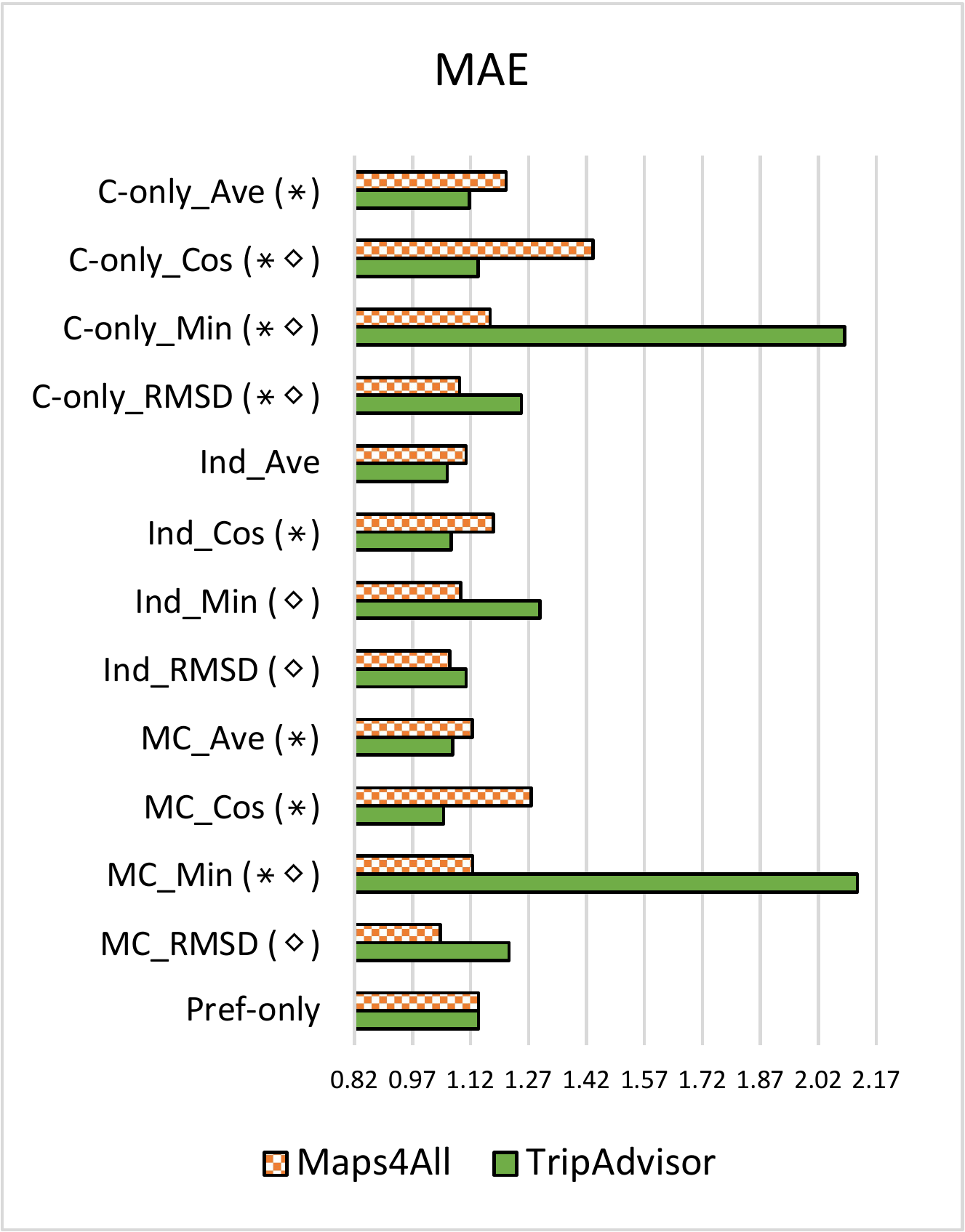}\hfill
    \includegraphics[width=.32\textwidth]{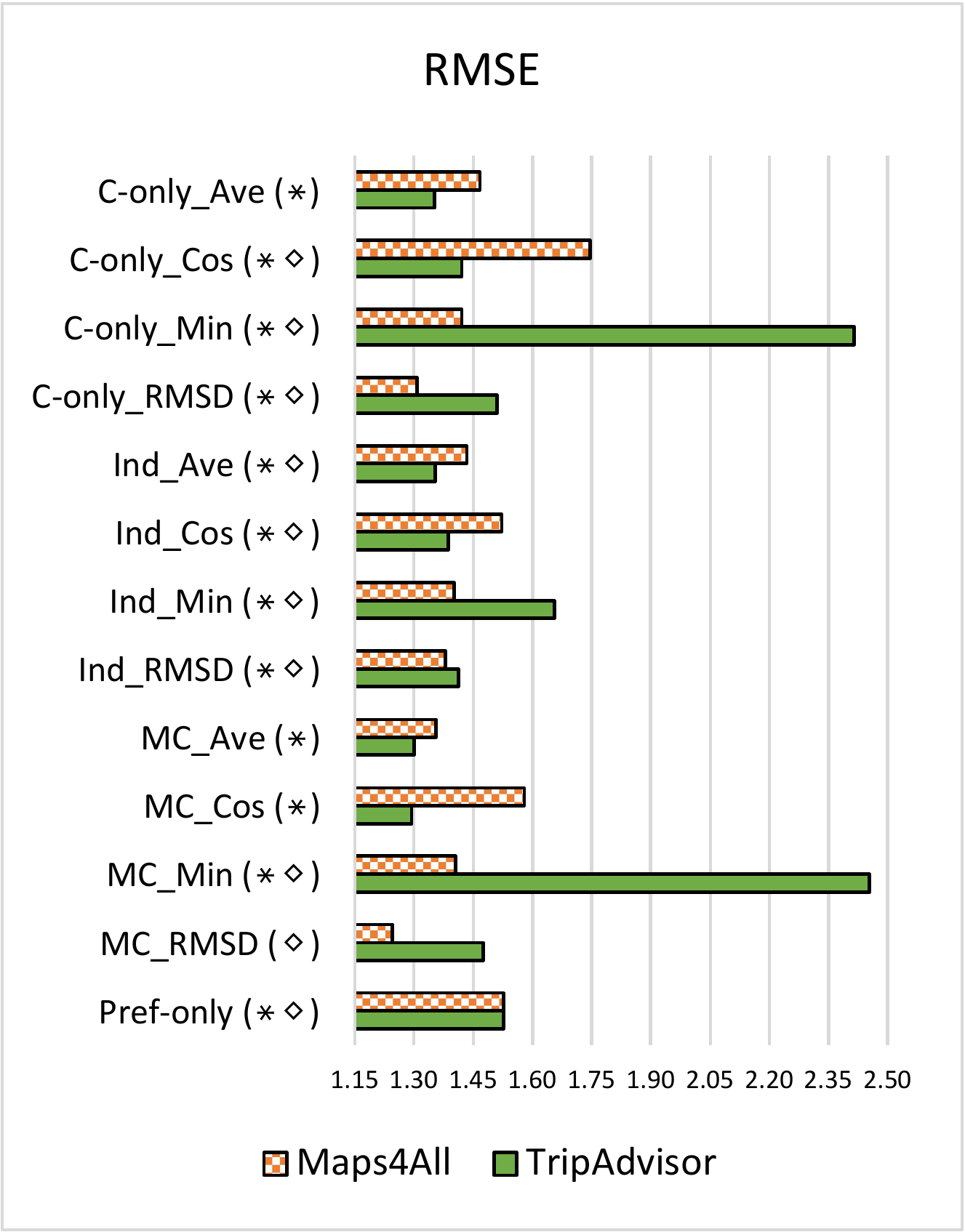}\hfill
    \includegraphics[width=.32\textwidth]{figures/empty.pdf}
    
\caption{Comparison of performance results using Maps4All and TripAdvisor on the NEU dataset, based on the 50 PoIs of set $\Pi$. We use the same notation as in Figure \ref{fig:results_aut}. See Table \ref{tab:neurotipici} for details.}
\label{fig:results_nor}
\end{figure}

\section{Evaluation results}
\label{sec:results}
\subsection{Comparing crowdsourced sensory information to consumer feedback}
\label{sec:comparison}
We first compare the recommendation performance of algorithms when they use either Maps4All or TripAdvisor for rating prediction, on the items of set $\Pi$ (50 PoIs in Torino city center). We evaluate the algorithms assuming that the recommendation list has length 5 because longer lists would overload people with autism, due to their attention problems \citep{murray2005attention}. 

Figures \ref{fig:results_aut} and \ref{fig:results_nor} graphically summarize the performance results concerning the users of the AUT and NEU datasets. See Tables \ref{tab:autistici} and \ref{tab:neurotipici} for details. We omit the results concerning user coverage because it is 100\% in all the cases.
The figures group results by accuracy, ranking capability, and error metrics. Notice that the \textsf{Pref-only} algorithm achieves the same results on both datasets because it only uses preference information. Therefore, it does not depend on how sensory data about places is retrieved. 

\subsubsection{Accuracy}
Most algorithms obtain higher accuracy on TripAdvisor than on Maps4All. In the AUT dataset, this happens to 10 algorithms regarding Recall and F1. Moreover, it happens to 11 regarding Precision. In the NEU dataset, this happens to 12 algorithms. 
This means that, by relying on sensory features extracted from consumer feedback, the recommender system suggests a larger number of PoIs that the user appreciates. This might be due to the fact that, compared to the low number of evaluations received by each place in Maps4All, the TripAdvisor reviews provide a more extensive amount of data about items. While this finding does not discriminate performance among algorithms, it encourages analyzing the online reviews collected from location-based services to build item profiles. 

We now compare the performance of individual algorithms on TripAdvisor, where they achieve superior results, to investigate the impact of item compatibility and user preferences on the accuracy of recommendations:
    \begin{itemize}
        \item 
        On the AUT dataset, \textsf{MC$_{Cos}$} has the highest F1 and Recall, and \textsf{C-only$_{Cos}$} is the second best. Moreover, \textsf{MC$_{Min}$} has maximum Precision, and \textsf{MC$_{Ave}$} is the second best. By focusing on F1, which summarizes accuracy, we can see that the difference between \textsf{MC$_{Cos}$} and most of the other \textsf{C-only} algorithms, which only use compatibility, is statistically significant. Similarly, the difference between \textsf{MC$_{Cos}$} and most of the other multi-criteria algorithms is significant. The accuracy of the \textsf{Ind} algorithms is lower but the results are not statistically significant.
        \item
        In NEU, \textsf{C-only$_{Cos}$} achieves better results than the other algorithms in the three metrics, and \textsf{MC$_{Cos}$} is the second best in Recall and F1. The difference between F1 of \textsf{C-only$_{Cos}$} and the other algorithms is statistically significant.
    \end{itemize} 
On both AUT and NEU, these algorithms have higher accuracy than \textsf{Pref-only}, which is agnostic with respect to compatibility information, with statistically significant differences on the NEU dataset.

Overall, the accuracy results support our hypothesis that compatibility information plays an important role in PoI recommendation.

\subsubsection{Ranking capability}
Most algorithms obtain better results when they use TripAdvisor than Maps4All. On the AUT dataset, this happens to 10 algorithms regarding MAP, and to 11 concerning MRR. On NEU, 11 algorithms have higher MAP and 9 have higher MRR. 
This finding supports the hypothesis that TripAdvisor is more effective than Maps4All in promoting items suitable for the user. Similar to the evaluation of accuracy, the algorithms that take both preferences and compatibility into account obtain higher results than \textsf{Pref-only}, which overlooks compatibility. However, the situation of the other algorithms is mixed and does not reveal a neat superiority of a specific way to combine these two types of information.

On the AUT dataset, \textsf{Ind$_{Cos}$} has the highest MAP and MRR on Maps4All, with a statistically significant difference of MAP compared to most \textsf{C-Only} and \textsf{MC} algorithms. However, on TripAdvisor, where algorithms perform better, the multi-criteria models achieve the best results: \textsf{MC$_{Cos}$} excels in MAP, and \textsf{MC$_{Min}$} in MRR (most results are not statistically significant).

On the NEU dataset, the \textsf{Ind} models achieve the best results on Maps4All. However, on TripAdvisor, \textsf{C-only$_{Cos}$} has the best MAP, with a statistically significant difference compared to most of the other algorithms. Moreover, \textsf{Ind$_{Cos}$} excels in MRR with poor statistical significance.

\subsubsection{Error in rating estimation}
Consumer feedback supports rating estimation in a less satisfactory way. On the AUT dataset, only 3 (respectively 5) algorithms obtain lower MAE (RMSE) when using TripAdvisor; the other ones work better on Maps4All. Moreover, on the NEU dataset, only 6 algorithms achieve lower rating estimation errors on TripAdvisor than on Maps4All.
    
The comparison between algorithms provides mixed results, as well. On the AUT dataset with Maps4All data, the best model is \textsf{Ind$_{Min}$} with statistically significant difference compared to the other ones. The second best is \textsf{Ind$_{RMSD}$} on both MAE and RMSE. 
We notice that the most pessimistic algorithms, which set item compatibility to the minimum one (e.g., \textsf{C-only$_{Min}$} and \textsf{MC$_{Min}$}), have low performance.

Differently, on NEU, multi-criteria models work better than the other ones. The best algorithms are \textsf{MC$_{RMSD}$} on Maps4All, and \textsf{MC$_{Cos}$} in TripAdvisor. In both cases, the results are statistically significant.
\textsf{Pref-only} is fairly good but, on both AUT and NEU, several algorithms that use compatibility information perform better than it.

\subsubsection{Overall performance}
Concerning the accuracy and ranking capabilities, the best algorithms are the multi-criteria ones. 
Notice that the promotion of good items at the top of a recommendation list is a prior goal to be achieved because a low number of items can be realistically proposed to users in the autism spectrum disorder. Thus, the improvement of ranking capability obtained by extracting sensory data about places from reviews is a particularly relevant result.  
The results concerning the error metrics are mixed but they show a superiority of the models that take both user preferences and item compatibility into account, compared to those that use a single type of information.

We found that rating estimation is not equally well-supported by consumer feedback.
Indeed, we believe that this weak performance might be caused by a lack of data about PoIs. As discussed in Section \ref{sec:feedback}, some sensory features, such as \texttt{smell}, are poorly covered in TripAdvisor. Moreover, only 34 places out of the 50 of set $\Pi$ are evaluated in TripAdvisor, against the 49 of Maps4All. This means that the algorithms we tested on TripAdvisor frequently worked blindly, assuming by default a maximum incompatibility between individual features and the user. 
This aspect is likely responsible for the bad rating prediction results of the algorithms that use the $Min$ aggregation strategy (\textsf{C-only$_{Min}$} and \textsf{MC$_{Min}$}) because, if a single feature value is unknown, they propagate the incompatibility to the whole item. However, as discussed in Section \ref{sec:perceptualNeeds}, when suggesting places to autistic people, we have to avoid any possible source of discomfort and stress. Thus, our pessimistic approach to the estimation of sensory feature compatibility is a must. At the same time, we believe that rating estimation might be improved by facing data sparsity. For instance, multiple consumer feedback sources might be integrated, such as different location-based services, with the aim of retrieving richer information about places.

\begin{figure}[!t]

     \begin{subfigure}[b]{1\textwidth}
         \centering
    \includegraphics[width=.32\textwidth]{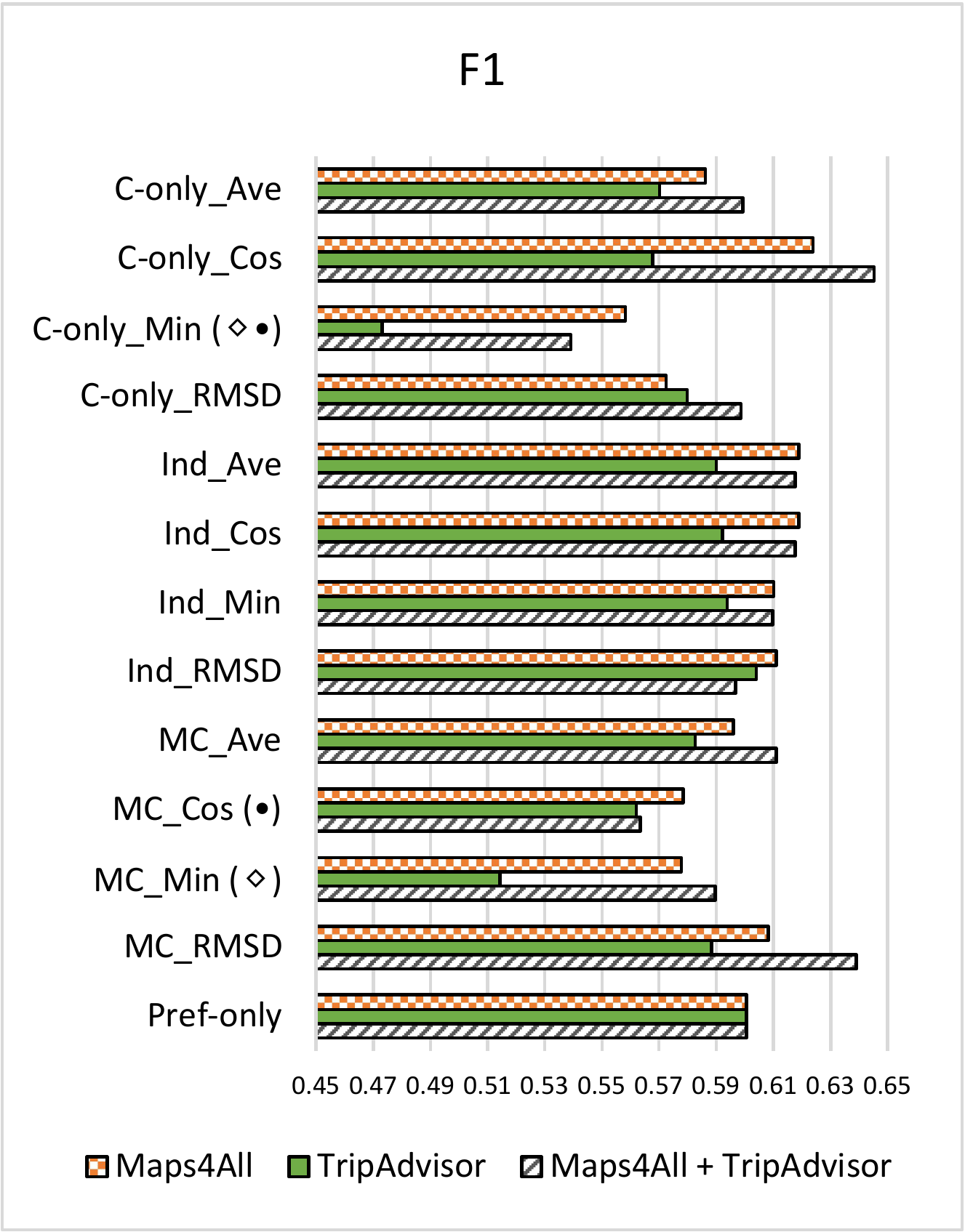}\hfill
    \includegraphics[width=.32\textwidth]{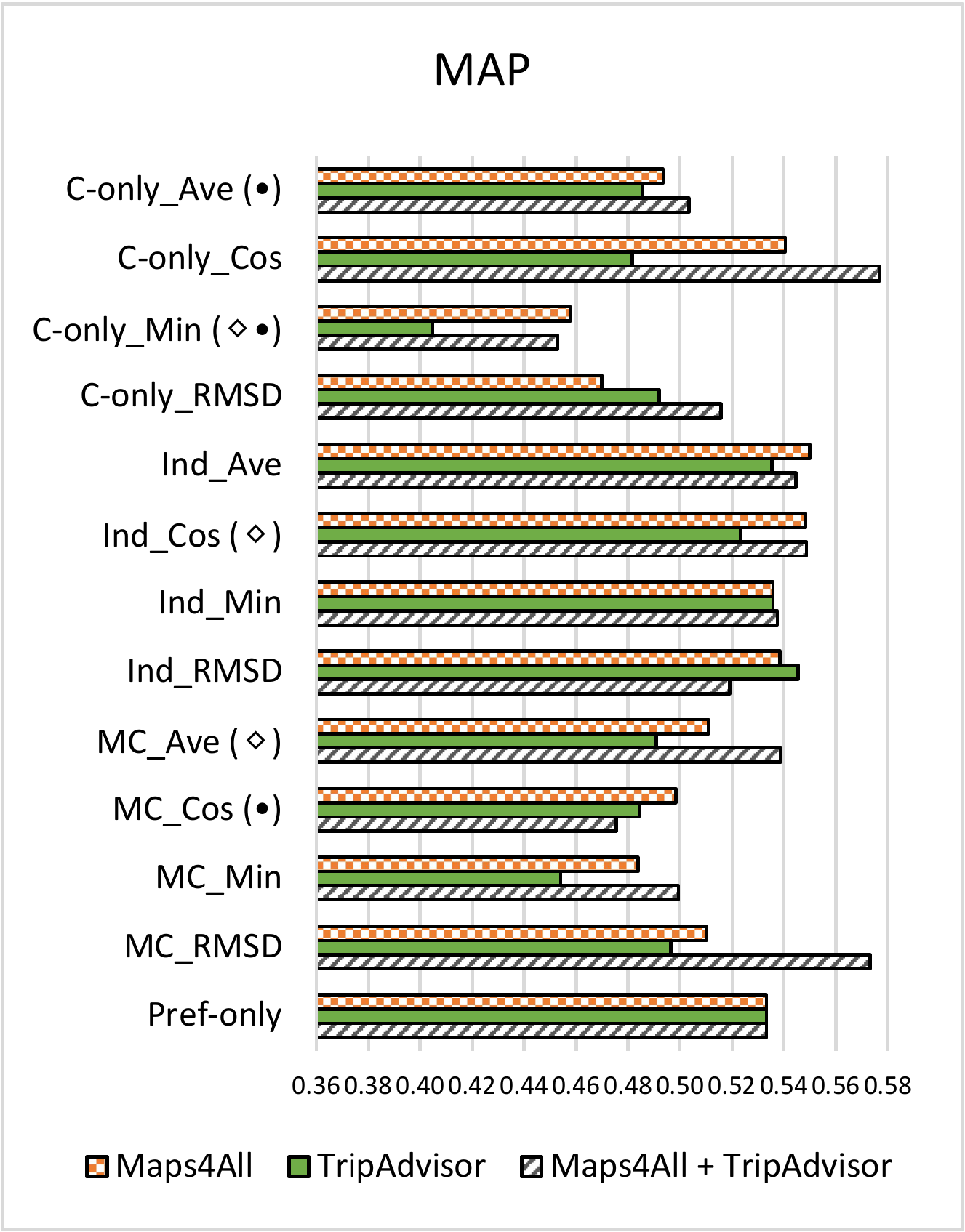}\hfill
    \includegraphics[width=.32\textwidth]{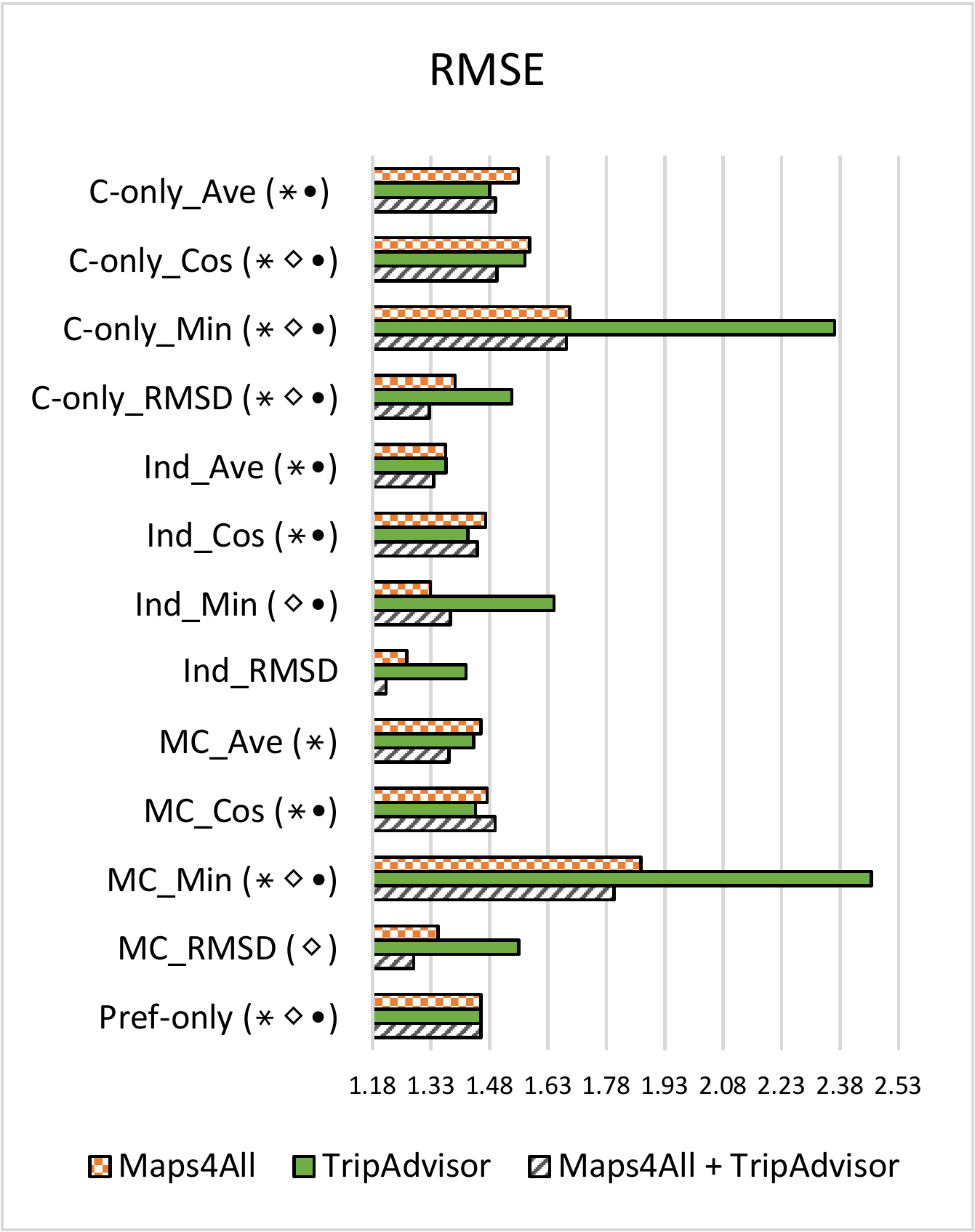}
    \caption{AUT dataset. See Tables \ref{tab:autistici_34} and \ref{tab:autistici_34_mix} for details.}
    \label{fig:results_aut_34}
     \end{subfigure}
   
   \vspace{5mm}
    \begin{subfigure}[b]{1\textwidth}
        \centering  
    \includegraphics[width=.32\textwidth]{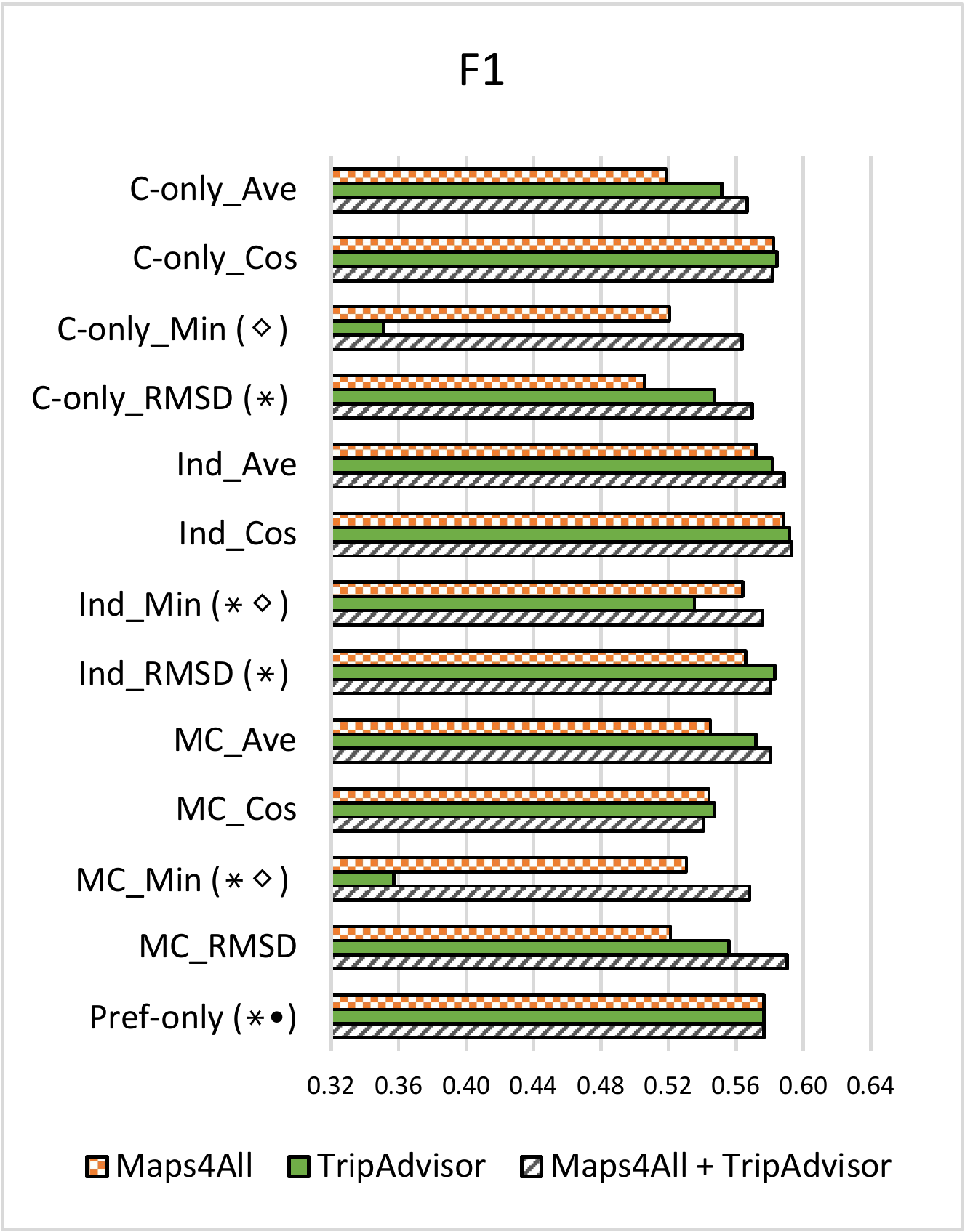}\hfill
    \includegraphics[width=.32\textwidth]{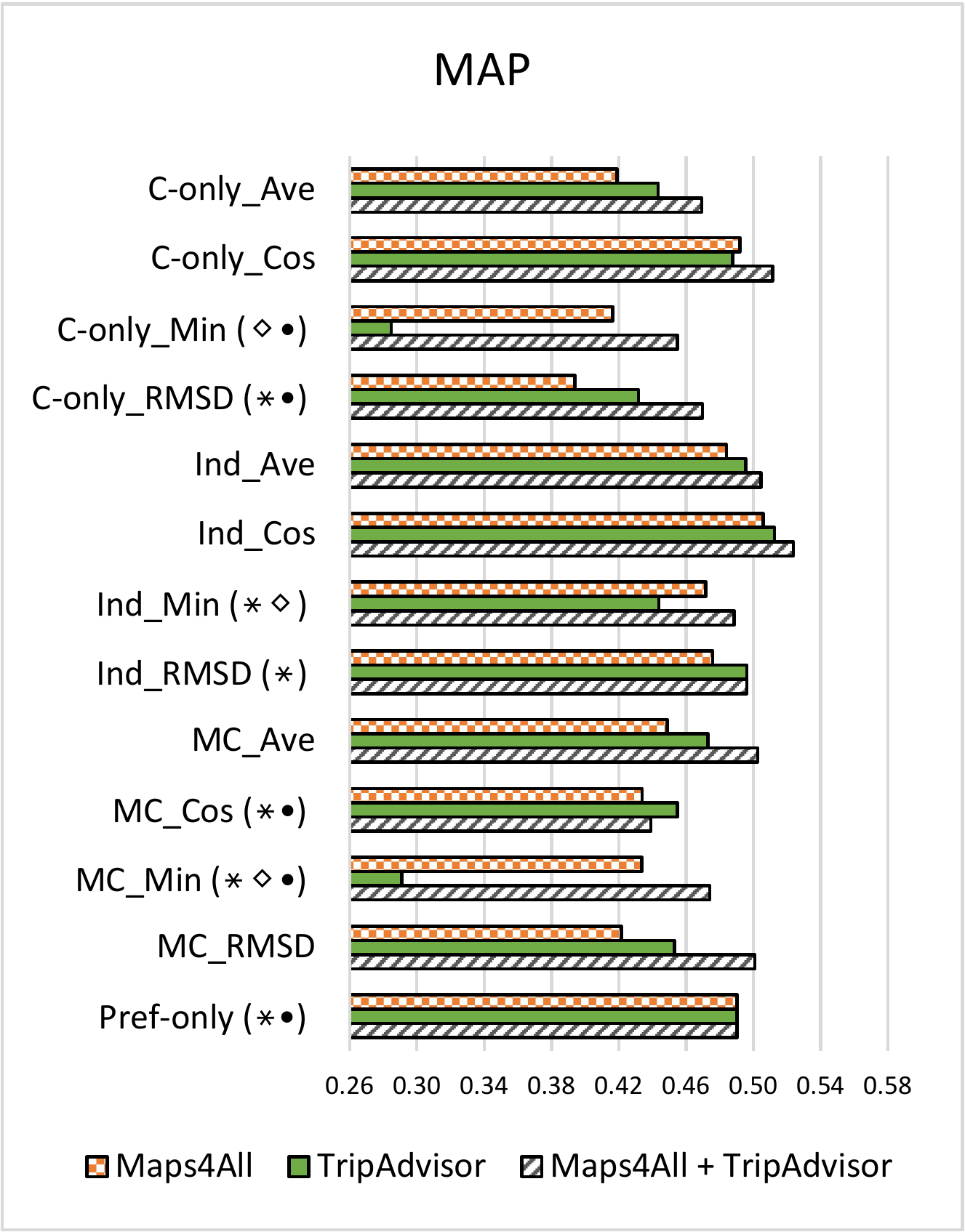}\hfill
    \includegraphics[width=.32\textwidth]{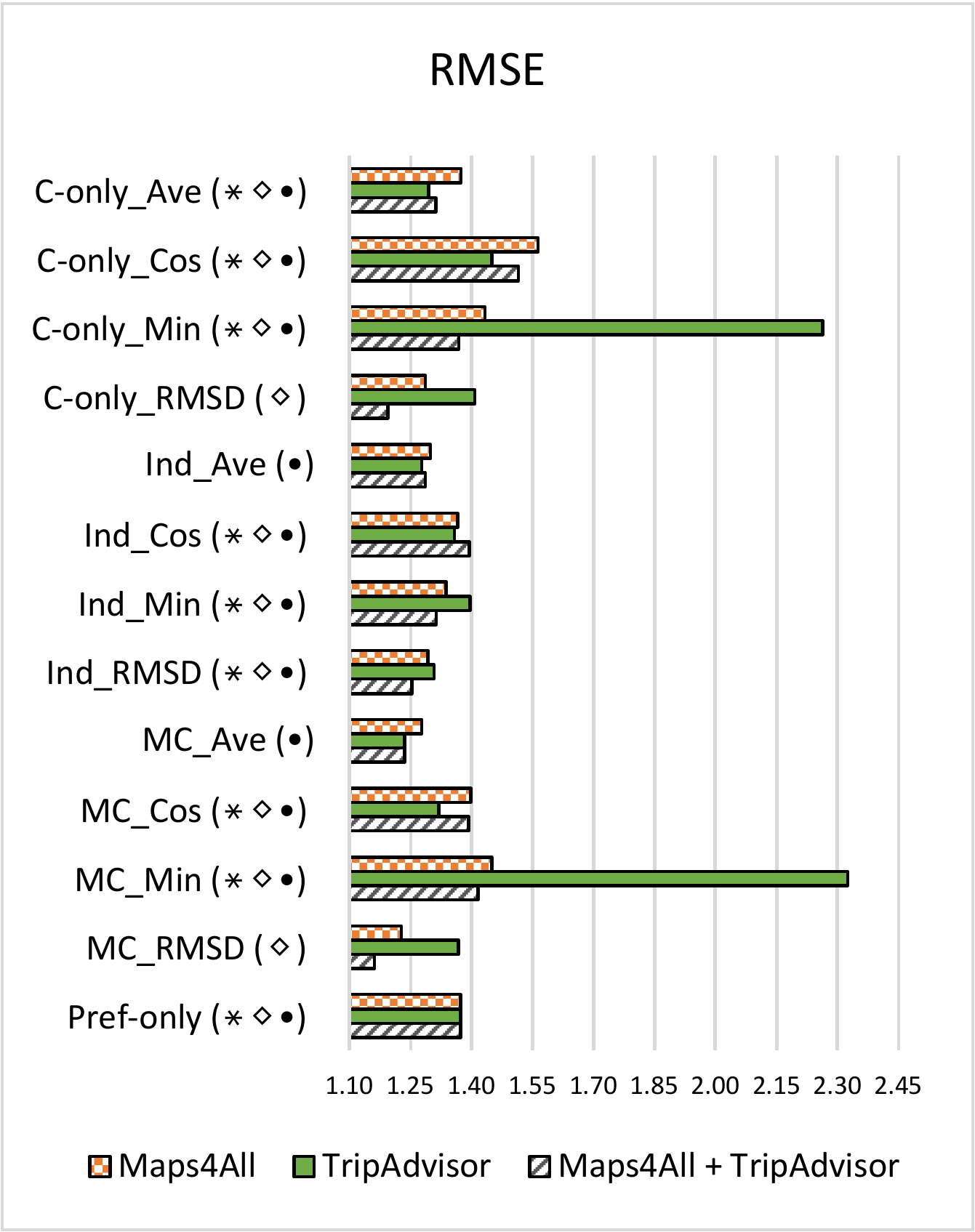}
        \caption{NEU dataset. See Tables \ref{tab:neurotipici_34} and \ref{tab:neurotipici_34_mix} for details.}
    \label{fig:results_nor_34}
     \end{subfigure}

\caption{Comparison of performance results using data about PoIs from Maps4All, TripAdvisor, or by fusing them in the \textsf{M+T} model. All the results concern the 34 places of set MA$\cap$TA. Symbol ``$\ast$" denotes the statistical significance (t-test, $p<0.05$) of the difference between the best performing algorithm and the other ones on Maps4All. Similarly, ``$\diamond$" (respectively ``$\bullet$") denotes significance on TripAdvisor (resp. fusion of Maps4All and TripAdvisor).}
\label{fig:results_34}
\end{figure}

\subsection{Integration of multiple data sources}
\label{sec:integration}
To assess the usefulness of sensory data extracted from consumer feedback, we shortly compare the performance achieved when separately using Maps4All or TripAdvisor data sources to that obtained when merging them in rating prediction (\textsf{M+T} model). In this case, instead of measuring performance on the whole set $\Pi$ of places, we focus on its 34 places that are mapped by both data sources, i.e., on set (MA$\cap$TA) of Section \ref{sec:intersection}. The reason for this choice is that we aim at understanding the usefulness of combining data sources when they can both provide at least partial information about places.

In the \textsf{M+T} model, we fuse data by computing the weighted average of feature values. In this way, we tune the impact of the two data sources in the estimation of feature values based the amount of available data about sensory features. Moreover, if a single data source provides information about a feature, it compensates the lack of knowledge affecting the other one. For each $i \in I$, for each $f \in F$:
\begin{equation}
    \vec{{\bf i}}_f^{\,} = \frac{n_1 val_{f_{Maps4All}} + n_2 val_{f_{TripAdvisor}}} {n_1+n_2}
    \label{eq:weightedMean}
\end{equation}
where $val_{f_{Maps4All}}$ (respectively $val_{f_{TripAdvisor}}$) represents the value of $f$ provided by Maps4All (resp. TripAdvisor) and $n_1$ (resp. $n_2$) is the number of feature evaluations on which this value is based. 

Figure \ref{fig:results_34} summarizes the accuracy, ranking capability, and rating prediction error of algorithms by considering F1, MAP, and RMSE in the three cases (only Maps4All, only TripAdvisor, \textsf{M+T}). We can see that 6 algorithms (AUT dataset) and 9 algorithms (NEU dataset) improve their F1 when the data retrieved from Maps4All and TripAdvisor is merged using Equation \ref{eq:weightedMean}.
Moreover, in that case, 8 algorithms (AUT) and 10 algorithms (NEU) improve their MAP.
Furthermore, 8 algorithms (AUT) and 7 algorithms (NEU) improve their RMSE. See Tables \ref{tab:autistici_34}, \ref{tab:autistici_34_mix}, \ref{tab:neurotipici_34}, and \ref{tab:neurotipici_34_mix} for details. 

Even though results are statistically significant in a few cases, they are consistent with the hypothesis that recommendation performance can be improved by combining different information sources to retrieve sensory feature evaluations. We can explain this finding with the fact that the recommender system leverages a larger amount of data and integrates missing information by retrieving it from the source that provides it.

\subsection{Analysis of the $\alpha$ weights for the \textsf{Ind} algorithms}
The optimization of the \textsf{Ind} algorithms, which personalize the balance of user preferences and compatibility to the individual user through the $\alpha$ weight of Equation \ref{eq:est_rating}, reveals interesting findings about the perception of places in the user population. As these algorithms achieve rather good performance in several evaluation metrics, they can provide evidence about how people weight these two types of information in the evaluation of places.

Table \ref{tab:weights} shows the average $\alpha$ weights for the different configurations of the \textsf{Ind} algorithm.
Surprisingly, in some cases, the $\alpha$ weights are higher on the NEU dataset than on the AUT one. 
This supports the hypothesis that, to some extent, both autistic and neurotypical people are susceptible to sensory features of places. At the same time, even though these features can cause uncomfortable feelings to people with autism, preferences are important as well, and sometimes users are willing to overcome their aversions if they really like a place. See \cite{Mauro-etal:20} for details about this.

\begin{table}[t]
\centering
\resizebox{\textwidth}{!}{%
\begin{tabular}{@{}lll|ll|ll|ll@{}}
\toprule
                       & \multicolumn{2}{c}{\textsf{Ind$_{Ave}$}}     & \multicolumn{2}{c}{\textsf{Ind$_{Cos}$}}  & \multicolumn{2}{c}{\textsf{Ind$_{Min}$}}     & \multicolumn{2}{c}{\textsf{Ind$_{RMSD}$}}     \\ \midrule
                       & \multicolumn{2}{c}{50 PoIs} & \multicolumn{2}{c}{50 PoIs} & \multicolumn{2}{c}{50 PoIs} & \multicolumn{2}{c}{50 PoIs} \\ \midrule
                       & AUT          & NEU          & AUT          & NEU          & AUT          & NEU          & AUT          & NEU          \\ \midrule
                       & Mean(SD)        & Mean(SD)        & Mean(SD)        & Mean(SD)        & Mean(SD)        & Mean(SD)        & Mean(SD)        & Mean(SD)        \\ \midrule
Maps4All               & 0.180(0.282)  & 0.327(0.406) & 0.230(0.365) & 0.322(0.401) & 0.200(0.304) & 0.347(0.401) & 0.180(0.271)   & 0.298(0.380) \\
TripAdvisor            & 0.260(0.299) & 0.363(0.397) & 0.310(0.373) & 0.416(0.428) & 0.285(0.328) & 0.341(0.381) & 0.305(0.329) & 0.373(0.393) \\ \midrule
                       & \multicolumn{2}{c}{34 PoIs} & \multicolumn{2}{c}{34 PoIs} & \multicolumn{2}{c}{34 PoIs} & \multicolumn{2}{c}{34 PoIs} \\ \midrule
Maps4All               & 0.275(0.392) & 0.251(0.376) & 0.320(0.440) & 0.322(0.417) & 0.235(0.356) & 0.281(0.385) & 0.280(0.375) & 0.264(0.384) \\
TripAdvisor            & 0.315(0.369) & 0.329(0.405) & 0.315(0.398) & 0.426(0.442) & 0.310(0.346) & 0.246(0.348) & 0.325(0.370) & 0.317(0.396) \\
Maps4All + TripAdvisor & 0.335(0.430) & 0.272(0.396) & 0.535(0.479) & 0.437(0.457) & 0.380(0.435) & 0.316(0.404) & 0.445(0.444) & 0.329(0.426) \\ \bottomrule
\end{tabular}%
}
\caption{Average $\alpha$ weights for the \textsf{Ind} algorithm.}
\label{tab:weights}
\end{table}

\section{Discussion}
\label{sec:discussion}
The experimental results allow us to positively answer our research questions. 
Concerning RQ1, we found that a relevant amount of sensory information about places can be extracted from the reviews collected in location-based services such as TripAdvisor, provided that they map the PoIs that the recommender system deals with. Especially for some features, such as \texttt{crowding} and \texttt{openness}, reviews offer rich information that can be reliably used to steer the suggestion of places to the individual user. Indeed, location-based services are particularly valuable because they represent a sustainable source of sensory data about PoIs, fed with a continuous, spontaneous reviewing activity concerning places distributed all over the world.

Concerning RQ2, we found that sensory data extracted from TripAdvisor reviews is useful because it improves accuracy and ranking capability in recommendation algorithms that only use compatibility information about items, or which combine it with user preferences.
Moreover, when merging this data with crowdsourced sensory information, the algorithms obtain better accuracy, ranking capability, and error minimization than when using a single data source. 
As all the results concern both users with autism and neurotypical ones, these findings show that consumer feedback is a precious type of information for the development of inclusive recommender systems.

These results have important practical implications. 
Regarding the specific target of our work, our approach supports the development of compatibility-aware recommender systems that can serve several locations, instead of being constrained to restricted areas where sensory information has been specified. Our model can be applied to large geographical areas, or to areas spread all over the world, because the knowledge base of the recommender system can be fed in an automatic way through a continuous analysis of the consumer feedback collected by social media and location-based services. In turn, this might dramatically help people with autism because it would extend the availability of a technological support while they are on the move, thus minimizing the level of stress and improving their quality of life.
On a different perspective, the applicability of our approach makes it adaptable to different targets. Even though we currently focus on autistic users, our approach can be useful to other fragile people, as well. In fact, the integration of compatibility in the evaluation of the suitability of items to the user makes it possible to deal with different sources of incompatibility between places and users, and thus with other types of disability. For instance, we might apply our approach to focus the recommendation algorithm not only on sensory aversions, but also on other specific user constraints and needs, such as trying to avoid architectural barriers for people with physical impairments (OpenStreetMap and other similar platforms provide some information about wheelchair access to places).

\section{Limitations and future work}
\label{sec:limitations}
The experiments we carried out show that our approach depends on the geographical coverage of the external data sources we exploit to retrieve sensory information about places. In this respect, we plan to extend our model in three ways. First, we will integrate in our feature extraction model further data sources, such Yelp and Google Maps, to retrieve sensory data about a larger number of places.
Second, we will extend the analysis of reviews to infer feature values by exploiting the correlations among sensory features that we found by analyzing the Maps4All and TripAdvisor datasets.
However, this inference is subject to uncertainty, which should be considered in the recommendation algorithms.
Third, we plan to investigate the use of generative models to address data sparsity.

Another limitation of our work is the fact that we recommend places by analyzing the user's interests in the categories of places, but not in their features. 
We plan to acquire fine-grained data from geographical servers such as OpenStreetMap, and to extract features of places from consumer feedback, to manage fine-grained user preferences in the user profiles.

Currently, we are integrating the approach described in this paper into the PIUMA mobile guide \citep{Cena-etal:20, DBLP:conf/interact/CenaRMMACBKT21} which suggests places to visit to people with autism.
We then plan to test our recommender systems in the field, by carrying out a user study with people from the Adult Autism Center of Torino. So far, we could only perform offline experiments because the center was closed due to Covid-19 pandemic and thus we could not interact with its guests. The development of this app will make it possible to acquire precise evaluation data about PoIs and to know the identities of the people who have provided feedback about sensory features. This opens a research avenue towards the exploitation of information diffusion models in recommender systems, similar to what has been done in \citep{Xiong-etal:20,Xiong-etal:20b} for Matrix Factorization.

Our future work also includes a cooperation with psychologists to develop novel recommendation algorithms that are robust with respect to individual biases in the evaluation of sensory features. In fact, as the perception of places is subjective, the feature values extracted from consumer feedback, or explicitly crowdsourced, might be biased. Thus, the evaluation of compatibility with a specific user might be affected by uncertainty.

\section{Conclusions}
\label{sec:conclusions}
Users with autism spectrum disorders are a particularly interesting target of PoI recommender systems because of their specific needs regarding places. To suggest PoIs that they can like and serenely experience, both their preferences and aversions to sensory features must be considered. In fact, the compatibility of items with a user's aversions can seriously affect her/his experience with places, causing negative feelings. 

Given the difficulties in retrieving sensory data from geographic information servers, we proposed a model to extract this type of information from the consumer feedback collected by location-based services. 
We compared the performance of a set of recommender systems on sensory data about places gathered in a crowdsourced campaign, from TripAdvisor reviews, or from both data sources. By using consumer feedback, the systems obtained higher accuracy and ranking capability. By fusing the two data sources, they achieved even higher accuracy, ranking capability, and they improved rating prediction. We also found that the algorithms that use compatibility in rating estimation outperform those that only rely on user preferences.

We conclude that the integration of user interests and sensory aversions is a promising approach to extend the target user groups of recommender systems. Concerning people with autism spectrum disorders, compatibility-aware recommender systems can reduce the level of stress perceived in moving within a city and increase autonomy.
Notice that the extraction of sensory feature evaluations from consumer feedback can be used when the sensory data is scarce to improve the quality of the suggestions. Moreover, it can be used to increase the number of places that can be mapped in a city, and it is more sustainable than a crowdsourced campaign. 

\section{Acknowledgments}
This work is supported by the Fondazione Compagnia di San Paolo. We thank the colleagues of our Department for the support in the work and the Adult Autism Center of the City of Torino for the recruitment of the subjects who participated in our experiments. We also thank the anonymous reviewers of this paper for their thoughtful comments and suggestions.


\appendix

\section{Detailed results using the 50 PoIs of set $\Pi$.}
\begin{table}[hbt!]
\centering
\caption{Top-N recommendation results on AUT dataset with N=5, using the information about the 50 PoIs of the $\Pi$ set. The lines of the table are ordered by MAP. The best value of each measure across all algorithms is printed in bold, the second best one is underlined. For each evaluation metric, ``$\ast$" denotes the statistical significance (t-test, $p<$0.05) of the difference between the best performing algorithm and the other ones on Maps4All. Similarly, ``$\diamond$" denotes significance on TripAdvisor.
}
\vspace{5pt}
\resizebox{0.8\columnwidth}{!}{%
{\def\arraystretch{0.9}
\begin{tabular}{lrrrrrrr}
\toprule
\multicolumn{8}{c}{\textbf{Maps4All}}                                          \\ \midrule
Algorithm       & \multicolumn{1}{c}{Prec.} & \multicolumn{1}{c}{Recall} & \multicolumn{1}{c}{F1} & \multicolumn{1}{c}{MAP} & \multicolumn{1}{c}{MRR} & \multicolumn{1}{c}{MAE} & \multicolumn{1}{c}{RMSE} \\ \midrule
\textsf{C-only$_{Ave}$}  & 0.5912             & $\ast$0.5154             & 0.5270             & $\ast$0.4142             & 0.7192             & $\ast$1.3045             & $\ast$1.6060             \\
\textsf{C-only$_{Cos}$}  & 0.6263             & \textbf{0.6224}    	& \underline{0.6001} & 0.4877             & 0.7583             & $\ast$1.3675             & $\ast$1.6948             \\
\textsf{C-only$_{Min}$}  & 0.6065             & $\ast$0.4999             & 0.5230             & $\ast$0.4166             & $\ast$0.7583             & $\ast$1.3675             & $\ast$1.6816             \\
\textsf{C-only$_{RMSD}$} & 0.5850 			  & $\ast$0.4970 			& 0.5134 			& $\ast$0.3996 				& $\ast$0.7142			 & $\ast$1.2025			 & $\ast$1.4400 \\
\textsf{Ind$_{Ave}$}     & 0.6118             & $\ast$0.5710             & 0.5736             & 0.4960             & 0.7667             & 0.9168             & 1.3659             \\
\textsf{Ind$_{Cos}$}     & 0.6290             & \underline{0.6207} & \textbf{0.6046}    & \textbf{0.5384}  		& \textbf{0.8095}    & 0.9927             & $\ast$1.4541             \\
\textsf{Ind$_{Min}$}     & \textbf{0.6328}    & 0.5832             & 0.5910             & \underline{0.5125}    & 0.7825             & \textbf{0.8691}    & \textbf{1.3020}   \\
\textsf{Ind$_{RMSD}$}    & 0.5968 			  & 0.5525 				& 	0.5561 			& 0.4787 			& $\ast$0.7537 & 			\underline{0.9018} & \underline{1.3295} \\
\textsf{MC$_{Ave}$}      & 0.6255             & $\ast$0.5383            & 0.5575             & 0.4489             & 0.7792             & $\ast$1.1902             & $\ast$1.4861             \\
\textsf{MC$_{Cos}$}      & 0.5917             & $\ast$0.5558             & 0.5459             & $\ast$0.4336             & $\ast$0.7217             & $\ast$1.3534             & $\ast$1.6236             \\
\textsf{MC$_{Min}$}      & \underline{0.6305} & $\ast$0.5057             & $\ast$0.5344             & $\ast$0.4352             & \underline{0.7950} & $\ast$1.4512             & $\ast$1.7943             \\
\textsf{MC$_{RMSD}$}     & 0.6105 & 			$\ast$0.5396 			& 0.5477 			& $\ast$0.4429 				& 0.7775			 & $\ast$1.1265			 & 1.3607 \\
\textsf{Pref-only}       & 0.6220             & 0.5912             & 0.5860             & $\ast$0.5114 				& 0.7858             & 0.9346			 & $\ast$1.4276 \\
 \midrule
\multicolumn{8}{c}{\textbf{TripAdvisor}}                                          \\ \midrule
\textsf{C-only$_{Ave}$}  & 0.6512          & $\diamond$0.6019          & $\diamond$0.5978          & 0.4855          & 0.7692          & $\diamond$1.2741          & $\diamond$1.5587          \\
\textsf{C-only$_{Cos}$}  & 0.6423          & \underline{0.6418}&\underline{0.6136} & \underline{0.5185} & 0.8003    & $\diamond$1.2638          & $\diamond$1.5513          \\
\textsf{C-only$_{Min}$}  & 0.6525          & $\diamond$0.5380          & $\diamond$0.5487          & $\diamond$0.4497          & 0.7900          & $\diamond$2.3017          & $\diamond$2.6292          \\
\textsf{C-only$_{RMSD}$} & 0.6453 			& $\diamond$0.5887 		& $\diamond$0.5881			 & $\diamond$0.4774 &		 0.7775 &			 $\diamond$1.3876 &			 $\diamond$1.6562 \\
\textsf{Ind$_{Ave}$}     & 0.6380          & 0.6007          & 0.6009          & 0.5116          & 0.7800          & \underline{0.9685}  & \textbf{1.3701}    \\
\textsf{Ind$_{Cos}$}     & 0.6113          & $\diamond$0.5745          & $\diamond$0.5704          & 0.4928          & 0.7783          & $\diamond$1.0072          & $\diamond$1.4244          \\
\textsf{Ind$_{Min}$}     & 0.6545          & $\diamond$0.5680          & 0.5881          & 0.4971          & 0.7900          & $\diamond$1.1948          & $\diamond$1.6535          \\
\textsf{Ind$_{RMSD}$}    & 0.6380 			& 0.6110 		& 0.6050 &			 0.5140 &		 0.7800 &			 0.9845 & $\diamond$\underline{1.3927} \\
\textsf{MC$_{Ave}$}      & \underline{0.6577} & $\diamond$0.6169          & 0.6059          & 0.5055        & 0.8148          & $\diamond$1.2010          & $\diamond$1.4810          \\
\textsf{MC$_{Cos}$}      & 0.6533          & \textbf{0.6666} & \textbf{0.6285} & \textbf{0.5306} & 0.7978          & $\diamond$1.1902          & 1.4237          \\
\textsf{MC$_{Min}$}      & \textbf{0.6585} & $\diamond$0.5836          & $\diamond$0.5768          & 0.4884          & \textbf{0.8170} & $\diamond$2.3241          & $\diamond$2.6586          \\
\textsf{MC$_{RMSD}$}     & 0.6473 			& $\diamond$0.5974		 & $\diamond$0.5900			 & 0.4905 & 	\underline{0.8153} & $\diamond$1.3427 & 		$\diamond$1.6144 \\
\textsf{Pref-only}       & 0.6220          & 0.5912          & 0.5860          & 0.5114          & 0.7858          & \textbf{0.9346} & 1.4276             \\ 
\bottomrule
\end{tabular}}
}
\label{tab:autistici}
\end{table}

\begin{table}[hbt!]
\centering
\caption{Top-N recommendation results on NEU dataset with N=5, using the 50 PoIs of the $\Pi$ set. We use the same notation as in Table \ref{tab:autistici}}.
\vspace{5pt}
\resizebox{0.9\columnwidth}{!}{%
{\def\arraystretch{0.9}
\begin{tabular}{lrrrrrrr}
\toprule
Algorithm       & \multicolumn{1}{c}{Prec.} & \multicolumn{1}{c}{Recall} & \multicolumn{1}{c}{F1} & \multicolumn{1}{c}{MAP} & \multicolumn{1}{c}{MRR} & \multicolumn{1}{c}{MAE} & \multicolumn{1}{c}{RMSE} \\ \midrule
\multicolumn{8}{c}{\textbf{Maps4All}}                                          \\ \midrule
\textsf{C-only$_{Ave}$}  & $\ast$0.5476             & $\ast$0.4936             & $\ast$0.4979             & $\ast$0.3701             & $\ast$0.7168             & $\ast$1.2122             & $\ast$1.4668             \\
\textsf{C-only$_{Cos}$}  & 0.5503             & \textbf{0.5414}    & 0.5255             & 0.4000             & 0.7189             & $\ast$1.4374             & $\ast$1.7456             \\
\textsf{C-only$_{Min}$}  & $\ast$0.5507             & $\ast$0.4769             & $\ast$0.4899             & $\ast$0.3673             & $\ast$0.7359             & $\ast$1.1704             & $\ast$1.4213             \\
\textsf{C-only$_{RMSD}$} & $\ast$0.5464    		& $\ast$0.4831				 & $\ast$0.4920			 & $\ast$0.3630 			& 0.7215 & 				$\ast$1.0908 & $\ast$\underline{1.3070} \\
\textsf{Ind$_{Ave}$}     & 0.5740             & 0.5261             & 0.5262             & 0.4108             & 0.7555             & 1.1085            & $\ast$1.4343             \\
\textsf{Ind$_{Cos}$}     & 0.5790             & \underline{0.5406} & \textbf{0.5349}    & \textbf{0.4139}    & 0.7475             & $\ast$1.1792             & $\ast$1.5232             \\
\textsf{Ind$_{Min}$}     & \underline{0.5791}  & 0.5225             & 0.5250             & \underline{0.4120}& \textbf{0.7688}    & 1.0950   			 & $\ast$1.4024 \\
\textsf{Ind$_{RMSD}$}    & 0.5817    		& 0.5272 &				 $\ast$0.5292 &			 0.4101 & 			0.7515 & 			\underline{1.0663} 	& $\ast$1.3796 \\
\textsf{MC$_{Ave}$}      & 0.5752             & 0.5154             & 0.5213             & 0.3995             & 0.7564             & $\ast$1.1238 &  			$\ast$1.3564      \\
\textsf{MC$_{Cos}$}      & $\ast$0.5274             & $\ast$0.5053             & $\ast$0.4974             & $\ast$0.3535             & $\ast$0.6591             & $\ast$1.2775             & $\ast$1.5795   \\
\textsf{MC$_{Min}$}      & 0.5664             & $\ast$0.4956             & $\ast$0.5053             & 0.3890             & \underline{0.7583} & $\ast$1.1249             & $\ast$1.4052             \\
\textsf{MC$_{RMSD}$}     & 0.5577   		 & 0.4809 			& $\ast$0.4953 &				 $\ast$0.3746 & 			0.7428 & 			\textbf{1.0417} & \textbf{1.2447} \\
\textsf{Pref-only}       & \textbf{0.5795} &  0.5408             & \underline{0.5347} & 0.4076 &			  $\ast$0.7304             & 1.1416             & $\ast$1.5270             \\

 \midrule
\multicolumn{8}{c}{\textbf{TripAdvisor}}                                          \\ \midrule
\textsf{C-only$_{Ave}$}  & 0.6230          & $\diamond$0.6016          & $\diamond$0.5866          & 0.4621          & 0.7563          & 1.1163          & 1.3521          \\
\textsf{C-only$_{Cos}$}  & \textbf{0.6310} & \textbf{0.6374} & \textbf{0.6063} & \textbf{0.4936} & \underline{0.7847} & $\diamond$1.1398          & $\diamond$1.4209          \\
\textsf{C-only$_{Min}$}  & 0.6110          & $\diamond$0.5336          & $\diamond$0.5315          & $\diamond$0.4116          & 0.7314          & $\diamond$2.0874          & $\diamond$2.4154          \\
\textsf{C-only$_{RMSD}$} & 0.6230   	 & $\diamond$0.5989 &			$\diamond$0.5846 &			 0.4569 &		 0.7527 &		 $\diamond$1.2508 & 				$\diamond$1.5106 \\
\textsf{Ind$_{Ave}$}     & $\diamond$0.6026         & $\diamond$0.5585          & $\diamond$0.5532          & $\diamond$0.4412          & 0.7809          & \underline{1.0608} & $\diamond$1.3538          \\
\textsf{Ind$_{Cos}$}     & $\diamond$0.6057         & $\diamond$0.5695          & $\diamond$0.5596          & $\diamond$0.4499          & \textbf{0.7878} & 1.0708         	 & $\diamond$1.3869          \\
\textsf{Ind$_{Min}$}     & $\diamond$0.5854         & $\diamond$0.5299          & $\diamond$0.5286          & $\diamond$0.4107          & $\diamond$0.7491          & $\diamond$1.3002         	 & $\diamond$1.6550           \\
\textsf{Ind$_{RMSD}$}    & $\diamond$0.6020   	 & $\diamond$0.5581 &			 $\diamond$0.5529 & 			$\diamond$0.4429 &		 0.7821 &			 $\diamond$1.1085 &			 $\diamond$1.4135 \\
\textsf{MC$_{Ave}$}      & \underline{0.6236} & $\diamond$0.6073       & 0.5891          & 0.4751          & 0.7801          & 1.0725          & \underline{1.3001}          \\
\textsf{MC$_{Cos}$}     & $\diamond$0.6159  & \underline{0.6158} & $\diamond$\underline{0.5898} & $\diamond$\underline{0.4786}  & 0.7758          & \textbf{1.0511}  & \textbf{1.2935}          \\
\textsf{MC$_{Min}$}      & 0.6067         &$\diamond$0.5555          & $\diamond$0.5447          & $\diamond$0.4259          & 0.7321          & $\diamond$2.1211          & $\diamond$2.4541  \\
\textsf{MC$_{RMSD}$}     & 0.6234    		& $\diamond$0.6031 		& 0.5872 &		 0.4736 &			 0.7825 &			 $\diamond$1.2202 & 		$\diamond$1.4749 \\
\textsf{Pref-only}       & $\diamond$0.5795         & $\diamond$0.5408          & $\diamond$0.5347          & $\diamond$0.4076          & $\diamond$0.7304          & 1.1416          & $\diamond$1.5270              \\ 
 \bottomrule
\end{tabular}}}
\label{tab:neurotipici}
\end{table}

\FloatBarrier

\section{Detailed results using the 34 PoIs mapped in both Maps4All and TripAdvisor (MA$\cap$TA).}

\begin{table}[hbt!]
\centering
\caption{Results on AUT dataset for N=5, using the information about places provided either by Maps4All, or by TripAdvisor, on the 34 PoIs of set $\Pi$ that are mapped by both data sources (MA$\cap$TA). We use the same notation as in Table \ref{tab:autistici}.}
\vspace{5pt}
\resizebox{0.9\columnwidth}{!}{%
{\def\arraystretch{0.8}
\begin{tabular}{lrrrrrrr}
\toprule
\multicolumn{8}{c}{\textbf{Maps4All}}                                          \\ \midrule
Algorithm       & \multicolumn{1}{c}{Prec.} & \multicolumn{1}{c}{Recall} & \multicolumn{1}{c}{F1} & \multicolumn{1}{c}{MAP} & \multicolumn{1}{c}{MRR} & \multicolumn{1}{c}{MAE} & \multicolumn{1}{c}{RMSE} \\ \midrule
C-only$_{Ave}$  & 0.6742          & 0.5705          & 0.5863         & 0.4933          & 0.7783          & $\ast$1.2441          & $\ast$1.5538          \\
C-only$_{Cos}$  & 0.6747          & \textbf{0.6393} & \textbf{0.624} & 0.5406          & 0.7775          & $\ast$1.2518          & $\ast$1.5839          \\
C-only$_{Min}$  & 0.6708          & 0.5303          & 0.5583         & 0.4578          & 0.7558          & $\ast$1.4010           & $\ast$1.6856          \\
C-only$_{RMSD}$ & 0.6727          & $\ast$0.5503          & 0.5725         & 0.4698          & 0.7617          & $\ast$1.1682          & $\ast$1.3918          \\
Ind$_{Ave}$     & 0.6830           & 0.6255          & 0.6189         & \textbf{0.5499} & 0.7900          & 0.9339          & $\ast$1.3682          \\
Ind$_{Cos}$     & $\ast$0.6680    & \underline{0.6363}    & \underline{0.6189}& \underline{0.5484}& 0.7800         & $\ast$1.0329          & $\ast$1.4697          \\
Ind$_{Min}$     & 0.6855          & 0.6022          & 0.6101         & 0.5357          & 0.7883          & \underline{0.9015} & \underline{1.3276}    \\
Ind$_{RMSD}$    & 0.6680           & 0.6155          & 0.6111         & 0.5385          & 0.7750           & \textbf{0.8885} & \textbf{1.2681} \\
MC$_{Ave}$      & 0.6887          & 0.5765          & 0.5960          & 0.5109          & \textbf{0.8083} & $\ast$1.1540           & $\ast$1.4572          \\
MC$_{Cos}$      & 0.6465          & 0.5920           & 0.5784         & 0.4982          & 0.7550           & $\ast$1.2027          & $\ast$1.4741          \\
MC$_{Min}$      & \textbf{0.7133} & 0.5453          & 0.5778         & 0.4838          & \underline{0.8083} & $\ast$1.5552          & $\ast$1.8692          \\
MC$_{RMSD}$     & \underline{0.6968} & 0.5882          & 0.6082        & 0.5100         & 0.8050           & $\ast$1.1107          & 1.3481          \\
Pref-only       & $\ast$0.6857          & 0.5938          & 0.6005         & 0.5333          & 0.7967          & $\ast$0.9484          & $\ast$1.4583       \\
 \midrule
\multicolumn{8}{c}{\textbf{TripAdvisor}}                                          \\ \midrule
C-only$_{Ave}$  & $\diamond$0.6395          & 0.5742          & 0.5701          & 0.4857          & $\diamond$0.7828          & $\diamond$1.2034          & 1.4800            \\
C-only$_{Cos}$  & $\diamond$0.6268          & 0.5828          & 0.5676          & 0.4816          & 0.7767          & $\diamond$1.2602          & $\diamond$1.5712          \\
C-only$_{Min}$  & 0.6528          & $\diamond$0.4570           & $\diamond$0.4730           & $\diamond$0.4046          & $\diamond$0.7275          & $\diamond$2.0903          & $\diamond$2.3653          \\
C-only$_{RMSD}$ & 0.6413          & 0.5820           & 0.5798          & 0.4919          & 0.7975          & $\diamond$1.3099          & $\diamond$1.5368          \\
Ind$_{Ave}$     & 0.6562          & 0.5897          & 0.5901          & 0.5353          & 0.7917          & \textbf{0.9459} & \textbf{1.3692} \\
Ind$_{Cos}$     & 0.6712          & 0.5922          & 0.5921          & $\diamond$0.5231          & 0.7850           & 0.9781          & 1.4254          \\
Ind$_{Min}$     & \textbf{0.6948} & 0.5755          & 0.5939          & \underline{0.5357} & 0.8067          & $\diamond$1.1957          & $\diamond$1.6454          \\
Ind$_{RMSD}$    & 0.6695   & \textbf{0.6013} &	 \textbf{0.6041} & 	\textbf{0.5454} &	 0.8100            & $\diamond$1.0091          & \underline{1.4196}    \\
MC$_{Ave}$      & 0.6652          & 0.5768          & 0.5827          & $\diamond$0.4909          & \underline{0.8145}    & $\diamond$1.1641        & 1.4396          \\
MC$_{Cos}$      & $\diamond$0.6310           & 0.5750           & 0.5621          & 0.4843          & 0.7900            & $\diamond$1.1871        & 1.4442          \\
MC$_{Min}$      & 0.6468          & 0.5452          & $\diamond$0.5142          & 0.4540           & $\diamond$0.7542          & $\diamond$2.1382          & $\diamond$2.4606          \\
MC$_{RMSD}$     & 0.6707          & 0.5710           & 0.5884          & 0.4964          & \textbf{0.8183} & $\diamond$1.3206          & $\diamond$1.5551          \\
Pref-on  & \underline{0.6857}   & \underline{0.5938}  & \underline{0.6005} & 0.5333      & 0.7967          & \underline{0.9484}    & $\diamond$1.4583              \\  
\bottomrule
\end{tabular}}
}
\label{tab:autistici_34}
\end{table}

\begin{table}[hbt!]
\centering
\caption{Results on AUT dataset for N=5, focusing on the places of set MA$\cap$TA. The data about places provided by Maps4All and TripAdvisor is fused by applying Equation \ref{eq:weightedMean}. For each evaluation metric, ``$\bullet$" denotes the statistical significance (t-test, $p<0.05$) of the difference between the best performing algorithm and the other ones.} 
\vspace{5pt}
\resizebox{0.9\columnwidth}{!}{%
{\def\arraystretch{0.9}
\begin{tabular}{lrrrrrrr}
\toprule
\multicolumn{8}{c}{\textbf{Maps4All + TripAdvisor (\textsf{M+T})}}                                          \\ \midrule
Algorithm       & \multicolumn{1}{c}{Prec.} & \multicolumn{1}{c}{Recall} & \multicolumn{1}{c}{F1} & \multicolumn{1}{c}{MAP} & \multicolumn{1}{c}{MRR} & \multicolumn{1}{c}{MAE} & \multicolumn{1}{c}{RMSE} \\ \midrule
C-only$_{Ave}$  & $\bullet$0.6670           & 0.6068          & 0.5993          & $\bullet$0.5034          & $\bullet$0.7600            & $\bullet$1.2185          & $\bullet$1.4951          \\
C-only$_{Cos}$  & \underline{0.7000} & \textbf{0.6427} & \textbf{0.6452} & \textbf{0.5767} & \underline{0.8350} & $\bullet$1.2144          & $\bullet$1.4993          \\
C-only$_{Min}$  & $\bullet$0.6217          & $\bullet$0.5258          & $\bullet$0.5392          & $\bullet$0.4528          & $\bullet$0.7417          & $\bullet$1.3864          & $\bullet$1.6779          \\
C-only$_{RMSD}$ & 0.6792          & 0.5930           & 0.5987          & 0.5158          & 0.7967          & $\bullet$1.1285          & $\bullet$1.3254          \\
Ind$_{Ave}$     & 0.6847          & 0.6163          & 0.6176          & 0.5447          & $\bullet$0.7917          & \underline{0.8985}   & $\bullet$1.3361          \\
Ind$_{Cos}$     & 0.6972          & 0.6038          & 0.6177          & 0.5487          & $\bullet$0.7950           & $\bullet$1.0227          & $\bullet$1.4484          \\
Ind$_{Min}$     & 0.6927          & 0.5997          & 0.6098          & 0.5375          & 0.7917          & $\bullet$0.9773          & $\bullet$1.3806          \\
Ind$_{RMSD}$    & $\bullet$0.6662          & 0.6047          & 0.5968          & 0.5192          & $\bullet$0.7733          & \textbf{0.8694} & \textbf{1.2152} \\
MC$_{Ave}$      & 0.6808          & 0.5983          & 0.6109          & 0.5387          & 0.8167          & $\bullet$1.0921          & 1.3754          \\
MC$_{Cos}$      & 0.6503          & $\bullet$0.5622          & $\bullet$0.5634          & $\bullet$0.4753          & 0.7958          & $\bullet$1.2220           &$\bullet$1.4940           \\
MC$_{Min}$      & 0.6895          & 0.5762          & 0.5897          & 0.4994          & 0.7792          & $\bullet$1.4909          & $\bullet$1.7999          \\
MC$_{RMSD}$  & \textbf{0.7073} & \underline{0.6373}& \underline{0.6390} & \underline{0.5731}& \textbf{0.8483} & $\bullet$1.076     & \underline{1.2856}    \\
Pref-only       & 0.6857          & 0.5938          & 0.6005          & 0.5333          & $\bullet$0.7967          & $\bullet$0.9484          & $\bullet$1.4583 \\\bottomrule
\end{tabular}}
}
\label{tab:autistici_34_mix}
\end{table}

\begin{table}[hbt!]
\centering
\caption{Results on NEU dataset for N=5, using the information about places provided either by Maps4All, or by TripAdvisor, on the 34 PoIs of set MA$\cap$TA. We use the same notation as in Table \ref{tab:autistici}.}
\vspace{5pt}
\resizebox{0.9\columnwidth}{!}{%
{\def\arraystretch{0.8}
\begin{tabular}{lrrrrrrr}
\toprule
Algorithm       & \multicolumn{1}{c}{Prec.} & \multicolumn{1}{c}{Recall} & \multicolumn{1}{c}{F1} & \multicolumn{1}{c}{MAP} & \multicolumn{1}{c}{MRR} & \multicolumn{1}{c}{MAE} & \multicolumn{1}{c}{RMSE} \\ \midrule
\multicolumn{8}{c}{\textbf{Maps4All}}                                          \\ \midrule
C-only$_{Ave}$  & $\ast$0.5580           & $\ast$0.5327         & 0.5186          & 0.4190           & $\ast$0.7131          & $\ast$1.1264          & $\ast$1.3742         \\
C-only$_{Cos}$  & 0.5885      & \textbf{0.6310} & \underline{0.5825}    & \underline{0.4919}    & 0.7566          & $\ast$1.2757          & $\ast$1.5642         \\
C-only$_{Min}$  & 0.5809          & $\ast$0.5221         & 0.5205          & 0.4163          & 0.7130           & $\ast$1.1720          & $\ast$1.4332         \\
C-only$_{RMSD}$ & 0.5642          & $\ast$0.4985         & $\ast$0.5059          & $\ast$0.3941          & $\ast$0.6973          & 1.0671          & 1.2868         \\
Ind$_{Ave}$     & 0.6090           & 0.5952         & 0.5717          & 0.4841          & 0.7576          & \underline{0.9804}    & 1.2989         \\
Ind$_{Cos}$ & \textbf{0.6201} & \underline{0.6215}  & \textbf{0.5884} & \textbf{0.5058} &	\textbf{0.7722} & $\ast$1.0363          & $\ast$1.3663         \\
Ind$_{Min}$     & 0.6082          & 0.5809         & $\ast$0.5638          & $\ast$0.4717          & 0.7487          & $\ast$1.0229          & $\ast$1.3379         \\
Ind$_{RMSD}$    & 0.6109          & 0.5807         & $\ast$0.5659          & $\ast$0.4759          & 0.7534          & \textbf{0.9746} & $\ast$1.2928         \\
MC$_{Ave}$      & 0.5812          & 0.5641         & 0.5449          & 0.4486          & 0.7422          & 1.0499          & \underline{1.2778}   \\
MC$_{Cos}$      & $\ast$0.5544          & $\ast$0.5847         & 0.5441          & $\ast$0.4340           & $\ast$0.6797          & $\ast$1.1190           & $\ast$1.3978         \\
MC$_{Min}$      & 0.5951          & $\ast$0.5333         & $\ast$0.5306          & $\ast$0.4337          & 0.7389          & $\ast$1.1520           & $\ast$1.4491         \\
MC$_{RMSD}$     & 0.5711          & $\ast$0.5217         & 0.5212          & 0.4217          & $\ast$0.7281          & 1.0214          & \textbf{1.2280} \\
Pref-only       & \underline{0.6139}    & 0.6046         & $\ast$0.5765   & $\ast$0.4902          & \underline{0.7577}    & 1.0012          & $\ast$1.3733     \\

 \midrule
\multicolumn{8}{c}{\textbf{TripAdvisor}}                                          \\ \midrule
C-only$_{Ave}$  & 0.5678          & 0.5831          & 0.5518         & 0.4435          & $\diamond$0.6993          & $\diamond$1.0753          & $\diamond$1.2955          \\
C-only$_{Cos}$  & 0.5878      & \textbf{0.6361} & \underline{0.5844}   & 0.4877          & $\diamond$0.7304          & $\diamond$1.1639          & $\diamond$1.4498          \\
C-only$_{Min}$  & $\diamond$0.5042          & $\diamond$0.3471          & $\diamond$0.3509         & $\diamond$0.2845          & $\diamond$0.5743          & $\diamond$1.9253          & $\diamond$2.2629          \\
C-only$_{RMSD}$ & 0.5725          & 0.5686          & 0.5470          & 0.4317          & $\diamond$0.7017          & $\diamond$1.1739          & $\diamond$1.4081          \\
Ind$_{Ave}$     & 0.6139          & 0.6091          & 0.5814         & 0.4954      & \underline{0.7642}    & \textbf{0.9815} & \underline{1.2781}    \\
Ind$_{Cos}$  & \textbf{0.6226} & \underline{0.6218} & \textbf{0.5920} & \textbf{0.5126} & \textbf{0.7852} & 1.0187          & $\diamond$1.3577          \\
Ind$_{Min}$     & $\diamond$0.6015          & $\diamond$0.5480           & $\diamond$0.5356         & $\diamond$0.4438          & $\diamond$0.7296          & $\diamond$1.0631          & $\diamond$1.3953          \\
Ind$_{RMSD}$ & \underline{0.6177}   & 0.6070      & 0.5831         & \underline{0.4961}    & 0.7624          & $\diamond$1.0093          & $\diamond$1.3086          \\
MC$_{Ave}$      & 0.5952          & 0.6023          & 0.5718         & 0.4729          & 0.7401          & $\diamond$1.0291          & \textbf{1.2358} \\
MC$_{Cos}$      & $\diamond$0.5681          & 0.5754          & 0.5470          & 0.4546          & 0.7209          & $\diamond$1.0673          & $\diamond$1.3207          \\
MC$_{Min}$      & $\diamond$0.4907           & $\diamond$0.3607          & $\diamond$0.3571         & $\diamond$0.2908          & $\diamond$0.5612          & $\diamond$1.9627          & $\diamond$2.3249          \\
MC$_{RMSD}$     & 0.5916          & 0.5731          & 0.5558         & 0.4530           & $\diamond$0.7291          & $\diamond$1.1336          & $\diamond$1.3682          \\
Pref-only       & 0.6139          & 0.6046          & 0.5765         & 0.4902          & $\diamond$0.7577          & \underline{1.0012}    & $\diamond$1.3733     \\ 
 \bottomrule
\end{tabular}}}
\label{tab:neurotipici_34}
\end{table}

\begin{table}[hbt!]
\centering
\caption{Results on NEU dataset for N=5, focusing on the places of set MA$\cap$TA. The data about places provided by Maps4All and TripAdvisor is fused by applying Equation \ref{eq:weightedMean}. We use the same notation as in Table \ref{tab:autistici_34_mix}.}
\vspace{5pt}
\resizebox{0.9\columnwidth}{!}{%
{\def\arraystretch{0.9}
\begin{tabular}{lrrrrrrr}
\toprule
Algorithm       & \multicolumn{1}{c}{Prec.} & \multicolumn{1}{c}{Recall} & \multicolumn{1}{c}{F1} & \multicolumn{1}{c}{MAP} & \multicolumn{1}{c}{MRR} & \multicolumn{1}{c}{MAE} & \multicolumn{1}{c}{RMSE} \\ \midrule
\multicolumn{8}{c}{\textbf{Maps4All + TripAdvisor (\textsf{M+T})}}                                          \\ \midrule
C-only$_{Ave}$  & 0.5968         & 0.5924          & 0.5665          & 0.4694          & $\bullet$0.7433          & $\bullet$1.0822          & $\bullet$1.3129          \\
C-only$_{Cos}$  & $\bullet$0.5882         & \textbf{0.6289} & 0.5819          & \underline{0.5112}    & $\bullet$0.7827          & $\bullet$1.2360       & $\bullet$1.5152          \\
C-only$_{Min}$  & 0.6029         & 0.5786          & 0.5635          & $\bullet$0.4546          & $\bullet$0.7202          & $\bullet$1.1196          & $\bullet$1.3694          \\
C-only$_{RMSD}$ & 0.6061         & 0.5870           & 0.5699          & $\bullet$0.4697          & $\bullet$0.7463          & 0.9951       & \underline{1.1946}    \\
Ind$_{Ave}$   & \textbf{0.6260} & 0.6136          & 0.5887          & 0.5047          & $\bullet$0.7717          & 0.9770           & $\bullet$1.2869          \\
Ind$_{Cos}$     & 0.6194 & \underline{0.6284} & \textbf{0.5933} & \textbf{0.5236} & \textbf{0.8017} & $\bullet$1.0786          & $\bullet$1.3941          \\
Ind$_{Min}$     & 0.6189         & 0.5948          & 0.5761          & 0.4887          & $\bullet$0.7684          & $\bullet$1.0115          & $\bullet$1.3141          \\
Ind$_{RMSD}$    & $\bullet$0.6207         & 0.6027          & 0.5809          & 0.4960           & $\bullet$0.7718       & \textbf{0.9589} & $\bullet$1.2533          \\
MC$_{Ave}$      & 0.6125         & 0.6000             & 0.5810       & 0.5025     & \underline{0.7911}    & 1.0182          & $\bullet$1.2358          \\
MC$_{Cos}$      & 0.5724         & $\bullet$0.5575          & 0.5405          & $\bullet$0.4390           & $\bullet$0.7142          & $\bullet$1.1223          & $\bullet$1.3936          \\
MC$_{Min}$      & 0.6099         & 0.5860           & 0.5683          & $\bullet$0.4738          & $\bullet$0.7569          & $\bullet$1.1377          & $\bullet$1.4157          \\
MC$_{RMSD}$     & \underline{0.6241}   & 0.6088  & \underline{0.5903}    & 0.5009    & 0.7828    & \underline{0.9679}    & \textbf{1.1615} \\
Pref-only       & 0.6139         & 0.6046          & $\bullet$0.5765          & $\bullet$0.4902          & $\bullet$0.7577          & $\bullet$1.0012          & $\bullet$1.3733  \\
 \bottomrule
\end{tabular}}}
\label{tab:neurotipici_34_mix}
\end{table}

\FloatBarrier
\bibliographystyle{elsarticle-harv} 
\bibliography{plan-rec,httpbib,mybib,sample,fede}

\begin{thebibliography}{72}
\expandafter\ifx\csname natexlab\endcsname\relax\def\natexlab#1{#1}\fi
\providecommand{\url}[1]{\texttt{#1}}
\providecommand{\href}[2]{#2}
\providecommand{\path}[1]{#1}
\providecommand{\DOIprefix}{doi:}
\providecommand{\ArXivprefix}{arXiv:}
\providecommand{\URLprefix}{URL: }
\providecommand{\Pubmedprefix}{pmid:}
\providecommand{\doi}[1]{\href{http://dx.doi.org/#1}{\path{#1}}}
\providecommand{\Pubmed}[1]{\href{pmid:#1}{\path{#1}}}
\providecommand{\bibinfo}[2]{#2}
\ifx\xfnm\relax \def\xfnm[#1]{\unskip,\space#1}\fi
\bibitem[{Adomavicius and Kwon(2007)}]{Adomavicius-Kwon:07}
\bibinfo{author}{Adomavicius, G.}, \bibinfo{author}{Kwon, Y.},
  \bibinfo{year}{2007}.
\newblock \bibinfo{title}{New recommendation techniques for multicriteria
  rating systems}.
\newblock \bibinfo{journal}{IEEE Intelligent Systems} \bibinfo{volume}{22},
  \bibinfo{pages}{48--55}.
\newblock \DOIprefix\doi{10.1109/MIS.2007.58}.
\bibitem[{Adomavicius and Tuzhilin(2015)}]{Adomavicius-Tuzhilin:15}
\bibinfo{author}{Adomavicius, G.}, \bibinfo{author}{Tuzhilin, A.},
  \bibinfo{year}{2015}.
\newblock \bibinfo{title}{Context-Aware Recommender Systems}.
  \bibinfo{publisher}{Springer US}, \bibinfo{address}{Boston, MA}.
\newblock pp. \bibinfo{pages}{191--226}.
\newblock \URLprefix \url{https://doi.org/10.1007/978-1-4899-7637-6\_6},
  \DOIprefix\doi{10.1007/978-1-4899-7637-6\_6}.
\bibitem[{Al-Ghossein et~al.(2018)Al-Ghossein, Murena, Abdessalem, Barr\'{e}
  and Cornu\'{e}jols}]{Al-Ghossein-etal:18}
\bibinfo{author}{Al-Ghossein, M.}, \bibinfo{author}{Murena, P.A.},
  \bibinfo{author}{Abdessalem, T.}, \bibinfo{author}{Barr\'{e}, A.},
  \bibinfo{author}{Cornu\'{e}jols, A.}, \bibinfo{year}{2018}.
\newblock \bibinfo{title}{Adaptive collaborative topic modeling for online
  recommendation}, in: \bibinfo{booktitle}{Proceedings of the 12th ACM
  Conference on Recommender Systems}, \bibinfo{publisher}{Association for
  Computing Machinery}, \bibinfo{address}{New York, NY, USA}. p.
  \bibinfo{pages}{338–346}.
\newblock \URLprefix \url{https://doi.org/10.1145/3240323.3240363},
  \DOIprefix\doi{10.1145/3240323.3240363}.
\bibitem[{Ardissono et~al.(2003)Ardissono, Goy, Petrone, Segnan and
  Torasso}]{Ardissono-etal:03}
\bibinfo{author}{Ardissono, L.}, \bibinfo{author}{Goy, A.},
  \bibinfo{author}{Petrone, G.}, \bibinfo{author}{Segnan, M.},
  \bibinfo{author}{Torasso, P.}, \bibinfo{year}{2003}.
\newblock \bibinfo{title}{{INTRIGUE}: personalized recommendation of tourist
  attractions for desktop and handset devices}.
\newblock \bibinfo{journal}{Applied Artificial Intelligence, Special Issue on
  Artificial Intelligence for Cultural Heritage and Digital Libraries}
  \bibinfo{volume}{17}, \bibinfo{pages}{687--714}.
\newblock \URLprefix
  \url{https://www.tandfonline.com/doi/abs/10.1080/713827254},
  \DOIprefix\doi{https://doi.org/10.1080/713827254}.
\bibitem[{Baltrunas et~al.(2011)Baltrunas, Kaminskas, Ludwig, Moling, Ricci,
  Aydin, L{\"u}ke and Schwaiger}]{Baltrunas2011}
\bibinfo{author}{Baltrunas, L.}, \bibinfo{author}{Kaminskas, M.},
  \bibinfo{author}{Ludwig, B.}, \bibinfo{author}{Moling, O.},
  \bibinfo{author}{Ricci, F.}, \bibinfo{author}{Aydin, A.},
  \bibinfo{author}{L{\"u}ke, K.H.}, \bibinfo{author}{Schwaiger, R.},
  \bibinfo{year}{2011}.
\newblock \bibinfo{title}{Incarmusic: Context-aware music recommendations in a
  car}, in: \bibinfo{editor}{Huemer, C.}, \bibinfo{editor}{Setzer, T.} (Eds.),
  \bibinfo{booktitle}{E-Commerce and Web Technologies},
  \bibinfo{publisher}{Springer Berlin Heidelberg}, \bibinfo{address}{Berlin,
  Heidelberg}. pp. \bibinfo{pages}{89--100}.
\newblock \URLprefix \url{https://doi.org/10.1007/978-3-642-23014-1_8},
  \DOIprefix\doi{10.1007/978-3-642-23014-1_8}.
\bibitem[{Banskota and Ng(2020)}]{banskota2020recommending}
\bibinfo{author}{Banskota, A.}, \bibinfo{author}{Ng, Y.K.},
  \bibinfo{year}{2020}.
\newblock \bibinfo{title}{Recommending video games to adults with autism
  spectrum disorder for social-skill enhancement}, in:
  \bibinfo{booktitle}{Proceedings of the 28th ACM Conference on User Modeling,
  Adaptation and Personalization}, \bibinfo{publisher}{Association for
  Computing Machinery}, \bibinfo{address}{New York, NY, USA}. p.
  \bibinfo{pages}{14–22}.
\newblock \URLprefix \url{https://doi.org/10.1145/3340631.3394867},
  \DOIprefix\doi{10.1145/3340631.3394867}.
\bibitem[{Bao et~al.(2014)Bao, Fang and Zhang}]{Bao-etal:14}
\bibinfo{author}{Bao, Y.}, \bibinfo{author}{Fang, H.}, \bibinfo{author}{Zhang,
  J.}, \bibinfo{year}{2014}.
\newblock \bibinfo{title}{Topic{MF}: simultaneously exploiting ratings and
  reviews for recommendation}, in: \bibinfo{booktitle}{Proceedings of the
  Twenty-Eighth AAAI Conference on Artificial Intelligence},
  \bibinfo{publisher}{AAAI Press}. p. \bibinfo{pages}{2–8}.
\newblock \DOIprefix\doi{10.5555/2893873.2893874}.
\bibitem[{Bernardes et~al.(2015)Bernardes, Barros, Simoes and
  Castelo-Branco}]{bernardes2015serious}
\bibinfo{author}{Bernardes, M.}, \bibinfo{author}{Barros, F.},
  \bibinfo{author}{Simoes, M.}, \bibinfo{author}{Castelo-Branco, M.},
  \bibinfo{year}{2015}.
\newblock \bibinfo{title}{A serious game with virtual reality for travel
  training with autism spectrum disorder}, in: \bibinfo{booktitle}{2015
  International Conference on Virtual Rehabilitation (ICVR)},
  \bibinfo{organization}{IEEE}. pp. \bibinfo{pages}{127--128}.
\newblock \DOIprefix\doi{10.1109/ICVR.2015.7358609}.
\bibitem[{Biancalana et~al.(2013)Biancalana, Gasparetti, Micarelli and
  Sansonetti}]{Biancalana2013}
\bibinfo{author}{Biancalana, C.}, \bibinfo{author}{Gasparetti, F.},
  \bibinfo{author}{Micarelli, A.}, \bibinfo{author}{Sansonetti, G.},
  \bibinfo{year}{2013}.
\newblock \bibinfo{title}{An approach to social recommendation for
  context-aware mobile services}.
\newblock \bibinfo{journal}{ACM Trans. Intell. Syst. Technol.}
  \bibinfo{volume}{4}, \bibinfo{pages}{10:1--10:31}.
\newblock \URLprefix \url{http://doi.acm.org/10.1145/2414425.2414435},
  \DOIprefix\doi{10.1145/2414425.2414435}.
\bibitem[{Blei and McAuliffe(2007)}]{Blei-etal:10}
\bibinfo{author}{Blei, D.M.}, \bibinfo{author}{McAuliffe, J.D.},
  \bibinfo{year}{2007}.
\newblock \bibinfo{title}{Supervised topic models}, in:
  \bibinfo{booktitle}{Proceedings of the 20th International Conference on
  Neural Information Processing Systems}, \bibinfo{publisher}{Curran Associates
  Inc.}, \bibinfo{address}{Red Hook, NY, USA}. p. \bibinfo{pages}{121–128}.
\newblock \URLprefix \url{https://dl.acm.org/doi/10.5555/2981562.2981578},
  \DOIprefix\doi{10.5555/2981562.2981578}.
\bibitem[{Boyd et~al.(2016)Boyd, Rangel, Tomimbang, Conejo-Toledo, Patel,
  Tentori and Hayes}]{boyd2016saywat}
\bibinfo{author}{Boyd, L.E.}, \bibinfo{author}{Rangel, A.},
  \bibinfo{author}{Tomimbang, H.}, \bibinfo{author}{Conejo-Toledo, A.},
  \bibinfo{author}{Patel, K.}, \bibinfo{author}{Tentori, M.},
  \bibinfo{author}{Hayes, G.R.}, \bibinfo{year}{2016}.
\newblock \bibinfo{title}{Say{WAT}: Augmenting face-to-face conversations for
  adults with autism}, in: \bibinfo{booktitle}{Proceedings of the 2016 CHI
  Conference on Human Factors in Computing Systems},
  \bibinfo{publisher}{Association for Computing Machinery},
  \bibinfo{address}{New York, NY, USA}. p. \bibinfo{pages}{4872–4883}.
\newblock \URLprefix \url{https://doi.org/10.1145/2858036.2858215},
  \DOIprefix\doi{10.1145/2858036.2858215}.
\bibitem[{Brailsford et~al.(1999)Brailsford, Potts and
  Smith}]{Brailsford-etal:99}
\bibinfo{author}{Brailsford, S.C.}, \bibinfo{author}{Potts, C.N.},
  \bibinfo{author}{Smith, B.M.}, \bibinfo{year}{1999}.
\newblock \bibinfo{title}{Constraint satisfaction problems: Algorithms and
  applications}.
\newblock \bibinfo{journal}{European Journal of Operational Research}
  \bibinfo{volume}{119}, \bibinfo{pages}{557 -- 581}.
\bibitem[{Burke(2002)}]{Burke:02}
\bibinfo{author}{Burke, R.}, \bibinfo{year}{2002}.
\newblock \bibinfo{title}{Hybrid recommender systems: survey and experiments}.
\newblock \bibinfo{journal}{User Modeling and User-Adapted Interaction}
  \bibinfo{volume}{12}, \bibinfo{pages}{331--370}.
\newblock \URLprefix \url{https://doi.org/10.1023/A:1021240730564},
  \DOIprefix\doi{10.1023/A:1021240730564}.
\bibitem[{Cantador et~al.(2011)Cantador, Castells and
  Bellog\'{\i}n}]{Cantador-etal:11}
\bibinfo{author}{Cantador, I.}, \bibinfo{author}{Castells, P.},
  \bibinfo{author}{Bellog\'{\i}n, A.}, \bibinfo{year}{2011}.
\newblock \bibinfo{title}{An enhanced semantic layer for hybrid recommender
  systems: Application to news recommendation}.
\newblock \bibinfo{journal}{Int. Journal on Semantic Web and Information
  Systems} \bibinfo{volume}{7}, \bibinfo{pages}{44--77}.
\newblock \URLprefix \url{https://dl.acm.org/doi/10.4018/jswis.2011010103},
  \DOIprefix\doi{10.4018/jswis.2011010103}.
\bibitem[{Cena et~al.(2020)Cena, Mauro, Ardissono, Mattutino, Rapp, Cocomazzi,
  Brighenti and Keller}]{Cena-etal:20}
\bibinfo{author}{Cena, F.}, \bibinfo{author}{Mauro, N.},
  \bibinfo{author}{Ardissono, L.}, \bibinfo{author}{Mattutino, C.},
  \bibinfo{author}{Rapp, A.}, \bibinfo{author}{Cocomazzi, S.},
  \bibinfo{author}{Brighenti, S.}, \bibinfo{author}{Keller, R.},
  \bibinfo{year}{2020}.
\newblock \bibinfo{title}{Personalized tourist guide for people with autism},
  in: \bibinfo{booktitle}{Adjunct Publication of the 28th ACM Conference on
  User Modeling, Adaptation and Personalization},
  \bibinfo{publisher}{Association for Computing Machinery},
  \bibinfo{address}{New York, NY, USA}. p. \bibinfo{pages}{347–351}.
\newblock \URLprefix \url{https://doi.org/10.1145/3386392.3399280},
  \DOIprefix\doi{10.1145/3386392.3399280}.
\bibitem[{Cena et~al.(2021)Cena, Rapp, Mattutino, Mauro, Ardissono, Cuccurullo,
  Brighenti, Keller and Tirassa}]{DBLP:conf/interact/CenaRMMACBKT21}
\bibinfo{author}{Cena, F.}, \bibinfo{author}{Rapp, A.},
  \bibinfo{author}{Mattutino, C.}, \bibinfo{author}{Mauro, N.},
  \bibinfo{author}{Ardissono, L.}, \bibinfo{author}{Cuccurullo, S.A.G.},
  \bibinfo{author}{Brighenti, S.}, \bibinfo{author}{Keller, R.},
  \bibinfo{author}{Tirassa, M.}, \bibinfo{year}{2021}.
\newblock \bibinfo{title}{A personalised interactive mobile app for people with
  autism spectrum disorder}, in: \bibinfo{booktitle}{Human-Computer Interaction
  - {INTERACT} 2021 - 18th {IFIP} {TC} 13 International Conference, Bari,
  Italy, August 30 - September 3, 2021, Proceedings, Part {V}},
  \bibinfo{publisher}{Springer}. pp. \bibinfo{pages}{313--317}.
\newblock \URLprefix \url{https://doi.org/10.1007/978-3-030-85607-6\_28},
  \DOIprefix\doi{10.1007/978-3-030-85607-6\_28}.
\bibitem[{Chen et~al.(2019)Chen, Qiu, Yang, Zhou, Huang, Li and
  Bao}]{Chen-etal:19}
\bibinfo{author}{Chen, C.}, \bibinfo{author}{Qiu, M.}, \bibinfo{author}{Yang,
  Y.}, \bibinfo{author}{Zhou, J.}, \bibinfo{author}{Huang, J.},
  \bibinfo{author}{Li, X.}, \bibinfo{author}{Bao, F.S.}, \bibinfo{year}{2019}.
\newblock \bibinfo{title}{Multi-domain gated {CNN} for review helpfulness
  prediction}, in: \bibinfo{booktitle}{The World Wide Web Conference},
  \bibinfo{publisher}{Association for Computing Machinery},
  \bibinfo{address}{New York, NY, USA}. p. \bibinfo{pages}{2630–2636}.
\newblock \URLprefix \url{https://doi.org/10.1145/3308558.3313587},
  \DOIprefix\doi{10.1145/3308558.3313587}.
\bibitem[{Chen et~al.(2015)Chen, Chen and Wang}]{Chen-etal:15}
\bibinfo{author}{Chen, L.}, \bibinfo{author}{Chen, G.}, \bibinfo{author}{Wang,
  F.}, \bibinfo{year}{2015}.
\newblock \bibinfo{title}{Recommender systems based on user reviews: the state
  of the art}.
\newblock \bibinfo{journal}{User Modeling and User-Adapted Interaction}
  \bibinfo{volume}{25}, \bibinfo{pages}{99--154}.
\newblock \URLprefix \url{https://doi.org/10.1007/s11257-015-9155-5},
  \DOIprefix\doi{10.1007/s11257-015-9155-5}.
\bibitem[{Costa et~al.(2017)Costa, Costa, Juli{\'a}n and
  Novais}]{costa2017task}
\bibinfo{author}{Costa, M.}, \bibinfo{author}{Costa, A.},
  \bibinfo{author}{Juli{\'a}n, V.}, \bibinfo{author}{Novais, P.},
  \bibinfo{year}{2017}.
\newblock \bibinfo{title}{A task recommendation system for children and youth
  with autism spectrum disorder}, in: \bibinfo{editor}{De~Paz, J.F.},
  \bibinfo{editor}{Juli{\'a}n, V.}, \bibinfo{editor}{Villarrubia, G.},
  \bibinfo{editor}{Marreiros, G.}, \bibinfo{editor}{Novais, P.} (Eds.),
  \bibinfo{booktitle}{Ambient Intelligence-- Software and Applications -- 8th
  International Symposium on Ambient Intelligence (ISAmI 2017)},
  \bibinfo{publisher}{Springer International Publishing},
  \bibinfo{address}{Cham}. pp. \bibinfo{pages}{87--94}.
\newblock \URLprefix \url{https://doi.org/10.1007/978-3-319-61118-1\_12},
  \DOIprefix\doi{10.1007/978-3-319-61118-1\_12}.
\bibitem[{Desrosiers and Karypis(2011)}]{Desrosiers-Karypis:11}
\bibinfo{author}{Desrosiers, C.}, \bibinfo{author}{Karypis, G.},
  \bibinfo{year}{2011}.
\newblock \bibinfo{title}{A comprehensive survey of neighborhood-based
  recommendation methods}, in: \bibinfo{editor}{Ricci, F.},
  \bibinfo{editor}{Rokach, L.}, \bibinfo{editor}{Shapira, B.},
  \bibinfo{editor}{Kantor, P.B.} (Eds.), \bibinfo{booktitle}{Recommender
  Systems Handbook}. \bibinfo{publisher}{Springer US},
  \bibinfo{address}{Boston, MA}, pp. \bibinfo{pages}{107--144}.
\newblock \URLprefix \url{https://doi.org/10.1007/978-0-387-85820-3\_4},
  \DOIprefix\doi{10.1007/978-0-387-85820-3\_4}.
\bibitem[{Dong et~al.(2016)Dong, O'mahony, Schaal, Mccarthy and
  Smyth}]{Dong-etal:16}
\bibinfo{author}{Dong, R.}, \bibinfo{author}{O'mahony, M.P.},
  \bibinfo{author}{Schaal, M.}, \bibinfo{author}{Mccarthy, K.},
  \bibinfo{author}{Smyth, B.}, \bibinfo{year}{2016}.
\newblock \bibinfo{title}{Combining similarity and sentiment in opinion mining
  for product recommendation}.
\newblock \bibinfo{journal}{Journal of Intelligent Information Systems}
  \bibinfo{volume}{46}, \bibinfo{pages}{285–312}.
\newblock \URLprefix \url{https://doi.org/10.1007/s10844-015-0379-y},
  \DOIprefix\doi{10.1007/s10844-015-0379-y}.
\bibitem[{Dong et~al.(2013)Dong, Schaal, O’Mahony and Smyth}]{Dong-etal:13}
\bibinfo{author}{Dong, R.}, \bibinfo{author}{Schaal, M.},
  \bibinfo{author}{O’Mahony, M.P.}, \bibinfo{author}{Smyth, B.},
  \bibinfo{year}{2013}.
\newblock \bibinfo{title}{Topic extraction from online reviews for
  classification and recommendation}, in: \bibinfo{booktitle}{Proceedings of
  the Twenty-Third International Joint Conference on Artificial Intelligence},
  \bibinfo{publisher}{AAAI Press}. p. \bibinfo{pages}{1310–1316}.
\bibitem[{Dragone et~al.(2018)Dragone, Pellegrini, Vescovi, Tentori and
  Passerini}]{Dragone-etal:18}
\bibinfo{author}{Dragone, P.}, \bibinfo{author}{Pellegrini, G.},
  \bibinfo{author}{Vescovi, M.}, \bibinfo{author}{Tentori, K.},
  \bibinfo{author}{Passerini, A.}, \bibinfo{year}{2018}.
\newblock \bibinfo{title}{No more ready-made deals: constructive recommendation
  for telco service bundling}, in: \bibinfo{booktitle}{Proceedings of the 12th
  ACM Conference on Recommender Systems}, \bibinfo{publisher}{ACM},
  \bibinfo{address}{New York, NY, USA}. pp. \bibinfo{pages}{163--171}.
\newblock \URLprefix \url{http://doi.acm.org/10.1145/3240323.3240348},
  \DOIprefix\doi{10.1145/3240323.3240348}.
\bibitem[{Elsabbagh et~al.(2012)Elsabbagh, Divan, Koh, Kim, Kauchali,
  Marc{\'\i}n, Montiel-Nava, Patel, Paula, Wang et~al.}]{elsabbagh2012global}
\bibinfo{author}{Elsabbagh, M.}, \bibinfo{author}{Divan, G.},
  \bibinfo{author}{Koh, Y.J.}, \bibinfo{author}{Kim, Y.S.},
  \bibinfo{author}{Kauchali, S.}, \bibinfo{author}{Marc{\'\i}n, C.},
  \bibinfo{author}{Montiel-Nava, C.}, \bibinfo{author}{Patel, V.},
  \bibinfo{author}{Paula, C.S.}, \bibinfo{author}{Wang, C.}, et~al.,
  \bibinfo{year}{2012}.
\newblock \bibinfo{title}{Global prevalence of autism and other pervasive
  developmental disorders}.
\newblock \bibinfo{journal}{Autism research} \bibinfo{volume}{5},
  \bibinfo{pages}{160--179}.
\newblock \URLprefix \url{https://doi.org/10.1002/aur.239},
  \DOIprefix\doi{10.1002/aur.239}.
\bibitem[{Gemmell et~al.(2012)Gemmell, Schimoler, Mobasher and
  Burke}]{Gemmel-etal:12}
\bibinfo{author}{Gemmell, J.}, \bibinfo{author}{Schimoler, T.},
  \bibinfo{author}{Mobasher, B.}, \bibinfo{author}{Burke, R.},
  \bibinfo{year}{2012}.
\newblock \bibinfo{title}{Resource recommendation in social annotation systems:
  A linear-weighted hybrid approach}.
\newblock \bibinfo{journal}{Journal of Computer and System Sciences}
  \bibinfo{volume}{78}, \bibinfo{pages}{1160 -- 1174}.
\newblock \URLprefix
  \url{http://www.sciencedirect.com/science/article/pii/S0022000011001127},
  \DOIprefix\doi{10.1016/j.jcss.2011.10.006}.
\bibitem[{Ghose and Ipeirotis(2011)}]{Ghose-Ipeirotis:11}
\bibinfo{author}{Ghose, A.}, \bibinfo{author}{Ipeirotis, P.G.},
  \bibinfo{year}{2011}.
\newblock \bibinfo{title}{Estimating the helpfulness and economic impact of
  product reviews: mining text and reviewer characteristics}.
\newblock \bibinfo{journal}{{IEEE} {T}ransactions on on {K}nowledge and {D}ata
  {E}ngineering} \bibinfo{volume}{23}, \bibinfo{pages}{1498--1512}.
\newblock \URLprefix \url{https://doi.org/10.1109/TKDE.2010.188},
  \DOIprefix\doi{10.1109/TKDE.2010.188}.
\bibitem[{Grynszpan et~al.(2014)Grynszpan, Weiss, Perez-Diaz and
  Gal}]{grynszpan2014innovative}
\bibinfo{author}{Grynszpan, O.}, \bibinfo{author}{Weiss, P.L.},
  \bibinfo{author}{Perez-Diaz, F.}, \bibinfo{author}{Gal, E.},
  \bibinfo{year}{2014}.
\newblock \bibinfo{title}{Innovative technology-based interventions for autism
  spectrum disorders: a meta-analysis}.
\newblock \bibinfo{journal}{Autism} \bibinfo{volume}{18},
  \bibinfo{pages}{346--361}.
\newblock \URLprefix \url{https://doi.org/10.1177/1362361313476767},
  \DOIprefix\doi{10.1177/1362361313476767}.
\bibitem[{Hern{\'a}ndez-Rubio et~al.(2019)Hern{\'a}ndez-Rubio, Cantador and
  Bellog{\'i}n}]{Rubio-etal:19}
\bibinfo{author}{Hern{\'a}ndez-Rubio, M.}, \bibinfo{author}{Cantador, I.},
  \bibinfo{author}{Bellog{\'i}n, A.}, \bibinfo{year}{2019}.
\newblock \bibinfo{title}{A comparative analysis of recommender systems based
  on item aspect opinions extracted from user reviews}.
\newblock \bibinfo{journal}{User Modeling and User-Adapted Interaction}
  \bibinfo{volume}{29}, \bibinfo{pages}{381--441}.
\newblock \URLprefix \url{https://doi.org/10.1007/s11257-018-9214-9},
  \DOIprefix\doi{10.1007/s11257-018-9214-9}.
\bibitem[{Hong et~al.(2012)Hong, Kim, Abowd and Arriaga}]{hong2012designing}
\bibinfo{author}{Hong, H.}, \bibinfo{author}{Kim, J.G.},
  \bibinfo{author}{Abowd, G.D.}, \bibinfo{author}{Arriaga, R.I.},
  \bibinfo{year}{2012}.
\newblock \bibinfo{title}{Designing a social network to support the
  independence of young adults with autism}, in:
  \bibinfo{booktitle}{Proceedings of the ACM 2012 Conference on Computer
  Supported Cooperative Work}, \bibinfo{publisher}{Association for Computing
  Machinery}, \bibinfo{address}{New York, NY, USA}. p.
  \bibinfo{pages}{627–636}.
\newblock \URLprefix \url{https://doi.org/10.1145/2145204.2145300},
  \DOIprefix\doi{10.1145/2145204.2145300}.
\bibitem[{Jannach et~al.(2014)Jannach, Zanker and Fuchs}]{Jannach-etal:14}
\bibinfo{author}{Jannach, D.}, \bibinfo{author}{Zanker, M.},
  \bibinfo{author}{Fuchs, M.}, \bibinfo{year}{2014}.
\newblock \bibinfo{title}{Leveraging multi-criteria customer feedback for
  satisfaction analysis and improved recommendations}.
\newblock \bibinfo{journal}{Information Technology {\&} Tourism}
  \bibinfo{volume}{14}, \bibinfo{pages}{119--149}.
\newblock \URLprefix \url{https://doi.org/10.1007/s40558-014-0010-z},
  \DOIprefix\doi{10.1007/s40558-014-0010-z}.
\bibitem[{Kientz et~al.(2013)Kientz, Goodwin, Hayes and
  Abowd}]{kientz2013interactive}
\bibinfo{author}{Kientz, J.A.}, \bibinfo{author}{Goodwin, M.S.},
  \bibinfo{author}{Hayes, G.R.}, \bibinfo{author}{Abowd, G.D.},
  \bibinfo{year}{2013}.
\newblock \bibinfo{title}{Interactive technologies for autism}.
\newblock \bibinfo{journal}{Synthesis Lectures on Assistive, Rehabilitative,
  and Health-Preserving Technologies} \bibinfo{volume}{2},
  \bibinfo{pages}{1--177}.
\newblock \URLprefix \url{https://doi.org/10.2200/S00533ED1V01Y201309ARH004},
  \DOIprefix\doi{10.2200/S00533ED1V01Y201309ARH004}.
\bibitem[{Koren and Bell(2011)}]{Koren-Bell:11}
\bibinfo{author}{Koren, Y.}, \bibinfo{author}{Bell, R.}, \bibinfo{year}{2011}.
\newblock \bibinfo{title}{Advances in Collaborative Filtering}.
  \bibinfo{publisher}{Springer US}, \bibinfo{address}{Boston, MA}.
\newblock pp. \bibinfo{pages}{145--186}.
\newblock \URLprefix \url{https://doi.org/10.1007/978-0-387-85820-3\_5},
  \DOIprefix\doi{10.1007/978-0-387-85820-3\_5}.
\bibitem[{Korfiatis et~al.(2019)Korfiatis, Stamolampros, Kourouthanassis and
  Sagiadinos}]{Korfiatis-etal:19}
\bibinfo{author}{Korfiatis, N.}, \bibinfo{author}{Stamolampros, P.},
  \bibinfo{author}{Kourouthanassis, P.}, \bibinfo{author}{Sagiadinos, V.},
  \bibinfo{year}{2019}.
\newblock \bibinfo{title}{Measuring service quality from unstructured data: A
  topic modeling application on airline passengers’ online reviews}.
\newblock \bibinfo{journal}{Expert Systems with Applications}
  \bibinfo{volume}{116}, \bibinfo{pages}{472 -- 486}.
\newblock \URLprefix
  \url{http://www.sciencedirect.com/science/article/pii/S0957417418306146},
  \DOIprefix\doi{10.1016/j.eswa.2018.09.037}.
\bibitem[{Li et~al.(2013)Li, Wu and Lai}]{Li-etal:13}
\bibinfo{author}{Li, Y.M.}, \bibinfo{author}{Wu, C.T.}, \bibinfo{author}{Lai,
  C.Y.}, \bibinfo{year}{2013}.
\newblock \bibinfo{title}{A social recommender mechanism for e-commerce:
  Combining similarity, trust, and relationship}.
\newblock \bibinfo{journal}{Decision Support Systems} \bibinfo{volume}{55},
  \bibinfo{pages}{740 -- 752}.
\newblock \URLprefix
  \url{http://www.sciencedirect.com/science/article/pii/S0167923613000705},
  \DOIprefix\doi{10.1016/j.dss.2013.02.009}.
\bibitem[{Loomes et~al.(2017)Loomes, Hull and Mandy}]{loomes2017male}
\bibinfo{author}{Loomes, R.}, \bibinfo{author}{Hull, L.},
  \bibinfo{author}{Mandy, W.P.L.}, \bibinfo{year}{2017}.
\newblock \bibinfo{title}{What is the male-to-female ratio in autism spectrum
  disorder? a systematic review and meta-analysis}.
\newblock \bibinfo{journal}{Journal of the American Academy of Child \&
  Adolescent Psychiatry} \bibinfo{volume}{56}, \bibinfo{pages}{466--474}.
\bibitem[{Lops et~al.(2011)Lops, {de Gemmis} and Semeraro}]{Lops-etal:11}
\bibinfo{author}{Lops, P.}, \bibinfo{author}{{de Gemmis}, M.},
  \bibinfo{author}{Semeraro, G.}, \bibinfo{year}{2011}.
\newblock \bibinfo{title}{Content-based recommender systems: state of the art
  and trends}. \bibinfo{publisher}{Springer US}, \bibinfo{address}{Boston, MA}.
\newblock pp. \bibinfo{pages}{73--105}.
\newblock \URLprefix \url{https://doi.org/10.1007/978-0-387-85820-3\_3},
  \DOIprefix\doi{10.1007/978-0-387-85820-3\_3}.
\bibitem[{Lu et~al.(2018)Lu, Dong and Smyth}]{Lu-etal:18}
\bibinfo{author}{Lu, Y.}, \bibinfo{author}{Dong, R.}, \bibinfo{author}{Smyth,
  B.}, \bibinfo{year}{2018}.
\newblock \bibinfo{title}{Coevolutionary recommendation model: Mutual learning
  between ratings and reviews}, in: \bibinfo{booktitle}{Proceedings of the 2018
  World Wide Web Conference}, \bibinfo{publisher}{International World Wide Web
  Conferences Steering Committee}, \bibinfo{address}{Republic and Canton of
  Geneva, CHE}. p. \bibinfo{pages}{773–782}.
\newblock \URLprefix \url{https://doi.org/10.1145/3178876.3186158},
  \DOIprefix\doi{10.1145/3178876.3186158}.
\bibitem[{Matsushima and Kato(2013)}]{matsushima2013}
\bibinfo{author}{Matsushima, K.}, \bibinfo{author}{Kato, T.},
  \bibinfo{year}{2013}.
\newblock \bibinfo{title}{Social interaction and atypical sensory processing in
  children with autism spectrum disorders}.
\newblock \bibinfo{journal}{Hong Kong Journal of Occupational Therapy}
  \bibinfo{volume}{23}, \bibinfo{pages}{89--96}.
\bibitem[{Mauro et~al.(2020)Mauro, Ardissono and Cena}]{Mauro-etal:20}
\bibinfo{author}{Mauro, N.}, \bibinfo{author}{Ardissono, L.},
  \bibinfo{author}{Cena, F.}, \bibinfo{year}{2020}.
\newblock \bibinfo{title}{Personalized recommendation of {PoIs} to people with
  autism}, in: \bibinfo{booktitle}{Proceedings of the 28th ACM Conference on
  User Modeling, Adaptation and Personalization}, \bibinfo{publisher}{ACM},
  \bibinfo{address}{New York, NY, USA}. pp. \bibinfo{pages}{163--172}.
\newblock \URLprefix \url{https://dl.acm.org/doi/10.1145/3340631.3394845},
  \DOIprefix\doi{10.1145/3340631.3394845}.
\bibitem[{Mauro et~al.(2022)Mauro, Ardissono and Cena}]{Mauro-etal:22}
\bibinfo{author}{Mauro, N.}, \bibinfo{author}{Ardissono, L.},
  \bibinfo{author}{Cena, F.}, \bibinfo{year}{2022}.
\newblock \bibinfo{title}{Supporting people with autism spectrum disorders in
  the exploration of {PoIs}: An inclusive recommender system}.
\newblock \bibinfo{journal}{Communications of the {ACM}} \bibinfo{volume}{65},
  \bibinfo{pages}{101–109}.
\newblock \URLprefix \url{https://doi.org/10.1145/3505267},
  \DOIprefix\doi{10.1145/3505267}.
\bibitem[{Mauro et~al.(2021)Mauro, Ardissono and Petrone}]{Mauro-etal:21}
\bibinfo{author}{Mauro, N.}, \bibinfo{author}{Ardissono, L.},
  \bibinfo{author}{Petrone, G.}, \bibinfo{year}{2021}.
\newblock \bibinfo{title}{User and item-aware estimation of review
  helpfulness}.
\newblock \bibinfo{journal}{Information Processing \& Management}
  \bibinfo{volume}{58}, \bibinfo{pages}{102434}.
\newblock \URLprefix
  \url{http://www.sciencedirect.com/science/article/pii/S0306457320309274},
  \DOIprefix\doi{doi.org/10.1016/j.ipm.2020.102434}.
\bibitem[{McAuley and Leskovec(2013)}]{McAuley-Leskovec:13}
\bibinfo{author}{McAuley, J.}, \bibinfo{author}{Leskovec, J.},
  \bibinfo{year}{2013}.
\newblock \bibinfo{title}{Hidden factors and hidden topics: understanding
  rating dimensions with review text}, in: \bibinfo{booktitle}{Proceedings of
  the 7th ACM Conference on Recommender Systems},
  \bibinfo{publisher}{Association for Computing Machinery},
  \bibinfo{address}{New York, NY, USA}. p. \bibinfo{pages}{165–172}.
\newblock \URLprefix \url{https://doi.org/10.1145/2507157.2507163},
  \DOIprefix\doi{10.1145/2507157.2507163}.
\bibitem[{Murray et~al.(2005)Murray, Lesser and Lawson}]{murray2005attention}
\bibinfo{author}{Murray, D.}, \bibinfo{author}{Lesser, M.},
  \bibinfo{author}{Lawson, W.}, \bibinfo{year}{2005}.
\newblock \bibinfo{title}{Attention, monotropism and the diagnostic criteria
  for autism}.
\newblock \bibinfo{journal}{Autism} \bibinfo{volume}{9},
  \bibinfo{pages}{139--156}.
\newblock \URLprefix \url{https://doi.org/10.1177/1362361305051398},
  \DOIprefix\doi{10.1177/1362361305051398}.
\bibitem[{Musat and Faltings(2015)}]{Musat-Faltings:15}
\bibinfo{author}{Musat, C.C.}, \bibinfo{author}{Faltings, B.},
  \bibinfo{year}{2015}.
\newblock \bibinfo{title}{Personalizing product rankings using collaborative
  filtering on opinion-derived topic profiles}, in:
  \bibinfo{booktitle}{Proceedings of the 24th International Conference on
  Artificial Intelligence}, \bibinfo{publisher}{AAAI Press}. p.
  \bibinfo{pages}{830–836}.
\newblock \URLprefix \url{https://dl.acm.org/doi/10.5555/2832249.2832364}.
\bibitem[{Musto et~al.(2017)Musto, de~Gemmis, Semeraro and
  Lops}]{Musto-etal:17b}
\bibinfo{author}{Musto, C.}, \bibinfo{author}{de~Gemmis, M.},
  \bibinfo{author}{Semeraro, G.}, \bibinfo{author}{Lops, P.},
  \bibinfo{year}{2017}.
\newblock \bibinfo{title}{A multi-criteria recommender system exploiting
  aspect-based sentiment analysis of users' reviews}, in:
  \bibinfo{booktitle}{Proceedings of the Eleventh ACM Conference on Recommender
  Systems}, \bibinfo{publisher}{ACM}, \bibinfo{address}{New York, NY, USA}. pp.
  \bibinfo{pages}{321--325}.
\newblock \URLprefix \url{http://doi.acm.org/10.1145/3109859.3109905},
  \DOIprefix\doi{10.1145/3109859.3109905}.
\bibitem[{Musto et~al.(2011)Musto, Semeraro, Lops and
  de~Gemmis}]{Musto-etal:11}
\bibinfo{author}{Musto, C.}, \bibinfo{author}{Semeraro, G.},
  \bibinfo{author}{Lops, P.}, \bibinfo{author}{de~Gemmis, M.},
  \bibinfo{year}{2011}.
\newblock \bibinfo{title}{Random indexing and negative user preferences for
  enhancing content-based recommender systems}, in: \bibinfo{editor}{Huemer,
  C.}, \bibinfo{editor}{Setzer, T.} (Eds.), \bibinfo{booktitle}{E-Commerce and
  Web Technologies}, \bibinfo{publisher}{Springer Berlin Heidelberg},
  \bibinfo{address}{Berlin, Heidelberg}. pp. \bibinfo{pages}{270--281}.
\newblock \URLprefix
  \url{https://link.springer.com/chapter/10.1007/978-3-642-23014-1\_23},
  \DOIprefix\doi{10.1007/978-3-642-23014-1\_23}.
\bibitem[{Ng and Pera(2018)}]{Ng-Pera:18}
\bibinfo{author}{Ng, Y.}, \bibinfo{author}{Pera, M.}, \bibinfo{year}{2018}.
\newblock \bibinfo{title}{Recommending social-interactive games for adults with
  autism spectrum disorders ({ASD})}, in: \bibinfo{booktitle}{Proceedings of
  the 12th ACM Conference on Recommender Systems}, \bibinfo{publisher}{ACM},
  \bibinfo{address}{New York, NY, USA}. pp. \bibinfo{pages}{209--213}.
\newblock \URLprefix \url{https://dl.acm.org/doi/abs/10.1145/3240323.3240405},
  \DOIprefix\doi{10.1145/3240323.3240405}.
\bibitem[{O'Mahony and Smyth(2018)}]{O'Mahony-Smyth:18}
\bibinfo{author}{O'Mahony, M.P.}, \bibinfo{author}{Smyth, B.},
  \bibinfo{year}{2018}.
\newblock \bibinfo{title}{From opinions to recommendations}, in:
  \bibinfo{editor}{Brusilovsky, P.}, \bibinfo{editor}{He, D.} (Eds.),
  \bibinfo{booktitle}{Social Information Access: Systems and Technologies}.
  \bibinfo{publisher}{Springer International Publishing},
  \bibinfo{address}{Cham}, pp. \bibinfo{pages}{480--509}.
\newblock \URLprefix \url{https://doi.org/10.1007/978-3-319-90092-6\_13},
  \DOIprefix\doi{10.1007/978-3-319-90092-6\_13}.
\bibitem[{Page et~al.(1999)Page, Brin, Motwani and Winograd}]{Page-etal:99}
\bibinfo{author}{Page, L.}, \bibinfo{author}{Brin, S.},
  \bibinfo{author}{Motwani, R.}, \bibinfo{author}{Winograd, T.},
  \bibinfo{year}{1999}.
\newblock \bibinfo{title}{The PageRank Citation Ranking: Bringing Order to the
  Web}.
\newblock \bibinfo{type}{Technical Report} \bibinfo{number}{1999-66}. Stanford
  InfoLab.
\newblock \URLprefix \url{http://ilpubs.stanford.edu:8090/422/}.
  \bibinfo{note}{previous number = SIDL-WP-1999-0120}.
\bibitem[{Paul et~al.(2017)Paul, Sarkar, Chelliah, Kalyan and
  Sinai~Nadkarni}]{Paul-etal:17}
\bibinfo{author}{Paul, D.}, \bibinfo{author}{Sarkar, S.},
  \bibinfo{author}{Chelliah, M.}, \bibinfo{author}{Kalyan, C.},
  \bibinfo{author}{Sinai~Nadkarni, P.P.}, \bibinfo{year}{2017}.
\newblock \bibinfo{title}{Recommendation of high quality representative reviews
  in e-commerce}, in: \bibinfo{booktitle}{Proceedings of the Eleventh ACM
  Conference on Recommender Systems}, \bibinfo{publisher}{Association for
  Computing Machinery}, \bibinfo{address}{New York, NY, USA}. pp.
  \bibinfo{pages}{311--315}.
\newblock \URLprefix \url{https://doi.org/10.1145/3109859.3109901},
  \DOIprefix\doi{10.1145/3109859.3109901}.
\bibitem[{Pe\~{n}a et~al.(2020)Pe\~{n}a, O'Reilly-Morgan, Tragos, Hurley,
  Duriakova, Smyth and Lawlor}]{Pena-etal:20}
\bibinfo{author}{Pe\~{n}a, F.J.}, \bibinfo{author}{O'Reilly-Morgan, D.},
  \bibinfo{author}{Tragos, E.Z.}, \bibinfo{author}{Hurley, N.},
  \bibinfo{author}{Duriakova, E.}, \bibinfo{author}{Smyth, B.},
  \bibinfo{author}{Lawlor, A.}, \bibinfo{year}{2020}.
\newblock \bibinfo{title}{Combining rating and review data by initializing
  latent factor models with topic models for top-n recommendation}, in:
  \bibinfo{booktitle}{Fourteenth ACM Conference on Recommender Systems},
  \bibinfo{publisher}{Association for Computing Machinery},
  \bibinfo{address}{New York, NY, USA}. p. \bibinfo{pages}{438–443}.
\newblock \URLprefix \url{https://doi.org/10.1145/3383313.3412207},
  \DOIprefix\doi{10.1145/3383313.3412207}.
\bibitem[{Premasundari and Yamini(2019)}]{premasundari2019food}
\bibinfo{author}{Premasundari, M.}, \bibinfo{author}{Yamini, C.},
  \bibinfo{year}{2019}.
\newblock \bibinfo{title}{Food and therapy recommendation system for autistic
  syndrome using machine learning techniques}, in: \bibinfo{booktitle}{2019
  IEEE International Conference on Electrical, Computer and Communication
  Technologies (ICECCT)}, \bibinfo{organization}{IEEE}. pp.
  \bibinfo{pages}{1--6}.
\newblock \URLprefix \url{https://doi.org/10.1109/ICECCT.2019.8868979},
  \DOIprefix\doi{10.1109/ICECCT.2019.8868979}.
\bibitem[{Putnam and Chong(2008)}]{10.1145/1414471.1414475}
\bibinfo{author}{Putnam, C.}, \bibinfo{author}{Chong, L.},
  \bibinfo{year}{2008}.
\newblock \bibinfo{title}{Software and technologies designed for people with
  autism: What do users want?}, in: \bibinfo{booktitle}{Proceedings of the 10th
  International ACM SIGACCESS Conference on Computers and Accessibility},
  \bibinfo{publisher}{Association for Computing Machinery},
  \bibinfo{address}{New York, NY, USA}. p. \bibinfo{pages}{3–10}.
\newblock \URLprefix \url{https://doi.org/10.1145/1414471.1414475},
  \DOIprefix\doi{10.1145/1414471.1414475}.
\bibitem[{Qi et~al.(2016)Qi, Zhang, Jeon and Zhou}]{Qi-etal:16}
\bibinfo{author}{Qi, J.}, \bibinfo{author}{Zhang, Z.}, \bibinfo{author}{Jeon,
  S.}, \bibinfo{author}{Zhou, Y.}, \bibinfo{year}{2016}.
\newblock \bibinfo{title}{Mining customer requirements from online reviews: A
  product improvement perspective}.
\newblock \bibinfo{journal}{Information \& Management} \bibinfo{volume}{53},
  \bibinfo{pages}{951 -- 963}.
\newblock \URLprefix
  \url{http://www.sciencedirect.com/science/article/pii/S0378720616300581},
  \DOIprefix\doi{10.1016/j.im.2016.06.002}.
\bibitem[{Qiu et~al.(2011)Qiu, Liu, Bu and Chen}]{Guang:2011}
\bibinfo{author}{Qiu, G.}, \bibinfo{author}{Liu, B.}, \bibinfo{author}{Bu, J.},
  \bibinfo{author}{Chen, C.}, \bibinfo{year}{2011}.
\newblock \bibinfo{title}{Opinion word expansion and target extraction through
  double propagation}.
\newblock \bibinfo{journal}{Computational Linguistics} \bibinfo{volume}{37},
  \bibinfo{pages}{9--27}.
\newblock \DOIprefix\doi{10.1162/coli\_a\_00034}.
\bibitem[{Rapp et~al.(2017)Rapp, Cena, Boella, Antonini, Calafiore, Buccoliero,
  Tirassa, Keller, Castaldo and Brighenti}]{DBLP:conf/chi/RappCBACBTKCB17}
\bibinfo{author}{Rapp, A.}, \bibinfo{author}{Cena, F.},
  \bibinfo{author}{Boella, G.}, \bibinfo{author}{Antonini, A.},
  \bibinfo{author}{Calafiore, A.}, \bibinfo{author}{Buccoliero, S.},
  \bibinfo{author}{Tirassa, M.}, \bibinfo{author}{Keller, R.},
  \bibinfo{author}{Castaldo, R.}, \bibinfo{author}{Brighenti, S.},
  \bibinfo{year}{2017}.
\newblock \bibinfo{title}{Interactive urban maps for people with autism
  spectrum disorder}, in: \bibinfo{booktitle}{Proceedings of the 2017 {CHI}
  Conference on Human Factors in Computing Systems, Denver, CO, USA, May 06-11,
  2017, Extended Abstracts}, pp. \bibinfo{pages}{1987--1992}.
\newblock \URLprefix \url{https://doi.org/10.1145/3027063.3053145},
  \DOIprefix\doi{10.1145/3027063.3053145}.
\bibitem[{Rapp et~al.(2018)Rapp, Cena, Castaldo, Keller and
  Tirassa}]{rapp2018designing}
\bibinfo{author}{Rapp, A.}, \bibinfo{author}{Cena, F.},
  \bibinfo{author}{Castaldo, R.}, \bibinfo{author}{Keller, R.},
  \bibinfo{author}{Tirassa, M.}, \bibinfo{year}{2018}.
\newblock \bibinfo{title}{Designing technology for spatial needs: Routines,
  control and social competences of people with autism}.
\newblock \bibinfo{journal}{International Journal of Human-Computer Studies}
  \bibinfo{volume}{120}, \bibinfo{pages}{49 -- 65}.
\newblock \URLprefix
  \url{http://www.sciencedirect.com/science/article/pii/S1071581918303859},
  \DOIprefix\doi{10.1016/j.ijhcs.2018.07.005}.
\bibitem[{Rapp et~al.(2020)Rapp, Cena, Schifanella and
  Boella}]{rapp2019spatial}
\bibinfo{author}{Rapp, A.}, \bibinfo{author}{Cena, F.},
  \bibinfo{author}{Schifanella, C.}, \bibinfo{author}{Boella, G.},
  \bibinfo{year}{2020}.
\newblock \bibinfo{title}{Finding a secure place: A map-based crowdsourcing
  system for people with autism}.
\newblock \bibinfo{journal}{IEEE Transactions on Human-Machine Systems}
  \bibinfo{volume}{50}, \bibinfo{pages}{424--433}.
\newblock \DOIprefix\doi{10.1109/THMS.2020.2984743}.
\bibitem[{Ricci et~al.(2011)Ricci, Rokach and Shapira}]{Ricci-etal:11}
\bibinfo{author}{Ricci, F.}, \bibinfo{author}{Rokach, L.},
  \bibinfo{author}{Shapira, B.}, \bibinfo{year}{2011}.
\newblock \bibinfo{title}{Introduction to Recommender Systems Handbook}.
  \bibinfo{publisher}{Springer US}, \bibinfo{address}{Boston, MA}.
\newblock pp. \bibinfo{pages}{1--35}.
\newblock \URLprefix \url{https://doi.org/10.1007/978-0-387-85820-3\_1},
  \DOIprefix\doi{10.1007/978-0-387-85820-3\_1}.
\bibitem[{Robertson and Simmons(2013)}]{robertson2013relationship}
\bibinfo{author}{Robertson, A.E.}, \bibinfo{author}{Simmons, D.R.},
  \bibinfo{year}{2013}.
\newblock \bibinfo{title}{The relationship between sensory sensitivity and
  autistic traits in the general population}.
\newblock \bibinfo{journal}{Journal of Autism and Developmental disorders}
  \bibinfo{volume}{43}, \bibinfo{pages}{775--784}.
\newblock \URLprefix \url{https://doi.org/10.1007/s10803-012-1608-7},
  \DOIprefix\doi{10.1007/s10803-012-1608-7}.
\bibitem[{Shalom et~al.(2019)Shalom, Uziel and Kantor}]{Shalom-etal:19}
\bibinfo{author}{Shalom, O.S.}, \bibinfo{author}{Uziel, G.},
  \bibinfo{author}{Kantor, A.}, \bibinfo{year}{2019}.
\newblock \bibinfo{title}{A generative model for review-based recommendations},
  in: \bibinfo{booktitle}{Proceedings of the 13th ACM Conference on Recommender
  Systems}, \bibinfo{publisher}{Association for Computing Machinery},
  \bibinfo{address}{New York, NY, USA}. p. \bibinfo{pages}{353–357}.
\newblock \URLprefix \url{https://doi.org/10.1145/3298689.3347061},
  \DOIprefix\doi{10.1145/3298689.3347061}.
\bibitem[{Simm et~al.(2016)Simm, Ferrario, Gradinar, Tavares~Smith, Forshaw,
  Smith and Whittle}]{simm2016anxiety}
\bibinfo{author}{Simm, W.}, \bibinfo{author}{Ferrario, M.A.},
  \bibinfo{author}{Gradinar, A.}, \bibinfo{author}{Tavares~Smith, M.},
  \bibinfo{author}{Forshaw, S.}, \bibinfo{author}{Smith, I.},
  \bibinfo{author}{Whittle, J.}, \bibinfo{year}{2016}.
\newblock \bibinfo{title}{Anxiety and autism: Towards personalized digital
  health}, in: \bibinfo{booktitle}{Proceedings of the 2016 CHI Conference on
  Human Factors in Computing Systems}, \bibinfo{publisher}{Association for
  Computing Machinery}, \bibinfo{address}{New York, NY, USA}. p.
  \bibinfo{pages}{1270–1281}.
\newblock \URLprefix \url{https://doi.org/10.1145/2858036.2858259},
  \DOIprefix\doi{10.1145/2858036.2858259}.
\bibitem[{Smith(2015)}]{smith2015spatial}
\bibinfo{author}{Smith, A.D.}, \bibinfo{year}{2015}.
\newblock \bibinfo{title}{Spatial navigation in autism spectrum disorders: a
  critical review}.
\newblock \bibinfo{journal}{Frontiers in Psychology} \bibinfo{volume}{6},
  \bibinfo{pages}{31}.
\newblock \URLprefix
  \url{https://www.frontiersin.org/article/10.3389/fpsyg.2015.00031},
  \DOIprefix\doi{10.3389/fpsyg.2015.00031}.
\bibitem[{Soccini et~al.(2020)Soccini, Cuccurullo and
  Cena}]{DBLP:conf/eurovr/SocciniCC20}
\bibinfo{author}{Soccini, A.M.}, \bibinfo{author}{Cuccurullo, S.A.G.},
  \bibinfo{author}{Cena, F.}, \bibinfo{year}{2020}.
\newblock \bibinfo{title}{Virtual reality experiential training for individuals
  with autism: The airport scenario}, in: \bibinfo{booktitle}{Proceedings of
  the 17th EuroVR International Conference, EuroVR 2020},
  \bibinfo{publisher}{Springer}. pp. \bibinfo{pages}{234--239}.
\newblock \URLprefix \url{https://doi.org/10.1007/978-3-030-62655-6\_16},
  \DOIprefix\doi{10.1007/978-3-030-62655-6\_16}.
\bibitem[{Sui et~al.(2013)Sui, Elwood and Goodchild}]{sui2012crowdsourcing}
\bibinfo{author}{Sui, D.}, \bibinfo{author}{Elwood, S.},
  \bibinfo{author}{Goodchild, M.}, \bibinfo{year}{2013}.
\newblock \bibinfo{title}{Crowdsourcing geographic knowledge: volunteered
  geographic information (VGI) in theory and practice}.
\newblock \bibinfo{publisher}{Springer}.
\newblock \DOIprefix\doi{10.1007/978-94-007-4587-2}.
\bibitem[{Tang et~al.(2019)Tang, Fu, Yao and Xu}]{Tang-etal:19}
\bibinfo{author}{Tang, F.}, \bibinfo{author}{Fu, L.}, \bibinfo{author}{Yao,
  B.}, \bibinfo{author}{Xu, W.}, \bibinfo{year}{2019}.
\newblock \bibinfo{title}{Aspect based fine-grained sentiment analysis for
  online reviews}.
\newblock \bibinfo{journal}{Information Sciences} \bibinfo{volume}{488},
  \bibinfo{pages}{190 -- 204}.
\newblock \URLprefix
  \url{http://www.sciencedirect.com/science/article/pii/S0020025519301872},
  \DOIprefix\doi{10.1016/j.ins.2019.02.064}.
\bibitem[{Tavassoli et~al.(2014)Tavassoli, Hoekstra and
  Baron-Cohen}]{tavassoli_sensoriality}
\bibinfo{author}{Tavassoli, T.}, \bibinfo{author}{Hoekstra, R.A.},
  \bibinfo{author}{Baron-Cohen, S.}, \bibinfo{year}{2014}.
\newblock \bibinfo{title}{The sensory perception quotient ({SPQ}): Development
  and validation of a new sensory questionnaire for adults with and without
  autism}.
\newblock \bibinfo{journal}{Molecular Autism} \bibinfo{volume}{5},
  \bibinfo{pages}{29}.
\newblock \URLprefix \url{https://doi.org/10.1186/2040-2392-5-29},
  \DOIprefix\doi{10.1186/2040-2392-5-29}.
\bibitem[{Xiong et~al.(2020a)Xiong, Shen, Chen, Pan, Wang and
  Yan}]{Xiong-etal:20b}
\bibinfo{author}{Xiong, F.}, \bibinfo{author}{Shen, W.}, \bibinfo{author}{Chen,
  H.}, \bibinfo{author}{Pan, S.}, \bibinfo{author}{Wang, X.},
  \bibinfo{author}{Yan, Z.}, \bibinfo{year}{2020}a.
\newblock \bibinfo{title}{Exploiting implicit influence from information
  propagation for social recommendation}.
\newblock \bibinfo{journal}{IEEE Transactions on Cybernetics}
  \bibinfo{volume}{50}, \bibinfo{pages}{4186--4199}.
\newblock \DOIprefix\doi{10.1109/TCYB.2019.2939390}.
\bibitem[{Xiong et~al.(2020b)Xiong, Wang, Pan, Yang, Wang and
  Zhang}]{Xiong-etal:20}
\bibinfo{author}{Xiong, F.}, \bibinfo{author}{Wang, X.}, \bibinfo{author}{Pan,
  S.}, \bibinfo{author}{Yang, H.}, \bibinfo{author}{Wang, H.},
  \bibinfo{author}{Zhang, C.}, \bibinfo{year}{2020}b.
\newblock \bibinfo{title}{Social recommendation with evolutionary opinion
  dynamics}.
\newblock \bibinfo{journal}{IEEE Transactions on Systems, Man, and Cybernetics:
  Systems} \bibinfo{volume}{50}, \bibinfo{pages}{3804--3816}.
\newblock \DOIprefix\doi{10.1109/TSMC.2018.2854000}.
\bibitem[{Xu et~al.(2017)Xu, Wang, Li and Haghighi}]{Xu-etal:17}
\bibinfo{author}{Xu, X.}, \bibinfo{author}{Wang, X.}, \bibinfo{author}{Li, Y.},
  \bibinfo{author}{Haghighi, M.}, \bibinfo{year}{2017}.
\newblock \bibinfo{title}{Business intelligence in online customer textual
  reviews: understanding consumer perceptions and influential factors}.
\newblock \bibinfo{journal}{International Journal of Information Management}
  \bibinfo{volume}{37}, \bibinfo{pages}{673 -- 683}.
\newblock \URLprefix
  \url{http://www.sciencedirect.com/science/article/pii/S0268401217301378},
  \DOIprefix\doi{10.1016/j.ijinfomgt.2017.06.004}.
\bibitem[{Zhao et~al.(2015)Zhao, McAuley and King}]{Zhao-etal:15}
\bibinfo{author}{Zhao, T.}, \bibinfo{author}{McAuley, J.},
  \bibinfo{author}{King, I.}, \bibinfo{year}{2015}.
\newblock \bibinfo{title}{Improving latent factor models via personalized
  feature projection for one class recommendation}, in:
  \bibinfo{booktitle}{Proceedings of the 24th ACM International on Conference
  on Information and Knowledge Management}, \bibinfo{publisher}{ACM},
  \bibinfo{address}{New York, NY, USA}. pp. \bibinfo{pages}{821--830}.
\newblock \URLprefix \url{http://doi.acm.org/10.1145/2806416.2806511},
  \DOIprefix\doi{10.1145/2806416.2806511}.
\bibitem[{Zheng(2017)}]{Zheng:17}
\bibinfo{author}{Zheng, Y.}, \bibinfo{year}{2017}.
\newblock \bibinfo{title}{Criteria chains: a novel multi-criteria
  recommendation approach}, in: \bibinfo{booktitle}{Proceedings of the 22Nd
  International Conference on Intelligent User Interfaces},
  \bibinfo{publisher}{ACM}, \bibinfo{address}{New York, NY, USA}. pp.
  \bibinfo{pages}{29--33}.
\newblock \URLprefix \url{http://doi.acm.org/10.1145/3025171.3025215},
  \DOIprefix\doi{10.1145/3025171.3025215}.

\end{thebibliography}

\end{document}